\documentclass[12pt]{iopart}

\usepackage{iopams}
\usepackage{graphicx}
\usepackage{mathrsfs}
\usepackage{color}
\usepackage[table]{xcolor}
\usepackage{hyperref}
\hypersetup{colorlinks, citecolor=black, filecolor=black, linkcolor=black, urlcolor=black}

\expandafter\let\csname equation*\endcsname\relax
\expandafter\let\csname endequation*\endcsname\relax
\usepackage{amsmath}

\definecolor{grayish}{RGB}{230,230,230}

\newcommand{\refEq}[1] {(\ref{#1})}
\newcommand{\superscript}[1]{\ensuremath{^{\textrm{#1}}}}

\begin{document}

\title{Intrinsic momentum transport in up-down asymmetric tokamaks}

\author{Justin Ball\superscript{1,2}, Felix I. Parra\superscript{1,2}, Michael Barnes\superscript{2,3}, William Dorland\superscript{4}, Gregory W. Hammett\superscript{5}, Paulo Rodrigues\superscript{6}, and Nuno F. Loureiro\superscript{6}}

\address{\superscript{1} Rudolf Peierls Centre for Theoretical Physics, Oxford University, Oxford, OX1 3NP, UK}
\address{\superscript{2} Plasma Science and Fusion Center, Massachusetts Institute of Technology, Cambridge, MA 02139, USA}
\address{\superscript{3} Institute for Fusion Studies, University of Texas, Austin, TX 78712, USA}
\address{\superscript{4} Department of Physics, University of Maryland, College Park, MD 20742, USA}
\address{\superscript{5} Princeton Plasma Physics Laboratory, Princeton University, P.O. Box 451, Princeton, NJ 08543, USA}
\address{\superscript{6} {Instituto de Plasmas e Fus\~{a}o Nuclear, Instituto Superior T\'{e}cnico, Universidade de Lisboa, 1049-001 Lisboa, Portugal}}
\ead{Justin.Ball@physics.ox.ac.uk}

\begin{abstract}

Recent work demonstrated that breaking the up-down symmetry of tokamak flux surfaces removes a constraint that limits intrinsic momentum transport, and hence toroidal rotation, to be small. We show, through MHD analysis, that ellipticity is most effective at introducing up-down asymmetry throughout the plasma. We detail an extension to GS2, a local $\delta f$ gyrokinetic code that self-consistently calculates momentum transport, to permit up-down asymmetric configurations. Tokamaks with tilted elliptical poloidal cross-sections were simulated to determine nonlinear momentum transport. The results, which are consistent with experiment in magnitude, suggest that a toroidal velocity gradient, $\left( \partial u_{\zeta i} / \partial \rho \right) / v_{th i}$, of 5\% of the temperature gradient, $\left(\partial T_{i} / \partial \rho \right) / T_{i}$, is sustainable. Here $v_{th i}$ is the ion thermal speed, $u_{\zeta i}$ is the ion toroidal mean flow, $\rho$ is the minor radial coordinate normalized to the tokamak minor radius, and $T_{i}$ is the ion temperature. Though other known core intrinsic momentum transport mechanisms scale poorly to larger machines, these results indicate that up-down asymmetry may be a feasible method to generate the current experimentally-measured rotation levels in reactor-sized devices.

\end{abstract}



\section{Introduction}
\label{sec:introduction}

Due to the symmetry of the tokamak, the plasma flow is constrained to be purely toroidal to lowest order in $\rho_{\ast} \equiv \rho_{i}/a$, the ratio of the ion gyroradius to the minor radius \cite{HintonToroidalRotation1985,CattoToroidalRotation1987}. This toroidal rotation has been experimentally proven \cite{StraitExpRWMstabilizationD3D1995,SabbaghExpRWMstabilizationNSTX2002,deVriesRotMHDStabilization1996,ReimerdesRWMmachineComp2006} to improve MHD stability by stabilizing the resistive wall mode. It has enabled sustained, reproducible plasmas that exceed the Troyon beta limit \cite{TroyonMHDLimit1984} by a factor of two \cite{GarofaloExpRWMstabilizationD3D2002}. This is important because the Troyon limit determines the maximum fusion power at a given minor radius, plasma current, and on-axis magnetic field. Additionally, one of the most promising strategies to reduce turbulent energy transport and increase energy confinement time relies on toroidal velocity, $u_{\zeta}$. Experiments \cite{RitzRotShearTurbSuppression1990,BurrellShearTurbStabilization1997} and theory \cite{BarnesFlowShear2011,HighcockRotationBifurcation2010,ParraMomentumTransitions2011,HighcockManifold2012} show that plasmas with a gradient in toroidal velocity, also called toroidal velocity shear, can exhibit a significant reduction in turbulence.

Toroidal rotation can be generated in a number of ways. Neutral particle beams are frequently used to heat the plasma, but can also generate rotation if injected toroidally \cite{SuckewerRotationNBI1979}. Similarly lower hybrid waves, primarily used to noninductively drive current, can induce rotation \cite{InceLowHybridRotation2009}. Both of these methods represent an external injection of momentum, however they are not expected to scale well to large devices \cite{JardinARIESATphysicsBasis2006}. It is unclear if the external momentum injection on ITER and future power plants will induce enough rotation to stabilize the resistive wall mode \cite{LiuITERrwmStabilization2004}.

An attractive alternative is intrinsic rotation, which refers to rotation that is observed in the absence of any external injection of momentum. The plasma can move momentum between flux surfaces, creating a nonzero rotation profile from an initially stationary state, as well as push off the vacuum vessel and external coils. This rotation comes for free, but it is poorly understood and measurements in current experiments reveal it to be rather small, often less than a tenth of the plasma sound speed \cite{RiceExpIntrinsicRotMeas2007}. Theoretically, in a conventional up-down symmetric tokamak, intrinsic rotation is constrained to be small in $\rho_{\ast}$, meaning it is on the order of the ion diamagnetic speed \cite{ParraUpDownSym2011,SugamaUpDownSym2011,CamenenPRLSim2009}. However, up-down asymmetry breaks this constraint and allows rotation to lowest order in $\rho_{\ast}$, permitting background flow velocities on the order of the sound speed.

The only other known mechanisms that break this constraint to lowest order are large, preexisting rotation or rotation shear \cite{PeetersMomTransOverview2011}. All other effects, such as background profile variation \cite{CamenenRadialVariation2011, WaltzMomRadialProfileVar2011, SungMomParallelGrad2013} and neoclassical flows \cite{ParraLowFlowMomTransport2010, ParraIntrinsicRotSources2011}, generate intrinsic rotation to next order in $\rho_{\ast}$. Therefore, unless a feasible method of scaling external momentum injection to reactor-sized devices is found, up-down asymmetry appears to be the most promising option. One caveat is that near the edge the ion gyroradius can be comparable to the background gradient scale length, meaning the $\rho_{\ast}$ scaling argument breaks down and formally small rotation drives may be larger than expected. Still, there is some evidence to suggest that the momentum flux near the edge may scale with $\rho_{\ast}$ \cite{LeeDiamagneticMom2014}.

Initial quasilinear gyrokinetic estimates of achievable rotation levels have been made for the up-down asymmetry present in existing tokamaks \cite{CamenenPRLSim2009,CamenenPhysPlasmas2009}. However, this work \cite{BallMastersThesis2013} will analyze the equilibrium and nonlinear momentum transport in new tokamak configurations that have been chosen to try to maximize rotation.

This section motivates investigation into up-down asymmetric configurations. Then, Section \ref{sec:MHD} presents the results of MHD equilibrium analysis, which demonstrates that the toroidal current distribution within the plasma has a significant effect on the flux surface shape. It is shown that hollow current profiles are optimal for supporting up-down asymmetry near the magnetic axis \cite{RodriguesMHDupDownAsym}. Furthermore, ellipticity, the lowest harmonic shaping effect, penetrates to the magnetic axis most effectively.

Section \ref{sec:GS2} details the necessary modifications to GS2 \cite{KotschenreutherGS21995}, a local $\delta f$ gyrokinetic code that self-consistently calculates momentum transport, to correctly simulate up-down asymmetric tokamak configurations. In Section \ref{sec:momTransport}, this modified code is applied to model the turbulent momentum transport in tilted elliptical tokamaks. The effects of tilting elliptical flux surfaces on turbulent energy transport is still unclear and is left for future investigation. However, the results of nonlinear gyrokinetic momentum flux simulations approximately agree with TCV experimental results \cite{CamenenTCVExp2010}. The velocity shear, $1/v_{th i} \left( \partial u_{\zeta i} / \partial \rho \right)$, inferred from assuming diffusive transport is approximately 5\% of $1/T_{i} \left(\partial T_{i} / \partial \rho \right)$ for elliptical flux surfaces with a $\pi/8$ tilt. Here $v_{th i}$ is the ion thermal velocity, $\rho$ is the normalized minor radial coordinate, and $T_{i}$ is the ion temperature. The introduction of this tilt in TCV was enough to change to core rotation by over 50\% \cite{CamenenTCVExp2010}. In larger tokamaks $\rho_{\ast}$ is smaller, so all sources of intrinsic rotation except up-down asymmetry should diminish. This means that, in a reactor with a $\pi/8$ tilt, we would still expect $1/v_{th i} \left( \partial u_{\zeta i} / \partial \rho \right)$ to be approximately 5\% of $1/T_{i} \left(\partial T_{i} / \partial \rho \right)$, which means the effects of up-down asymmetry would dominate the rotation profile. In a reactor, up-down asymmetry is a possible means to obtain intrinsic rotation levels similar to those observed in current experiments.

\section{Up-down asymmetric MHD equilibrium}
\label{sec:MHD}

Since we are ultimately interested in achieving high levels of intrinsic rotation in fusion devices, we should start by identifying practical up-down asymmetric configurations. To do this, we will use the ideal MHD model \cite{FreidbergIdealMHD1987} to find equilibrium geometries that maximize up-down asymmetry. Since external Poloidal Field (PF) coils set the shape of the outermost closed flux surface, it is a free parameter. However, we must determine if up-down asymmetry introduced at the edge effectively propagates through the plasma to the magnetic axis.

\subsection{Expansion of the Grad-Shafranov equation}

To determine how the flux surface shape changes within the tokamak, we begin by writing the Grad-Shafranov equation \cite{GradGradShafranovEq1958,ShafranovGradShafranovEq1966}
\begin{eqnarray}
   R^{2} \vec{\nabla} \cdot \left( \frac{\vec{\nabla} \psi}{R^{2}} \right) = -\mu_{0} R^{2} \frac{dp}{d \psi} - I \frac{d I}{d \psi} , \label{eq:gradShafranov}
\end{eqnarray}
where $I \equiv R B_{\zeta}$ and the plasma pressure, $p$, are free flux functions to be specified in this calculation. We note that $R$ is the major radial coordinate, $\zeta$ is the toroidal angle, $\psi$ is the poloidal magnetic flux divided by $2 \pi$, $\mu_{0}$ is the vacuum permeability, $\vec{B}_{p} = \vec{\nabla} \zeta \times \vec{\nabla} \psi$ is the poloidal magnetic field, and $\vec{B} = I \vec{\nabla} \zeta + \vec{B}_{p}$. Though there has been work on general \cite{KuiroukidisAnalyticGradShaf2012} and up-down asymmetric \cite{RodriguesMHDupDownAsym} solutions to the Grad-Shafranov equation, we only want simple, approximate solutions to several specific cases to develop our intuition. Thus, we take the orderings in the inverse aspect ratio, $\epsilon \equiv a / R_{0} \ll 1$, typical for an ohmically heated tokamak \cite{FreidbergIdealMHD1987pg126}
\begin{eqnarray}
   \frac{B_{p}}{B_{0}} \sim \epsilon, \hspace{10pt}
   \frac{2 \mu_{0} p}{B_{0}^{2}} \sim \epsilon^{2} , \label{eq:gradShafranovOrderings}
\end{eqnarray}
where $B_{0}$ is the on-axis toroidal magnetic field.

Next we must expand $\psi = \psi_{0} + \psi_{1} + \ldots$, $I = I_{0} + I_{1} + I_{2} + \ldots$, and $p = p_{2} + \ldots$, where $I_{0} = R_{0} B_{0}$ is a constant. Each subscript indicates the quantity's order in $\epsilon$. We also let $\psi_{0} \sim a R_{0} B_{p}$, $R = R_{0} + R_{1}$, and $R_{1} = r \cos \left( \theta \right)$, where $r$ is the toroidal minor radius and $\theta$ is the poloidal angle measured from the outboard midplane. We find from the $O \left( \epsilon^{- 1} B_{0} \right)$ Grad-Shafranov equation that $I_{1} = 0$. Consequently, to $O \left( B_{0} \right)$, the Grad-Shafranov equation becomes
\begin{eqnarray}
   \frac{1}{r} \frac{\partial }{\partial r} \left( r \frac{\partial \psi_{0}}{\partial r} \right) + \frac{1}{r^{2}} \frac{\partial^{2} \psi_{0}}{\partial \theta^{2}} = -\mu_{0} R_{0}^{2} \frac{dp_{2}}{d \psi_{0}} - I_{0} \frac{dI_{2}}{d \psi_{0}} . \label{eq:gradShafranovNextOrder}
\end{eqnarray}

\subsection{Solutions to the $O \left( B_{0} \right)$ Grad-Shafranov equation}
\label{subsec:lowestOrderGradShafSol}

\begin{figure}
 \centering
 \includegraphics[width=0.6\textwidth]{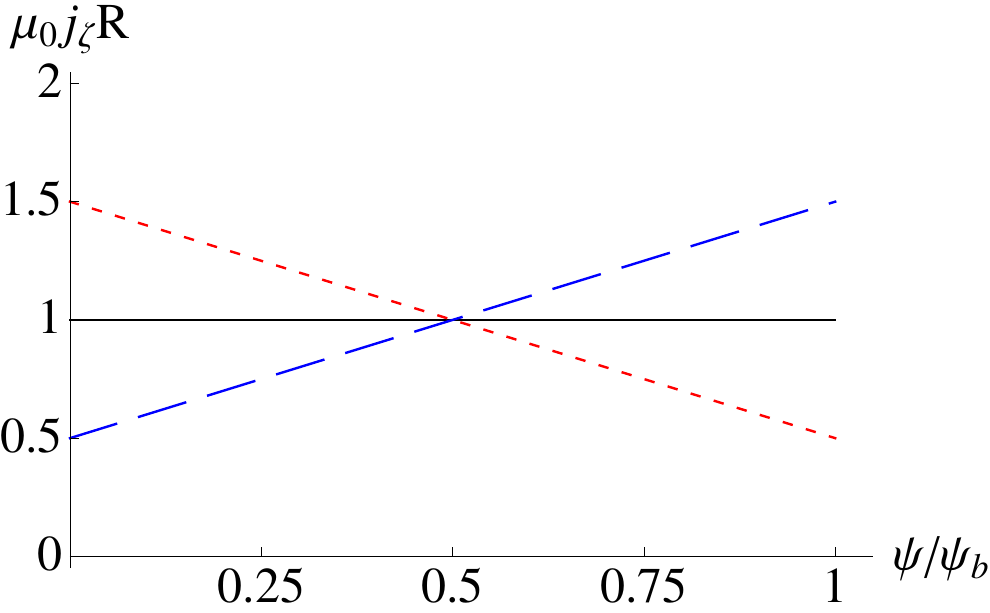}
 \caption{Normalized radial profiles of the plasma current used to produce the constant (black, solid), linear hollow (blue, dashed), and linear peaked  (red, dotted) flux surface shapes, where $\psi_{b}$ is the poloidal magnetic flux at the plasma boundary.}
 \label{fig:gradShafCurrentProfiles}
\end{figure}

The left side of eq. \refEq{eq:gradShafranovNextOrder} is solved by cylindrical harmonics. Furthermore, since $p_{2}$ and $I_{2}$ are free flux functions, we can choose them to get simple forms for the right side of eq. \refEq{eq:gradShafranovNextOrder} and still illuminate the physics of the problem. Using Ampere's law and $\vec{B} = I \vec{\nabla} \zeta + \vec{\nabla} \zeta \times \vec{\nabla} \psi$, one can show that the right side is related to the toroidal current as
\begin{align}
   -\mu_{0} R^{2} \frac{d p}{d \psi} - I \frac{d I}{d \psi} = \mu_{0} j_{\zeta} R .
\end{align}
So we will choose to study a constant toroidal current profile $\mu_{0} j_{\zeta} R_{0} = A$, a linear hollow profile $\mu_{0} j_{\zeta} R_{0} = A_{h} + A'_{h} \psi_{0}$, and a linear peaked profile $\mu_{0} j_{\zeta} R_{0} = A_{c} - A'_{c} \psi_{0}$, where $A$, $A_{h}$, $A'_{h}$, $A_{c}$, and $A'_{c}$ are positive constants of our choosing (see fig. \ref{fig:gradShafCurrentProfiles}). Then, eq. \refEq{eq:gradShafranovNextOrder} becomes
\begin{eqnarray} \label{eq:gradShafranovNextOrderCases}
   \frac{1}{r} \frac{\partial }{\partial r} \left( r \frac{\partial \psi_{0}}{\partial r} \right) + \frac{1}{r^{2}} \frac{\partial^{2} \psi_{0}}{\partial \theta^{2}} &= A , \\
   \frac{1}{r} \frac{\partial }{\partial r} \left( r \frac{\partial \psi_{0}}{\partial r} \right) + \frac{1}{r^{2}} \frac{\partial^{2} \psi_{0}}{\partial \theta^{2}} &= A_{h} + A'_{h} \psi_{0} , \\
   \frac{1}{r} \frac{\partial }{\partial r} \left( r \frac{\partial \psi_{0}}{\partial r} \right) + \frac{1}{r^{2}} \frac{\partial^{2} \psi_{0}}{\partial \theta^{2}} &= A_{c} - A'_{c} \psi_{0}
\end{eqnarray}
for each case respectively. These equations are solved by
\begin{eqnarray}
   \psi_{0} \left( r, \theta \right) &= \frac{A}{4} r^{2} + \sum _{m=0}^{\infty} r^{m} \left( C_{m} \cos \left( m \theta \right) + D_{m} \sin \left( m \theta \right) \right) \label{eq:gradShafranovNextOrderSolsConst} , \\
   \psi_{0} \left( r, \theta \right) &= \frac{A_{h}}{A'_{h}} \left( I_{0} \left( \sqrt{A'_{h}} r \right) - 1 \right) \nonumber \\
   &+ \sum _{m=0}^{\infty} I_{m} \left( \sqrt{A'_{h}} r \right) \left( C_{h m} \cos \left( m \theta \right) + D_{h m} \sin \left( m \theta \right) \right) \label{eq:gradShafranovNextOrderSolsHollow} , \\
   \psi_{0} \left( r, \theta \right) &= - \frac{A_{c}}{A'_{c}} \left( J_{0} \left( \sqrt{A'_{c}} r \right) - 1 \right) \nonumber \\
   &+ \sum _{m=0}^{\infty} J_{m} \left( \sqrt{A'_{c}} r \right) \left( C_{c m} \cos \left( m \theta \right) + D_{c m} \sin \left( m \theta \right) \right) \label{eq:gradShafranovNextOrderSolsPeaked} ,
\end{eqnarray}
respectively. Here $m$ is the poloidal mode number, $J_{m}$ is the m\superscript{th} Bessel function of the first kind, and $I_{m}$ is the m\superscript{th} modified Bessel function of the first kind. The coefficients $C_{m}$, $D_{m}$, $C_{h m}$, $D_{h m}$, $C_{c m}$, and $D_{c m}$ are Fourier harmonic coefficients determined by the boundary condition at the plasma edge. Note that, close enough to the magnetic axis, any toroidal current profile can be considered a constant, meaning the solution reduces to the constant current case.

\begin{figure}
 \centering
 (a) \hspace{0.95\textwidth}

 \includegraphics[width=0.3\textwidth]{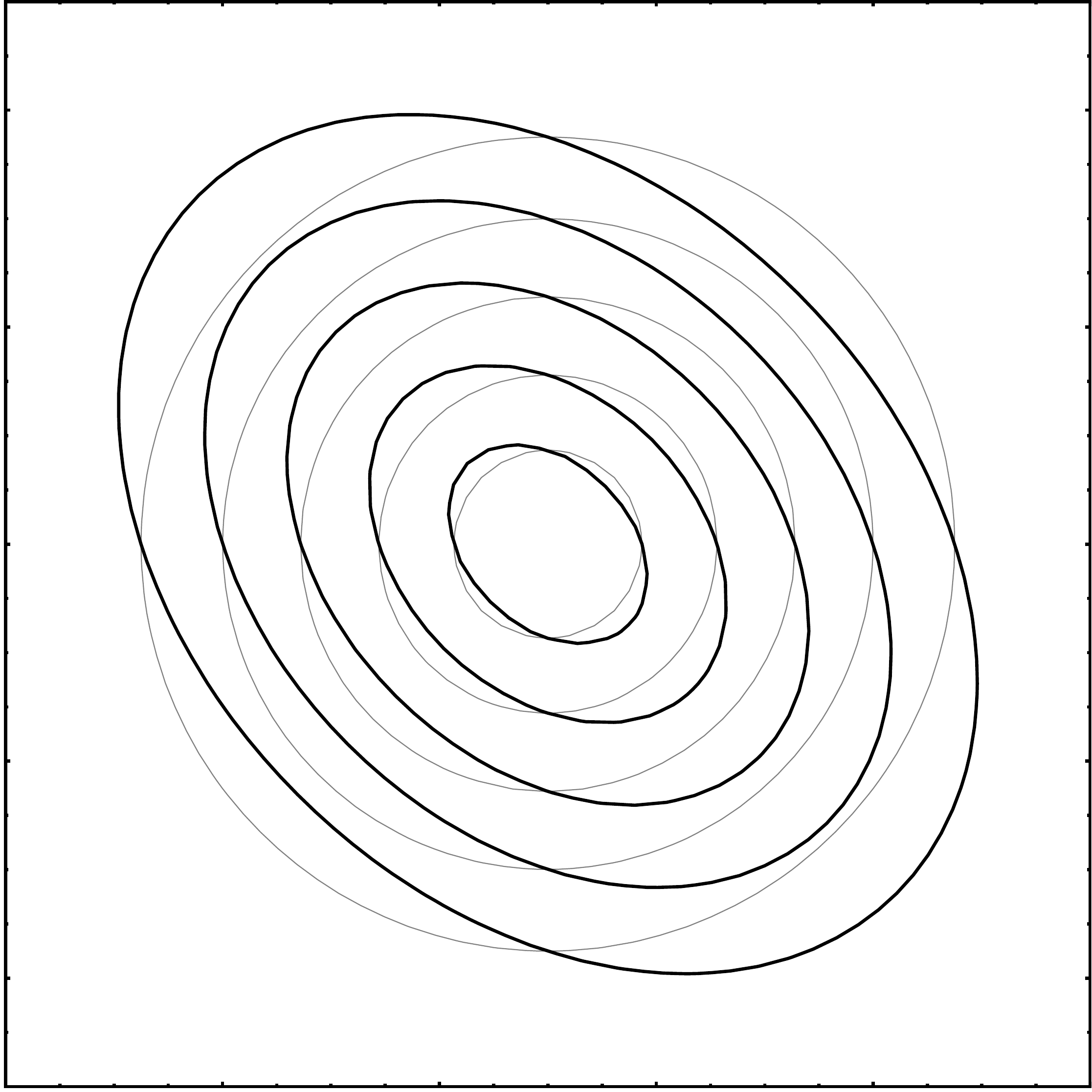}
 \includegraphics[width=0.3\textwidth]{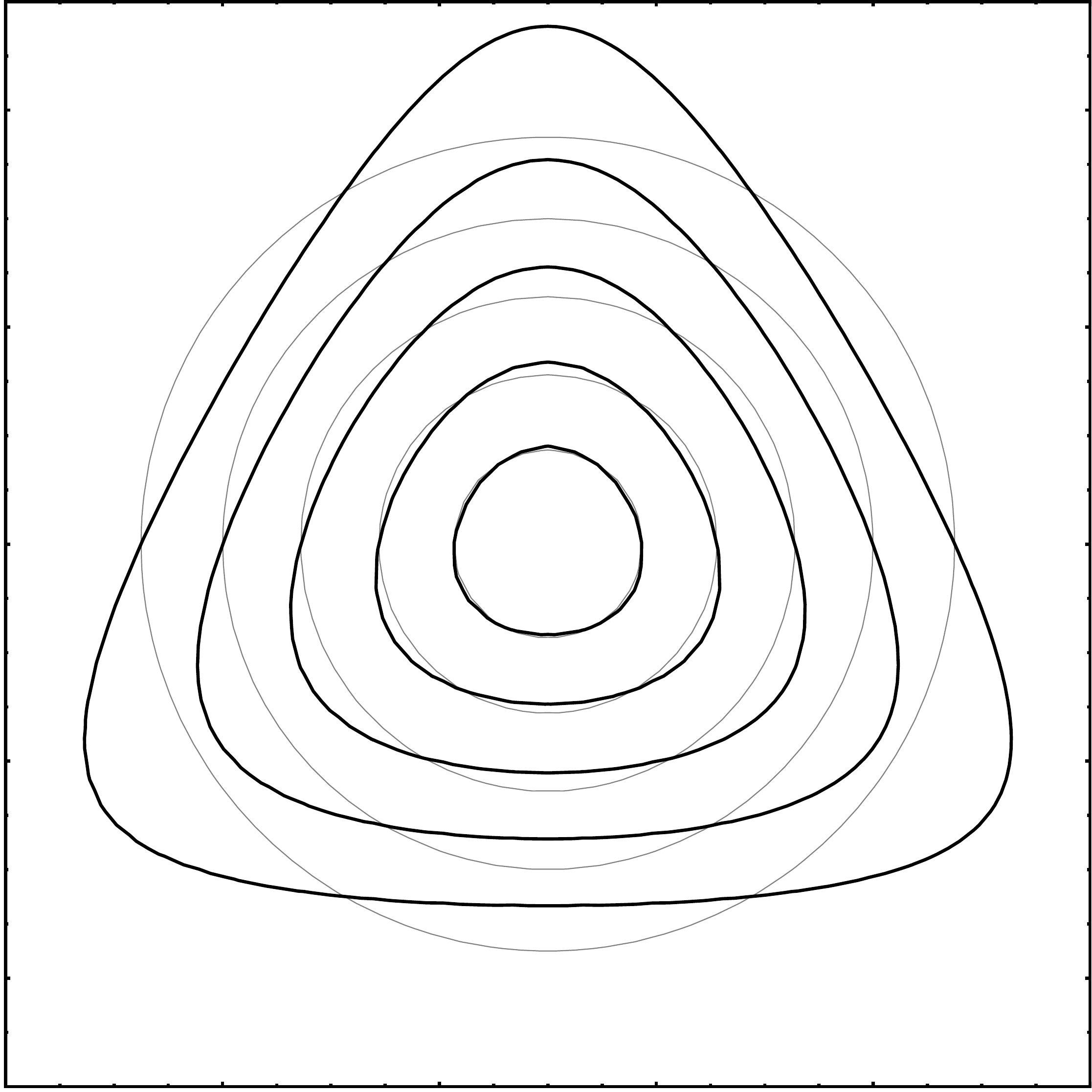}
 \includegraphics[width=0.3\textwidth]{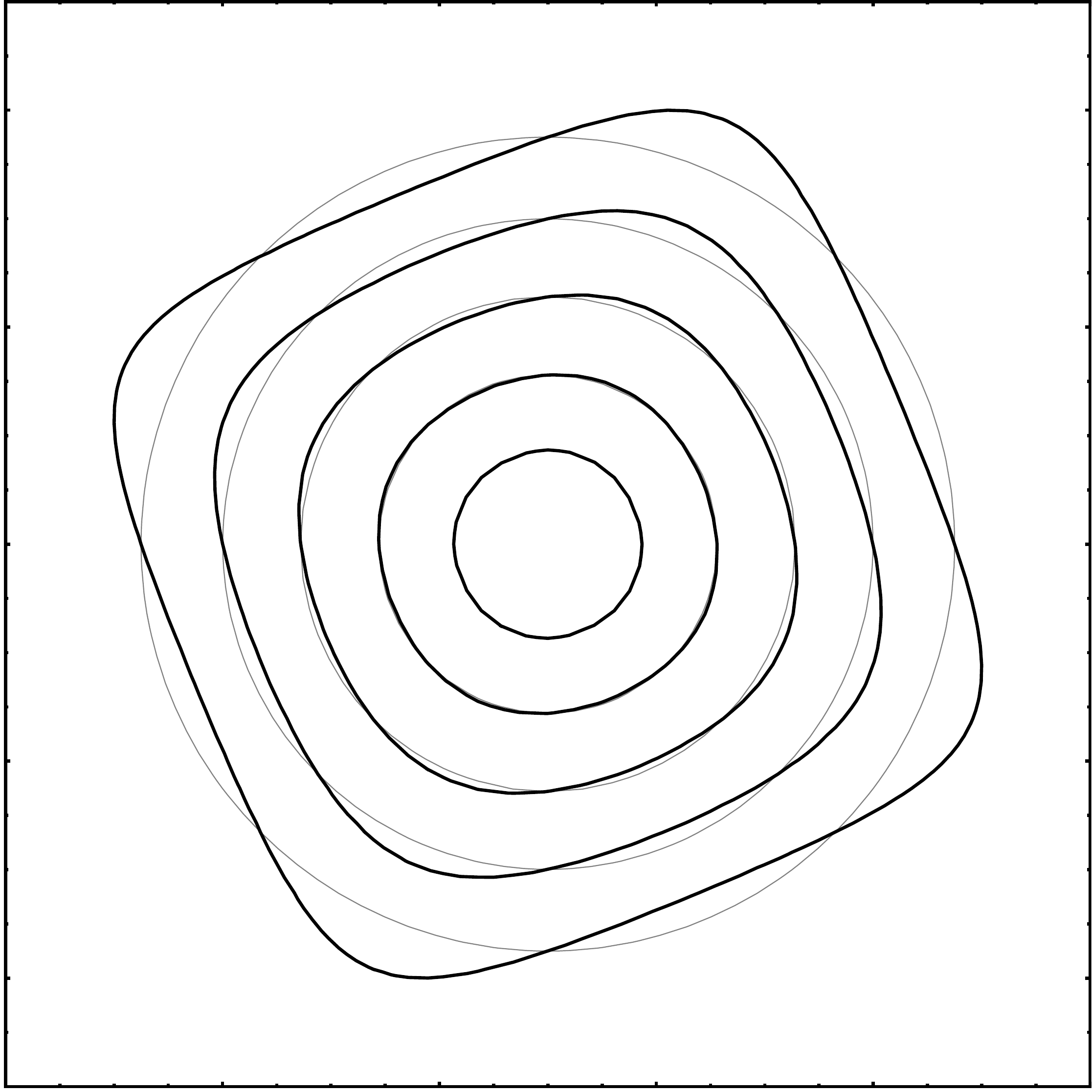}

 (b) \hspace{0.95\textwidth}

 \includegraphics[width=0.3\textwidth]{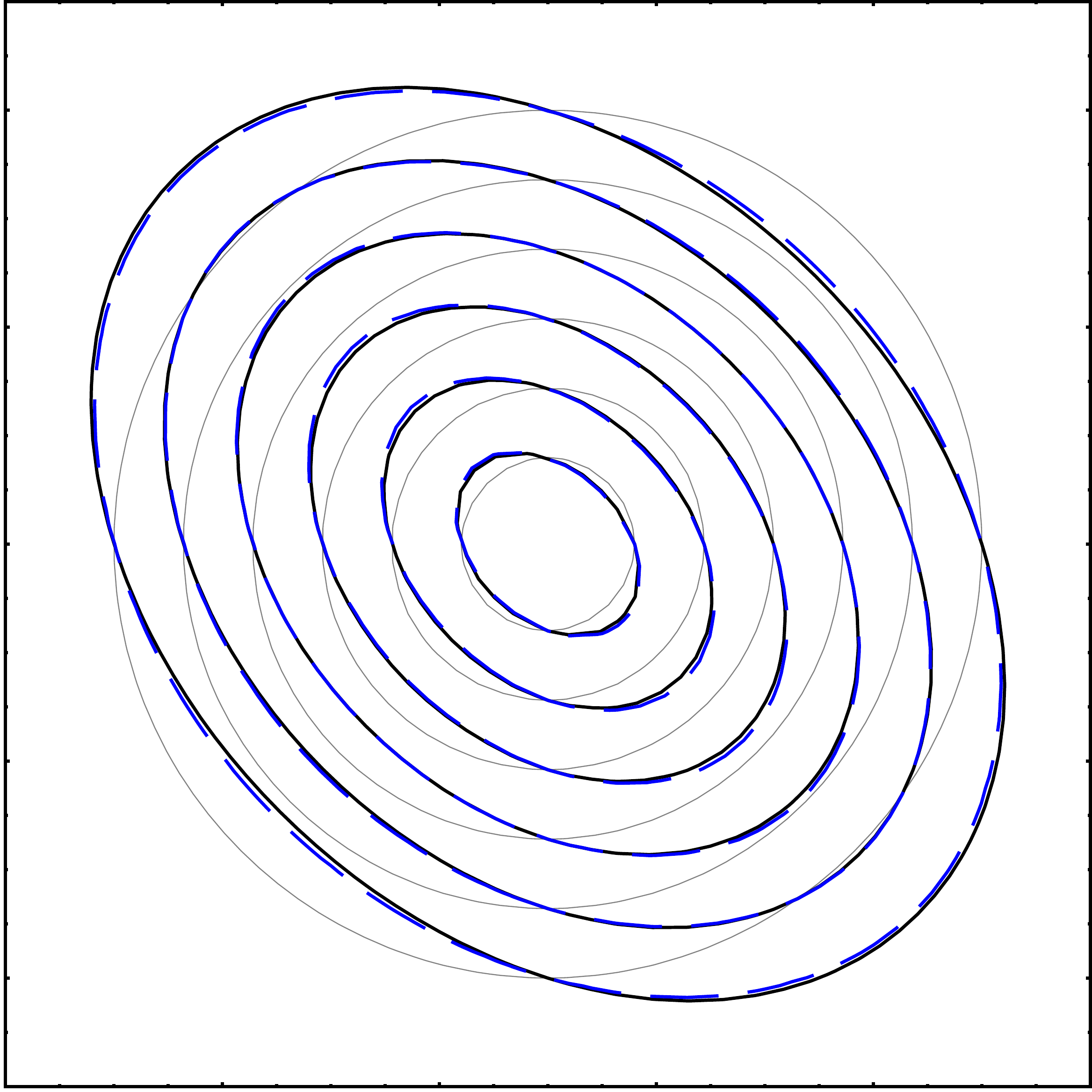}
 \includegraphics[width=0.3\textwidth]{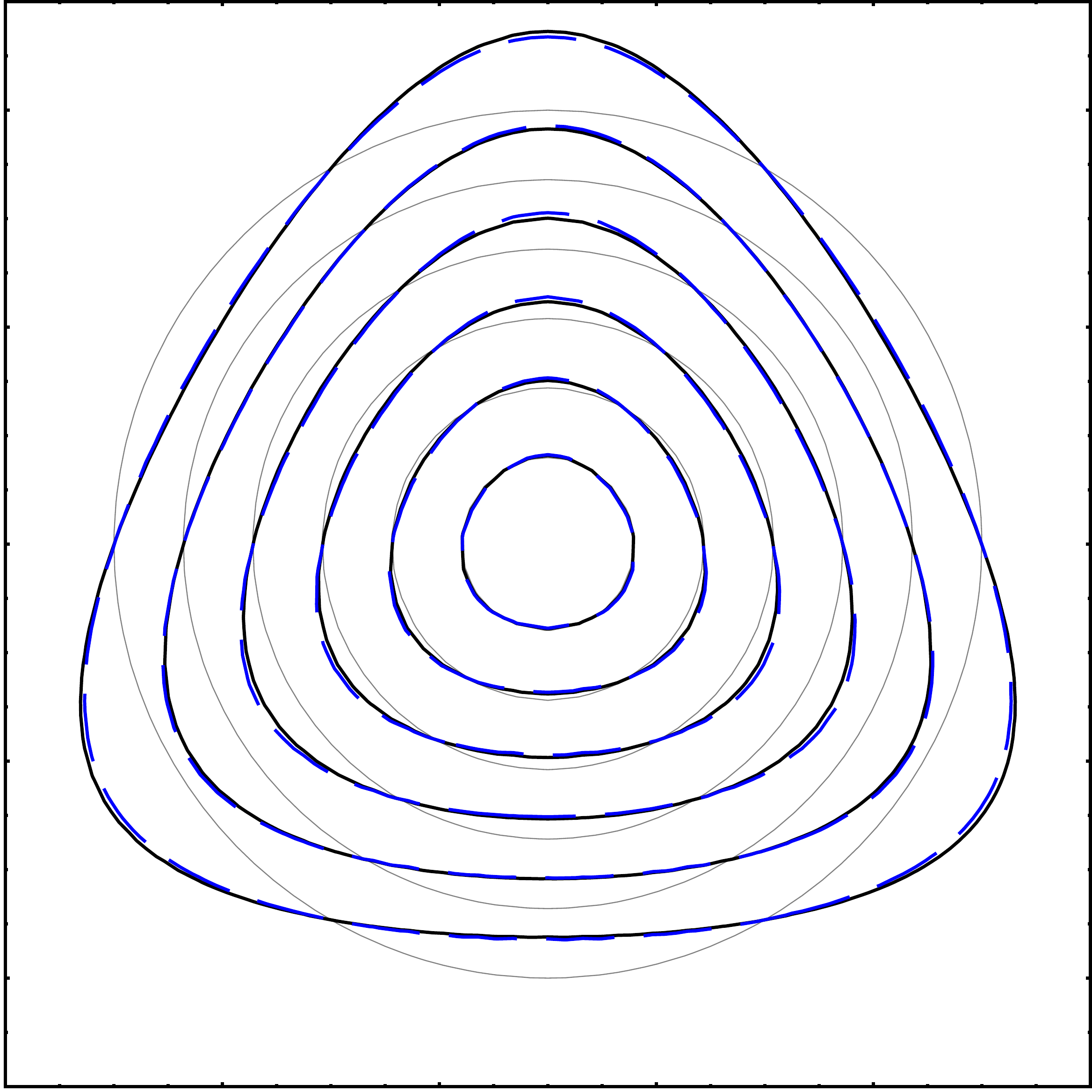}
 \includegraphics[width=0.3\textwidth]{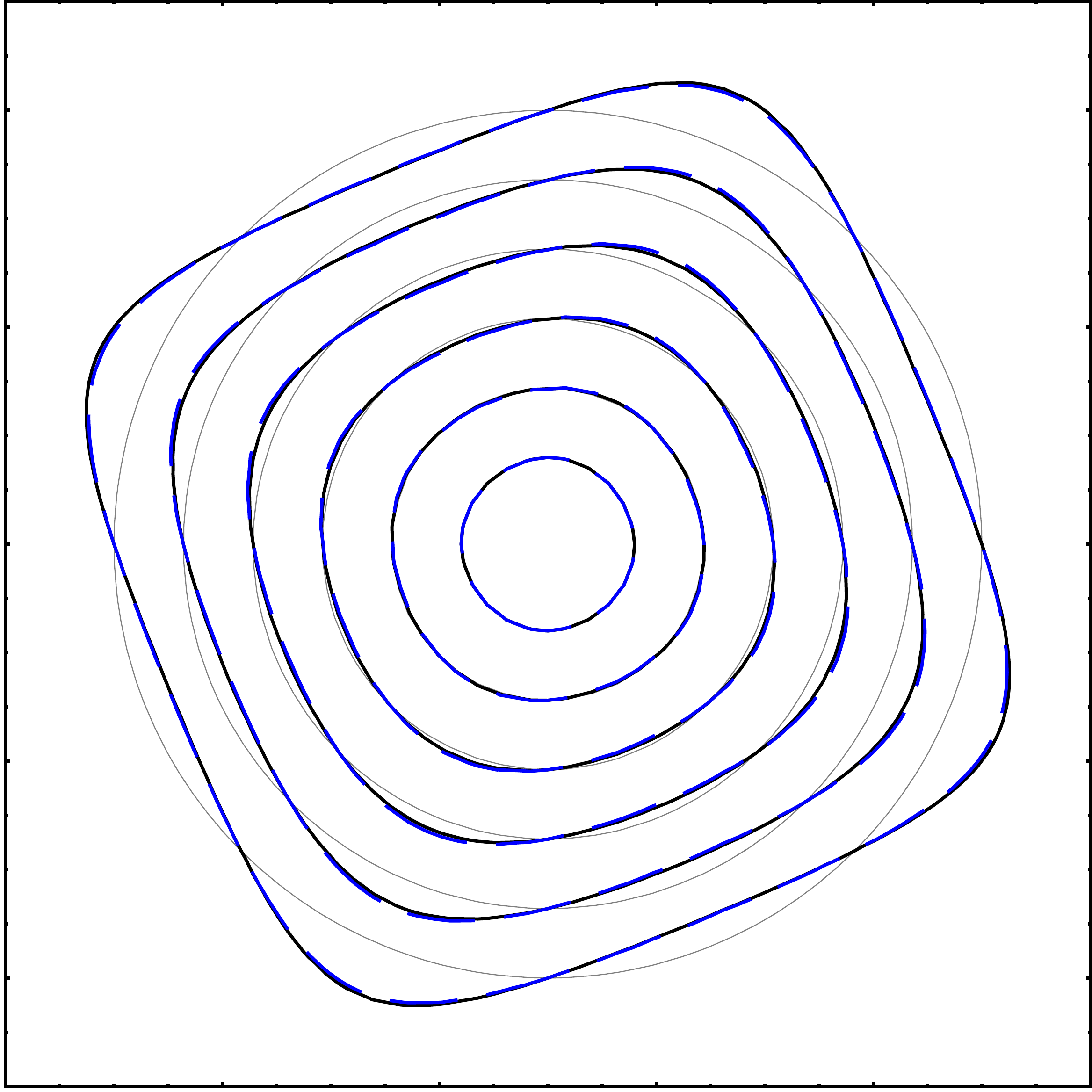}

 (c) \hspace{0.95\textwidth}

 \includegraphics[width=0.3\textwidth]{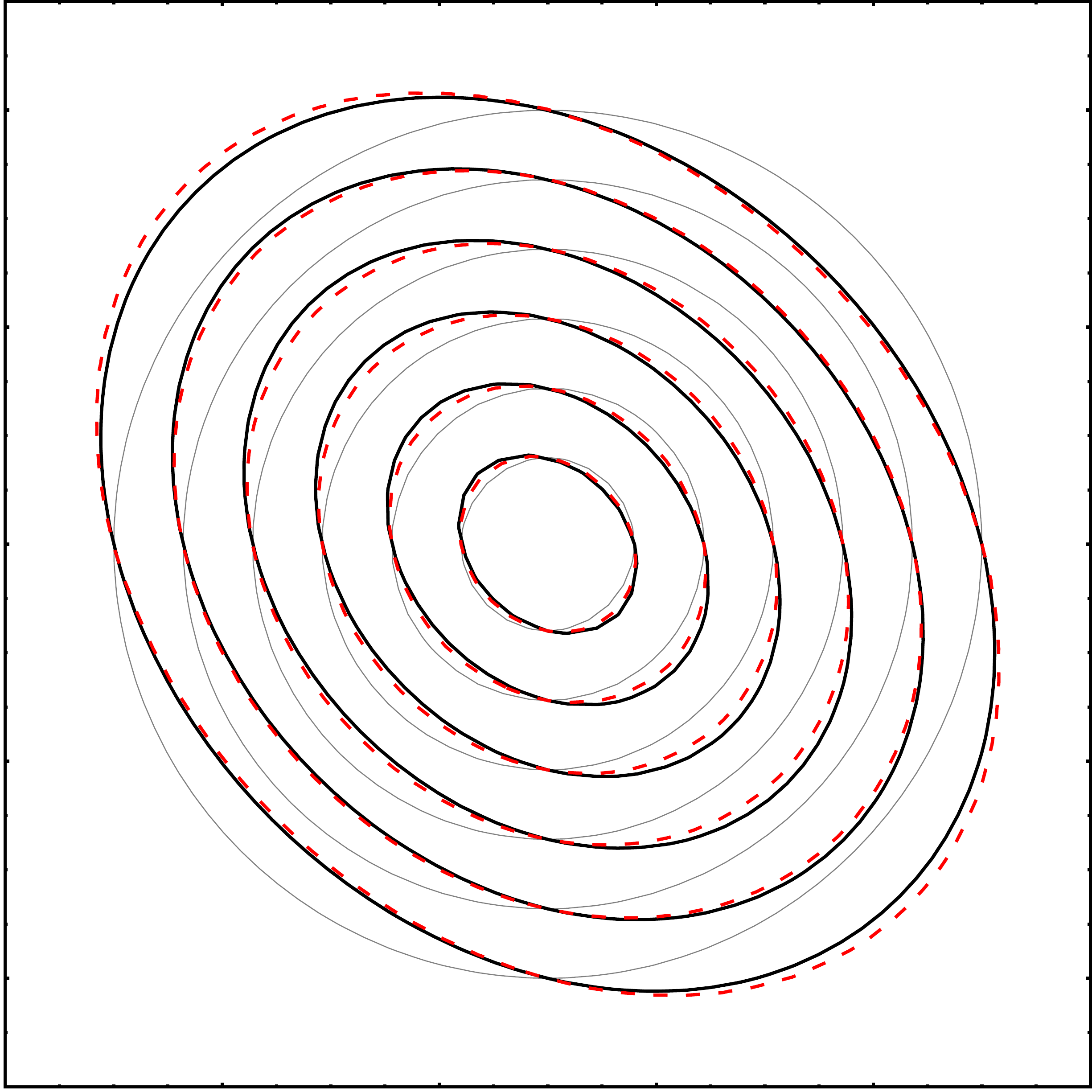}
 \includegraphics[width=0.3\textwidth]{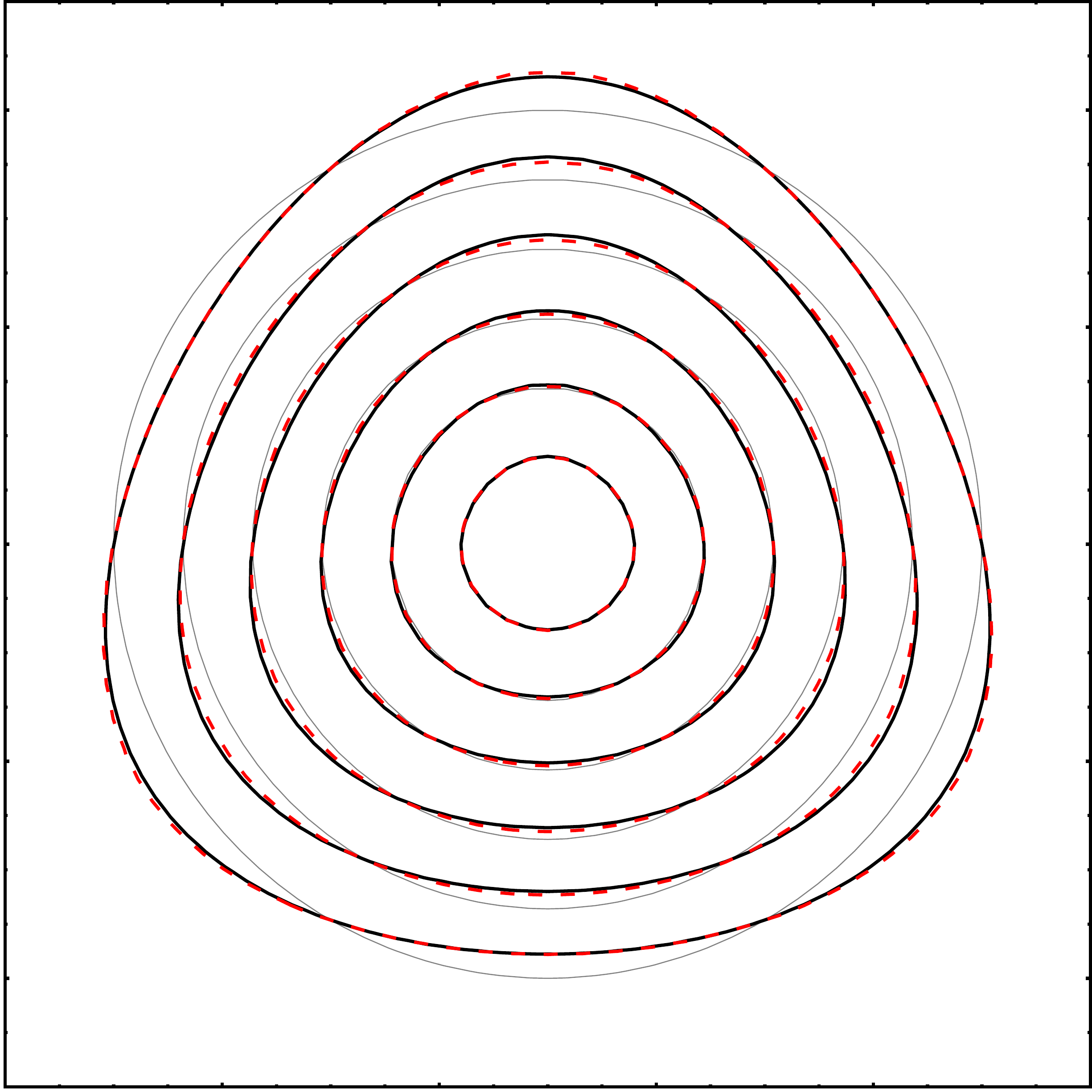}
 \includegraphics[width=0.3\textwidth]{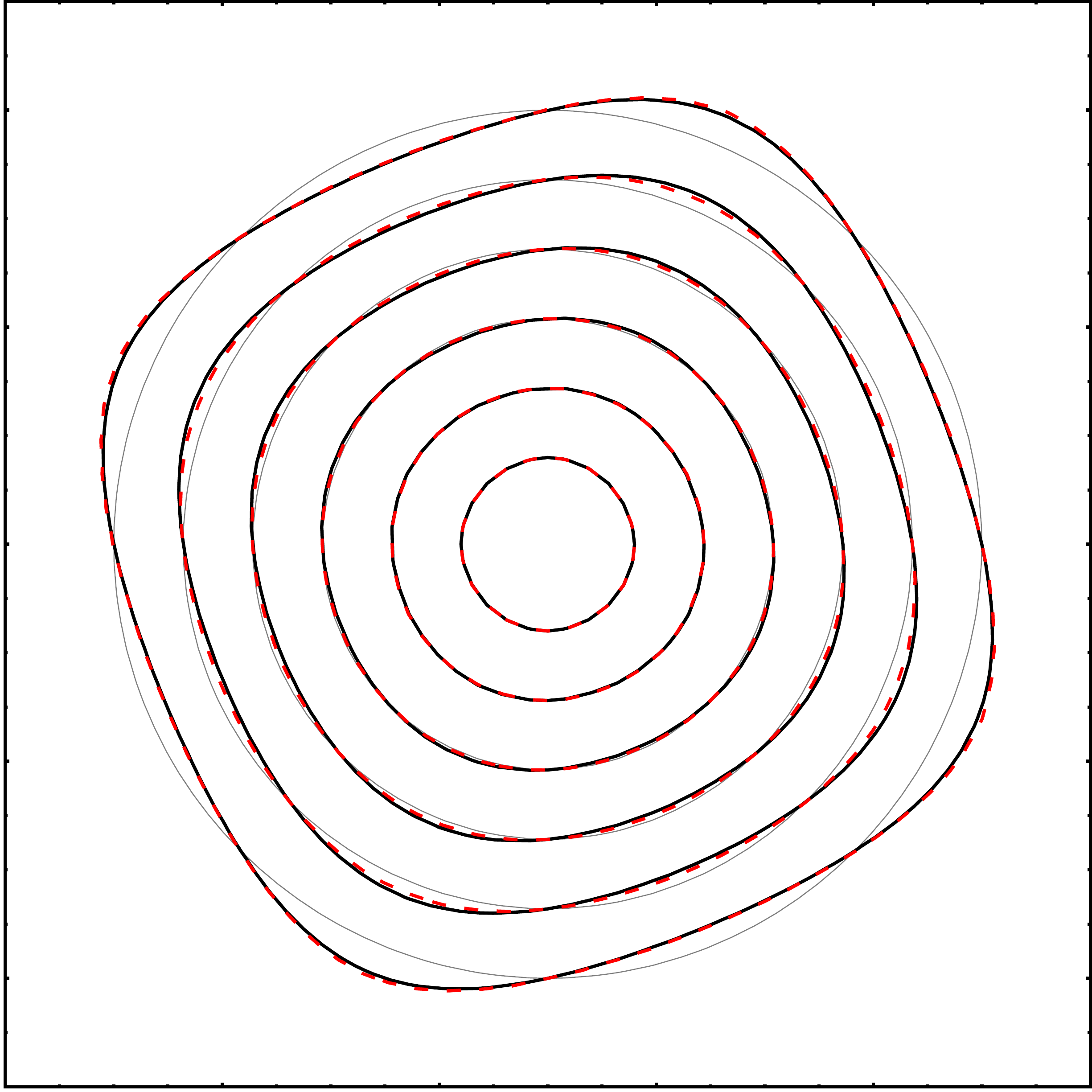}
 \caption{Example $\psi_{0}$ contours with pure $m=2$, $m=3$, and $m=4$ outer boundary conditions respectively for the (a) constant (black, solid), (b) linear hollow (blue, dashed), and (c) linear peaked (red, dotted) current profiles shown in fig. \ref{fig:gradShafCurrentProfiles}. Circular (gray, solid) and constant current (black, solid) flux surfaces are shown for comparison.}
 \label{fig:gradShafSolLowestOrder}
\end{figure}

From studying the plots in fig. \ref{fig:gradShafSolLowestOrder} we can obtain the results of this calculation. First of all, the $m=2$ mode roughly corresponds to plasma elongation, $\kappa$, the ratio of the maximum chord length, $2 b$, to the minimum chord length, $2 a$. The $m=3$ mode roughly corresponds to triangularity, $\delta$, and the $m=4$ mode to squareness. Also, we observe that the flux surfaces near the magnetic axis in all the $m=3$ cases are circular. This can be confirmed by taking the limit of eqs. \refEq{eq:gradShafranovNextOrderSolsConst}, \refEq{eq:gradShafranovNextOrderSolsHollow}, and \refEq{eq:gradShafranovNextOrderSolsPeaked} as $r \rightarrow 0$. For example, the constant current case becomes $\left( A / 4 \right) r^{2} + r^{3} \left( C_{3} \cos \left( 3 \theta \right) + D_{3} \sin \left( 3 \theta \right) \right) \rightarrow \left( A / 4 \right) r^{2}$, which has no dependence on $\theta$. In these cases, the tokamak is only up-down asymmetric near the plasma edge. This effect only gets more pronounced with higher $m$ modes. Therefore, if we want to make the tokamak as asymmetric as possible, we should use low $m$ modes.

For the constant current case, the $m=1$ mode, in the absence of higher modes, is purely a translation and does not introduce any asymmetry into the flux surface shape. For the two linear current cases, the $m=1$ mode is not purely a translation. In addition to translation, which is given to lowest order by a term $\propto r$, it introduces a flux surface shaping effect that decays as $r^{3}$ with $r \rightarrow 0$, so it is limited to the edge. This can be seen by noting that the Taylor expansion of either $m=1$ Bessel function has no $r^{2}$ component. This means the $m=2$ mode, which introduces elongation, appears optimal for getting penetration of up-down asymmetry into the core.

For the constant current pure $m=2$ mode case, one can use trigonometric identities and rearrange the solution
\begin{eqnarray}
   \psi_{0} \left( r, \theta \right) &= \frac{A}{4} r^{2} + r^{2} \left( C_{2} \cos \left( 2 \theta \right) + D_{2} \sin \left( 2 \theta \right) \right) \label{eq:fourierN2LowestOrderSol}
\end{eqnarray}
to show that the flux surfaces are exactly elliptical. Furthermore, one can translate the Fourier coefficients to the elongation,
\begin{eqnarray}
   \kappa \left( \psi_{0} \right) &= \kappa_{b} \equiv \sqrt{\frac{\frac{A}{4} + \sqrt{C_{2}^{2} + D_{2}^{2}}}{\frac{A}{4} - \sqrt{C_{2}^{2} + D_{2}^{2}}}} \label{eq:fourierElongation} ,
\end{eqnarray}
the tilt angle of the elongation (see fig. \ref{fig:tiltAngleSpecification}a),
\begin{eqnarray}
   \theta_{\kappa} \left( \psi_{0} \right) &= \theta_{\kappa b} \equiv - \frac{1}{2} \arctan \left( \frac{D_{2}}{C_{2}} \right) \label{eq:fourierElongationTiltAngle},
\end{eqnarray}
and the minor radius of the flux surface,
\begin{eqnarray}
   r_{\psi} \left( \psi_{0} \right) &\equiv a \rho \left( \psi_{0} \right) = \sqrt{\frac{\psi_{0}}{\frac{A}{4} + \sqrt{C_{2}^{2} + D_{2}^{2}}}} . \label{eq:fourierMinorRadius}
\end{eqnarray}
Here $\rho \equiv \sqrt{\psi/\psi_{b}}$ is the normalized flux surface label and the subscript $b$ indicates a value at the plasma boundary. It should be mentioned that the tilt angle of the ellipse, $\theta_{\kappa}$, is defined to be a left-handed rotation with respect to $\hat{e}_{\zeta}$ (see fig. \ref{fig:tiltAngleSpecification}a), whereas $\hat{e}_{\theta}$ is in the right-handed direction. These definitions give rise to the negative sign appearing in eq. \refEq{eq:fourierElongationTiltAngle}.

\begin{figure}
 \centering
 \includegraphics[width=0.6\textwidth]{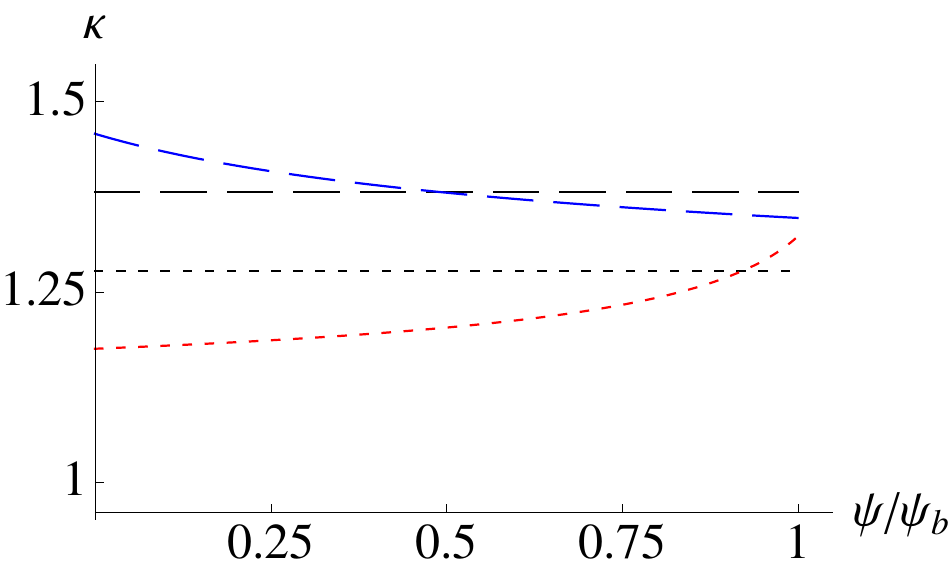}
 \caption{Plasma elongation from fig. \ref{fig:gradShafSolLowestOrder}b (dashed) for the constant (black) and linear hollow (blue) current profiles as well as fig. \ref{fig:gradShafSolLowestOrder}c (dotted) for the constant (black) and linear peaked (red) current profiles.}
 \label{fig:gradShafElongationPenetration}
\end{figure}

Crucially, we see in eqs. \refEq{eq:fourierElongation} and \refEq{eq:fourierElongationTiltAngle} that the elongation and elongation tilt angle are independent of the radial coordinate. This means that, for a constant current profile, the elongation and elongation tilt at the plasma boundary, $\kappa_{b}$ and $\theta_{\kappa b}$, will uniformly penetrate throughout the plasma. We can also numerically calculate elongation at different flux surfaces for the two other current distributions to produce fig. \ref{fig:gradShafElongationPenetration}. The important trend to notice is that hollow current profiles exaggerate elongation for flux surfaces near the magnetic axis, while peaked profiles tend to limit elongation to the plasma edge. In order to demonstrate this point, the hollow current flux surface boundary condition for fig. \ref{fig:gradShafSolLowestOrder}b was chosen to be more circular than the constant current flux surfaces at the edge. Nevertheless, we see that it is more strongly shaped than the constant current surfaces near the magnetic axis. On the other hand, the peaked flux surface boundary condition in fig. \ref{fig:gradShafSolLowestOrder}c was chosen to be more shaped at the edge and we see the opposite trend. The flux surfaces become more circular than in the constant current case near the axis.

There are three general points that are illuminated by the specific cases in this calculation. First, external PF coils only exert direct control over the flux surface shape at the plasma-vacuum boundary. Second, low order Fourier harmonics, specifically elongation, penetrate to the core most effectively. Higher order modes will only cause up-down asymmetry near the plasma edge. Lastly, a hollow toroidal current profile will more readily permit asymmetry to penetrate into the plasma core and can even amplify the asymmetry applied to the boundary. From this analysis, we identify tilted elliptical flux surfaces as the most promising geometry to create a significantly up-down asymmetric tokamak and maximize intrinsic rotation. 

\section{Modifications to GS2}
\label{sec:GS2}

GS2 \cite{KotschenreutherGS21995}, a local $\delta f$ gyrokinetic code, was modified to simulate the up-down asymmetric configurations that are investigated in this work. First of all, new input parameters were added to the Miller geometry package to allow for a more general specification of the tokamak geometry. Also, for reasons of computational efficiency, several numerical derivatives assumed the up-down symmetry of flux surfaces and the calculation of these numerical derivatives had to be modified. Lastly, in its treatment of trapped particles, GS2 assumed that the poloidal location of the maximum magnetic field was at $\pm \pi$, which is not necessarily the case for up-down asymmetric flux surfaces. Note that all modifications occurred within the Miller geometry specification and GS2's capability to read numerical equilibrium was not used.

\subsection{Gyrokinetics}

Gyrokinetics \cite{CattoLinearizedGyrokinetics1978,FriemanNonlinearGyrokinetics1982,ParraGyrokineticLimitations2008,AbelGyrokineticsDeriv2012} is a theoretical framework to study plasma behavior with perpendicular wavenumbers comparable to the gyroradius ($k_{\perp} \rho_{i} \sim 1$) and timescales much slower than the particle cyclotron frequencies ($\omega \ll \Omega_{i} \ll \Omega_{e}$). These particular scales have been experimentally shown to be appropriate for modeling turbulence \cite{McKeeTurbulenceScale2001}. To derive the gyrokinetic equations, we expand the distribution function, $f_{s} = f_{s 0} + f_{s 1} + \ldots$, and assume the perturbation is small compared to the background ($f_{s 1} \ll f_{s 0}$) \cite{HowesAstroGyro2006}. For tokamak plasmas, axisymmetry implies radially confined orbits. In addition, the transport timescale usually exceeds the collisional timescale. As a result, the lowest order distribution function is a Maxwellian ($f_{s 0} = F_{Ms}$). Here
\begin{align}
   F_{M s} \equiv n_{s} \left( \frac{m_{s}}{2 \pi T_{s}} \right)^{3/2} \text{exp} \left( - \frac{m_{s} \left( \vec{v} - \vec{u}_{s} \right)^2}{2 T_{s}} \right) \label{eq:maxwellianDef}
\end{align}
is the Maxwellian distribution function, $n_{s}$ and $T_{s}$ are the density and temperature of species $s$, and $m_{s}$ is the particle mass. Since we are interested in the momentum redistribution that occurs in an initially stationary plasma, for most of this paper we will take $\vec{u}_{s}$, the mean plasma flow, to be small in order to determine the momentum flux in the absence of rotation. It will become necessary to introduce finite rotation in order to determine the momentum diffusivity. The equations in this section take $\vec{u}_{\zeta} = R \vec{\Omega}_{\zeta} \neq 0$.

We will start with the electrostatic Fokker-Plank equation,
\begin{align}
   \frac{\partial f_{s}}{\partial t} + \vec{v} \cdot \vec{\nabla}f_{s} + \frac{Z_{s} e}{m_{s}} \left( -\vec{\nabla} \phi + \vec{v} \times \vec{B} \right) \cdot \vec{\nabla}_{v} f_{s} = \sum_{s'} C_{ss'} , \label{eq:FokkerPlank}
\end{align}
and quasineutrality equation,
\begin{align}
   \sum_{s} Z_{s} \int d^{3}v f_{s} = 0 , \label{eq:quasineutrality}
\end{align}
assuming weak electromagnetic effects. Here $Z_{s}$ is the charge number, $e$ is the charge of the proton, $\phi$ is the scalar electric potential, and $\sum_{s'} C_{ss'}$ is the collision operator. Both equations can be expanded order by order in $\rho_{\ast}$ and simplified. In doing so, we change real-space coordinates to the guiding center position
\begin{align}
   \vec{R}_{gc} \equiv \vec{r}_{p} + \frac{\vec{w} \times \hat{b}}{\Omega_{s}} , \label{eq:guidingCenterPosDef}
\end{align}
specified by the poloidal flux, $\psi$, a poloidal angle, $\theta$, and
\begin{align}
   \alpha \equiv \zeta - I \left( \psi \right) \left. \int_{0}^{\theta} \right|_{\psi} d \theta' \left( R^{2} \vec{B} \cdot \vec{\nabla} \theta \right)^{-1} - \Omega_{\zeta} \left( \psi \right) t , \label{eq:alphaDef}
\end{align}
which parameterizes the direction perpendicular to the magnetic field line, but still within the flux surface. We also change velocity-space coordinates to the parallel velocity, $w_{||}$, the magnetic moment, $\mu \equiv m_{s} w_{\perp}^{2} / \left( 2 B \right)$, and the gyrophase angle,
\begin{align}
   \varphi \equiv \text{arctan} \left( \frac{\left( \vec{w} \times \hat{b} \right) \cdot \vec{\nabla} \psi}{\vec{w} \cdot \vec{\nabla} \psi} \right) . \label{eq:gyroangleDef}
\end{align}
Here $\Omega_{s} \equiv Z_{s} e B / m_{s}$ is the species cyclotron frequency and $\vec{w} = \vec{v} - \vec{u}_{s}$ is the particle velocity in the frame rotating with the plasma. We make use of the substitution
\begin{align}
   \bar{h}_{s} \left( \vec{R}_{gc}, w_{||}, \mu, t \right) \equiv f_{s 1} + \frac{Z_{s} e \phi}{T_{s}} F_{M s} \label{eq:nonadiabaticDistFnDef}
\end{align}
and average over the gyrophase angle. Instead of gyrating charged particles, our equations now govern the evolution of charged rings around a guiding center position. Because of the strong anisotropy introduced by the magnetic field, the perpendicular scale of the turbulence is much smaller than the parallel variation. Furthermore, two of the assumptions of gyrokinetics, $k_{\perp} \rho_{i} \sim 1$ and $\rho_{i} \ll l_{p}$, show that the perpendicular turbulence length scale is smaller than the characteristic scale lengths of the background radial gradients, $k_{\perp} l_{p} \gg 1$. This allows us to Fourier analyze using
\begin{align}
   \bar{h}_{s} \left( \psi, \alpha, \theta, w_{||}, \mu, t \right) = \sum_{k_{\psi}, k_{\alpha}} h_{s} \left( k_{\psi}, k_{\alpha}, \theta, w_{||}, \mu, t \right) \text{exp} \left( i k_{\psi} \psi + i k_{\alpha} \alpha \right) . \label{eq:FourierAnalyzedDistFn}
\end{align}
This produces the Fourier-analyzed gyrokinetic equation in $\mu$ and $w_{||}$ velocity variables, \cite{ParraUpDownSym2011},
\begin{align}
   \frac{\partial h_{s}}{\partial t} &+ w_{||} \hat{b} \cdot \vec{\nabla} \theta \left. \frac{\partial h_{s}}{\partial \theta} \right|_{w_{||}, \mu} + i \left( k_{\psi} v_{d s \psi} + k_{\alpha} v_{d s \alpha} \right) h_{s} \nonumber \\
&+ a_{s ||} \left. \frac{\partial h_{s}}{\partial w_{||}} \right|_{\theta, \mu} - \sum_{s'} \langle C_{ss'}^{\left( l \right)} \rangle_{\varphi} + \left\{ \langle \phi \rangle_{\varphi}, h_{s} \right\} = \frac{Z_{s} e F_{M s}}{T_{s}} \frac{\partial \langle \phi \rangle_{\varphi}}{\partial t} \label{eq:gyrokineticEq} \\
 &- v_{\phi s \psi} F_{M s} \left[ \frac{1}{n_{s}} \frac{\partial n_{s}}{\partial \psi} + \frac{m_{s} I w_{||}}{B T_{s}} \frac{\partial \Omega_{\zeta}}{\partial \psi} - \frac{m_{s} R \Omega_{\zeta}^{2}}{T_{s}} \frac{\partial R}{\partial \psi} + \left( \frac{m_{s} w^{2}}{2 T_{s}} - \frac{3}{2} \right) \frac{1}{T_{s}} \frac{\partial T_{s}}{\partial \psi} \right] , \nonumber
\end{align}
and the Fourier-analyzed quasineutrality equation,
\begin{align}
      \sum_{s} \frac{2 \pi Z_{s}}{m_{s}} B \int dw_{||} \int d \mu J_{0} \left( \frac{k_{\perp} \sqrt{2 \mu B}}{\Omega_{s} \sqrt{m_{s}}} \right) h_{s} = \sum_{s} \frac{Z_{s}^{2} e n_{s}}{T_{s}} \phi , \label{eq:gyroQuasineutrality}
\end{align}
where $J_{0} \left( \ldots \right)$ is the zeroth order Bessel function of the first kind. In eq. \refEq{eq:gyrokineticEq}, the guiding center background magnetic drift velocity
is split up into
\begin{eqnarray}
   v_{d s \psi} &\equiv \vec{v}_{d s} \cdot \vec{\nabla} \psi \label{eq:radialGCdriftvelocity} \\
   &= \left( - \frac{I \left( m_{s} w_{||}^{2} + \mu B \right)}{m_{s} \Omega_{s} B} \frac{\partial B}{\partial \theta} + \frac{2 B R \Omega_{\zeta} w_{||}}{\Omega_{s}} \frac{\partial R}{\partial \theta} + \frac{I R \Omega_{\zeta}^{2}}{\Omega_{s}} \frac{\partial R}{\partial \theta} \right) \hat{b} \cdot \vec{\nabla} \theta \nonumber
\end{eqnarray}
and
\begin{eqnarray}
   v_{d s \alpha} &\equiv \vec{v}_{d s} \cdot \vec{\nabla} \alpha = - \frac{m_{s} w_{||}^{2} + \mu B}{m_{s} \Omega_{s}} \left[ \frac{\partial B}{\partial \psi} - \frac{\partial B}{\partial \theta} \frac{ \hat{b} \cdot \left( \vec{\nabla} \theta \times \vec{\nabla} \alpha \right) }{B} \right] \nonumber \\
   &- \frac{\mu_{0} w_{||}^{2}}{B \Omega_{s}} \left( \frac{\partial p}{\partial \psi} - \sum_{s} n_{s} m_{s} \Omega_{\zeta}^{2} R \frac{\partial R}{\partial \psi} \right) \label{eq:alphaGCdriftvelocity} \\
   &+ \frac{2 \Omega_{\zeta} w_{||}}{\Omega_{s}} \left( \vec{\nabla} R \times \hat{e}_{\zeta} \right) \cdot \vec{\nabla} \alpha + \frac{m_{s} R \Omega_{\zeta}^{2}}{Z_{s} e} \left( \frac{\partial R}{\partial \psi} - \frac{\partial R}{\partial \theta} \frac{\hat{b} \cdot \left( \vec{\nabla} \theta \times \vec{\nabla} \alpha \right)}{B} \right). \nonumber
\end{eqnarray}
The acceleration parallel to the magnetic field line is given by
\begin{eqnarray}
   a_{s ||} = \left( - \frac{\mu}{m_{s}} \frac{\partial B}{\partial \theta} + R \Omega_{\zeta}^{2} \frac{\partial R}{\partial \theta} \right) \hat{b} \cdot \vec{\nabla} \theta \label{eq:parallelAcceleration}
\end{eqnarray}
and $\langle \cdots \rangle_{\varphi}$ denotes an average over the gyromotion holding $\vec{R}_{gc}$ fixed. Importantly,
\begin{eqnarray}
   \left\{ \langle \phi \rangle_{\varphi}, h_{s} \right\} &\equiv \sum_{k'_{\psi}, k'_{\alpha}} \left( k'_{\psi} k_{\alpha} - k_{\psi} k'_{\alpha} \right) \langle \phi \rangle_{\varphi} \left( k'_{\psi}, k'_{\alpha} \right) h_{s} \left( k_{\psi} - k'_{\psi}, k_{\alpha} - k'_{\alpha} \right) \label{eq:nonlinearProductDef}
\end{eqnarray}
is the nonlinear term that represents the $\vec{E} \times \vec{B}$ motion of the fluctuations, where 
\begin{eqnarray}
   \langle \phi \rangle_{\varphi} &= J_{0} \left( \frac{k_{\perp} \sqrt{2 \mu B}}{\Omega_{s} \sqrt{m_{s}}} \right) \phi
\end{eqnarray}
is the gyroaveraged potential, $J_{0} \left( \ldots \right)$ is the zeroth order Bessel function of the first kind, and the perpendicular wavenumber can be written as
\begin{eqnarray}
   k_{\perp} &= \sqrt{k_{\psi}^{2} \left| \vec{\nabla} \psi \right|^{2} + 2 k_{\psi} k_{\alpha} \vec{\nabla} \psi \cdot \vec{\nabla} \alpha + k_{\alpha}^{2} \left| \vec{\nabla} \alpha \right|^{2}} . \label{eq:kperpForm}
\end{eqnarray}
Finally,
\begin{eqnarray}
   v_{\phi s \psi} \equiv i k_{\alpha} \langle \phi \rangle_{\varphi} \label{eq:radialEcrossBvel}
\end{eqnarray}
is the turbulent $\vec{E} \times \vec{B}$ drift normal to the flux surface. For most of this article we will assume $\Omega_{\zeta} = 0$ and $\partial \Omega_{\zeta} / \partial \psi = 0$ to determine how a tokamak generates rotation from rest.

Solving the gyrokinetic and quasineutrality equations, given in eqs. \refEq{eq:gyrokineticEq} and \refEq{eq:gyroQuasineutrality}, for $h_{s}$ and $\phi$ allows us to calculate the turbulent fluxes of particles, momentum, and energy given by
\begin{eqnarray}
   \Gamma_{tot} &= \sum_{s} \sum_{k_{\psi}, k_{\alpha}} \left\langle \int d^{3}v v_{\phi \psi} h_{s}\left( k_{\psi}, k_{\alpha}, \theta, w_{||}, \mu \right) e^{i \vec{k}_{\perp} \cdot \left( \vec{v} \times \hat{b} \right) / \Omega_{s}} \right\rangle_{\psi} \label{eq:particleFlux} \\
   \Pi_{\zeta tot} &= \sum_{s} \sum_{k_{\psi}, k_{\alpha}} \left\langle m_{s} R \int d^{3}v w_{\zeta} v_{\phi \psi} h_{s} \left( k_{\psi}, k_{\alpha}, \theta, w_{||}, \mu \right) e^{i \vec{k}_{\perp} \cdot \left( \vec{v} \times \hat{b} \right) / \Omega_{s}} \right\rangle_{\psi} \label{eq:momFlux} \\
   Q_{tot} &= \sum_{s} \sum_{k_{\psi}, k_{\alpha}} \left\langle \frac{m_{s}}{2} \int d^{3}v w^{2} v_{\phi \psi} h_{s}\left( k_{\psi}, k_{\alpha}, \theta, w_{||}, \mu \right) e^{i \vec{k}_{\perp} \cdot \left( \vec{v} \times \hat{b} \right) / \Omega_{s}} \right\rangle_{\psi} \label{eq:energyFlux}
\end{eqnarray}
respectively, where $d^{3}v = \left( B / m_{s} \right) dw_{||} d\mu d\varphi$, $v_{\phi \psi} \equiv \left( - i \vec{k}_{\perp} \phi \times \vec{B} / B^{2} \right) \cdot \vec{\nabla} \psi$ is the Fourier transformed turbulent $\vec{E} \times \vec{B}$ velocity evaluated at $- k_{\psi}$ and $- k_{\alpha}$, and $\left\langle \ldots \right\rangle_{\psi} \equiv \left( d V / d \psi \right)^{-1} \int_{0}^{2 \pi} d \theta \int_{0}^{2 \pi} d \zeta \left( \ldots \right) / \left| \vec{B} \cdot \vec{\nabla} \theta \right|$ denotes the flux surface average. Here $\vec{k}_{\perp}$ is the perpendicular wavenumber, $V$ is the volume contained by a flux surface, and $d V / d \psi = \int_{0}^{2 \pi} d \theta \int_{0}^{2 \pi} d \zeta \left| \vec{B} \cdot \vec{\nabla} \theta \right|^{-1}$. The momentum flux tells with $\Omega_{\zeta} = d \Omega_{\zeta} / d \psi = 0$ us how strongly a particular tokamak configuration will redistribute momentum to create nonzero rotation from an initially stationary plasma. Internally, GS2 manipulates the particle energy, $\mathscr{E} \equiv m_{s} w^{2}/2$, rather than $w_{||}$. However, we choose to write the gyrokinetic equation using $w_{||}$ because the symmetry constraining the momentum flux in up-down symmetric geometries is in $w_{||} \rightarrow - w_{||}$.

\subsection{Normalizations}

A common source of confusion regarding gyrokinetic codes comes from the different conventions each code uses to normalize physical quantities. We have thus explicitly given GS2 normalizations for quantities pertinent to this work in tables \ref{tab:GS2InputParameters} and \ref{tab:GS2Normalizations}, where the subscript $r$ indicates a reference quantity. GS2 allows several different ways of specifying the physical geometry of the simulation, however this work exclusively uses Miller equilibrium specification. Many of the conventions and definitions GS2 employs depend on the method of geometry specification. Thus, significant portions of this work may only be valid when using Miller geometry.

\begin{table}
  \centering
  \begin{tabular}{ r c c c }
    Quantity & Miller Parameter & GS2 Parameter & GS2 Variable \\
    \hline
    Minor radius\superscript{$\ast$} & $r_{\psi M}$ & $r_{\psi N} \equiv r_{\psi}/l_{r}$ & \texttt{rhoc} \\
    Ref. magnetic field\superscript{$\ast \dagger$} & $B_{0}$ & $B_{r}$ &  \\
    Major radius & $R_{0}/ r_{\psi M}$ & $R_{0 N} \equiv R_{0}/l_{r}$ & \texttt{Rmaj} \\
    Shafranov shift & $d R_{0} / d r_{\psi M}$ & $d R_{0 N} / d r_{\psi N}$ & \texttt{shift} \\
    Safety factor & $q$ & $q$ & \texttt{qinp} \\
    Magnetic shear & $d q / d r_{\psi M}$ & $\hat{s} \equiv \frac{r_{\psi N}}{q} \frac{d q}{d r_{\psi N}}$ & \texttt{s\_hat\_input} \\
    Elongation & $\kappa$ & $\kappa$ & \texttt{akappa} \\
    Elongation derivative & $d \kappa / d r_{\psi M}$ & $d \kappa / d r_{\psi N}$ & \texttt{akappri} \\
    Triangularity & $\delta_{M}$ & $\delta \equiv \text{sin}^{-1} \left( \delta_{M} \right)$ & \texttt{tri} \\
    Triangularity derivative & $d \delta_{M} / d r_{\psi M}$ & $d \delta / d r_{\psi N}$ & \texttt{tripri} \\
    Pressure derivative & $d p / d r_{\psi M}$ & $\frac{d p_{N}}{d r_{\psi N}} = \frac{2 \mu_{0}}{B_{r}^{2}} \frac{d p}{d r_{\psi N}}$ & \texttt{beta\_prime\_input} \\
    Magnetic field ref. point & & $R_{geo N} \equiv R_{geo}/l_{r}$ & \texttt{R\_geo} \\
    Ref. macroscopic length\superscript{$\dagger$} & & $l_{r}$ &
  \end{tabular}
  \caption{Miller and GS2 geometry input parameters, where \superscript{$\ast$} denotes a Miller normalization parameter and \superscript{$\dagger$} denotes a GS2 normalization parameter.}
  \label{tab:GS2InputParameters}
\end{table}

The traditional Miller equilibrium model is specified by the seven parameters and two normalization parameters listed in table \ref{tab:GS2InputParameters}. Normalization parameters are not specified to the model, but they must be kept consistent between input parameters and when connecting output back to reality. The GS2 implementation of Miller geometry, on the other hand, is specified by nine parameters and two normalization parameters. The extra parameters, $R_{geo}$ and $l_{r}$, are redundant and are only present for convenience \cite{BarnesTrinityThesis2008}. The major radial location $R_{geo}$ allows the user to specify the reference magnetic field at any major radial position, instead of forcing the reference magnetic field to be at $R_{0}$. The reference length $l_{r}$ allows the user to use any arbitrary length, such as a meter, to normalize the macroscopic lengths in the simulation, rather than forcing the reference length to be the minor radius. Also, note the quantity $r_{\psi}$ is a flux function and is used to specify the flux surface, not the traditional radius of circular flux surfaces.

\begin{table}
  \centering
  \begin{tabular}{ r l l }
    Name & Definition & \texttt{GS2} Variable \\
    \hline
    Mass & $m_{N s} \equiv m_{s}/m_{r}$ & \texttt{mass} \\
    Temperature & $T_{N s} \equiv T_{s}/T_{r}$ & \texttt{temp} \\
    Charge & $Z_{N s} \equiv Z_{s}/Z_{r}$ & \texttt{z} \\
    Thermal velocity & $v_{th N s} \equiv v_{th s}/v_{th r} = \sqrt{T_{N s}/m_{N s}}$ & \texttt{stm} \\
    Equilibrium dist. fn. & $F_{M N s} \equiv \left( v_{th s}^{3}/n_{s} \right) F_{M s}$ & \\
    Nonadiabatic dist. fn. & $h_{Ns} \equiv \left( l_{r} / \rho_{r} \right) \left(1/F_{M s} \right) h_{s}$ & \\
    Complementary dist. fn. & $g_{Ns} \equiv \left( l_{r} / \rho_{r} \right) \left(1/F_{M s} \right) g_{s}$ & \texttt{g} \\
    Perturbed electric potential & $\phi_{N} \equiv \left( l_{r} / \rho_{r} \right) \left( Z_{r} e/T_{r} \right) \phi$ & \texttt{phi} \\
    Time & $t_{N} \equiv \left( v_{th r}/l_{r} \right) t$ & \texttt{time} \\
    Parallel velocity & $w_{|| N} \equiv w_{||}/v_{th s}$ & \texttt{vpa} \\
    Perp. velocity squared & $w_{\perp N}^{2} \equiv w_{\perp}^{2}/v_{th s}^{2}$ & \texttt{vperp2} \\
    Radial perp. coordinate & $x_{N} \equiv x/ \rho_{r}$ & \\
    Poloidal perp. coordinate & $y_{N} \equiv y/ \rho_{r}$ & \\
    Parallel wavenumber & $k_{|| N} \equiv l_{r} k_{||}$ & \\
    Major radial coordinate & $R_{N} \equiv R/l_{r} $ & \texttt{Rpos} \\
    Vertical coordinate & $Z_{N} \equiv Z/l_{r} $ & \texttt{Zpos} \\
    Radial perp. wavenumber & $k_{x N} \equiv \rho_{r} k_{x}$ & \texttt{akx} \\
    Poloidal perp. wavenumber & $k_{y N} \equiv \rho_{r} k_{y}$ & \texttt{aky} \\
    Magnetic field magnitude & $B_{N} \equiv B/B_{r}$ & \texttt{bmag} \\
    Magnetic flux & $\psi_{N} \equiv \psi / \left( l_{r}^{2} B_{r} \right)$ & \\
    Poloidal current flux function & $I_{N} \equiv I / \left( l_{r} B_{r} \right) = R_{geo N}$ & \\
    Flow & $u_{\zeta N} \equiv u_{\zeta} / v_{th r} = R \Omega_{\zeta} / v_{th r}$ & \\
    Angular flow & $\Omega_{\zeta N} \equiv \left( l_{r}/v_{th r} \right)\Omega_{\zeta} = u_{\zeta N}/R_{N}$ & \texttt{mach} \\
    Angular flow shear & $\gamma_{E N} \equiv \left( r_{\psi N} / q \right) \left( d \Omega_{\zeta N} / d r_{\psi N} \right)$ & \texttt{g\_exb} \\
    Energy & $\mathscr{E}_{N} \equiv \mathscr{E}/T_{s}$ & \texttt{energy} \\
    Magnetic moment & $\mu_{N} \equiv w_{\perp N}^{2} / B_{N} = \left( B_{r} / T_{s} \right) \mu$ & \\
    Lambda & $\lambda_{N} \equiv \mu_{N}/\mathscr{E}_{N} $ & \texttt{al} \\
    Density & $n_{N s} \equiv n_{s}/n_{r}$ & \texttt{dens} \\
    Temperature gradient & $1/L_{T N s} \equiv - \left( l_{r} / T_{s} \right) \partial T_{s} / \partial r_{\psi}$ & \texttt{tprim} \\
    Density gradient & $1/L_{n N s} \equiv - \left( l_{r} / n_{s} \right) \partial n_{s} / \partial r_{\psi}$ & \texttt{fprim} \\
    Mode angular frequency & $\omega_{N} \equiv \left( l_{r}/v_{th r} \right) \text{Real} \left[ \omega \right]$ & \texttt{omega} \\
    Mode growth rate & $\gamma_{N} \equiv \left( l_{r}/v_{th r} \right) \text{Imag} \left[ \omega \right]$ & \texttt{omega} \\
    Particle flux & $\Gamma_{N s} \equiv \Gamma_{s}/\Gamma_{gB r}$ & \texttt{part\_fluxes} \\
    Angular momentum flux & $\Pi_{N s} \equiv \Pi_{s}/\Pi_{gB r}$ & \texttt{mom\_fluxes} \\
    Heat flux & $Q_{N s} \equiv Q_{s}/Q_{gB r}$ & \texttt{heat\_fluxes}
  \end{tabular}
  \caption{GS2 normalized quantities and their corresponding variable names within the code (table adapted from ref. \cite{HighcockManifold2012}).}
  \label{tab:GS2Normalizations}
\end{table}

The reference temperature, the reference mass, and the reference thermal velocity are related by $v_{th r} \equiv \sqrt{2 T_{r} / m_{r}}$. This means the process of normalizing equations frequently creates factors of $\sqrt{2}$ that other normalizations do not have. Also, since the velocity space coordinate normalizations are species dependent, factors of $\sqrt{T_{s} / T_{r}}$ and $\sqrt{m_{s} / m_{r}}$ can be created. The $x$ and $y$ wavenumbers used in GS2 are related to the $\psi$ and $\alpha$ wavenumbers appearing in the gyrokinetic equation as
\begin{align}
   k_{\psi} &\equiv \frac{q}{r_{\psi N}} \frac{k_{x}}{l_{r} B_{r}} , \label{eq:kPsiDef} \\
   k_{\alpha} &\equiv \frac{d \psi_{N}}{d r_{\psi N}} l_{r} k_{y} . \label{eq:kAlphaDef}
\end{align}

Generally, parameters are normalized to be roughly $O \left( 1 \right)$, so many must be scaled up by $\rho_{r} \equiv v_{th r} / \Omega_{r}$, where $\Omega_{r} \equiv Z_{r} e B_{r} / m_{r}$. The reference temperature, density, and mass are completely arbitrary and left to the user. When using adiabatic electrons, the reference charge is taken to be the elementary charge, otherwise $Z_{r}$ is also left to the user. The reference magnetic field magnitude is defined as $B_{r} \equiv I \left( \psi \right) / R_{geo}$ on the flux surface of interest. The reference macroscopic length, $l_{r}$, is not necessarily the minor radius, but is any arbitrary length, similar to $T_{r}$, $n_{r}$, and $m_{r}$. Lastly, all fluxes are normalized to their gyroBohm values of
\begin{align}
   \Gamma_{gB r} &\equiv \frac{\rho_{r}^{2}}{l_{r}^{2}} n_{r} v_{th r} , \\
   \Pi_{gB r} &\equiv \frac{\rho_{r}^{2}}{l_{r}^{2}} n_{r} l_{r} m_{r} v_{th r}^{2} , \label{eq:gyroBohmMomFlux} \\
   Q_{gB r} &\equiv \frac{\rho_{r}^{2}}{l_{r}^{2}} n_{r} T_{r} v_{th r} .
\end{align}

\subsection{Geometry specification}
\label{subsec:GS2GeoSpec}

\begin{figure}
 \centering
 (a) \hspace{0.47\textwidth} (b) \hspace{0.39\textwidth}

 \includegraphics[width=0.4\textwidth]{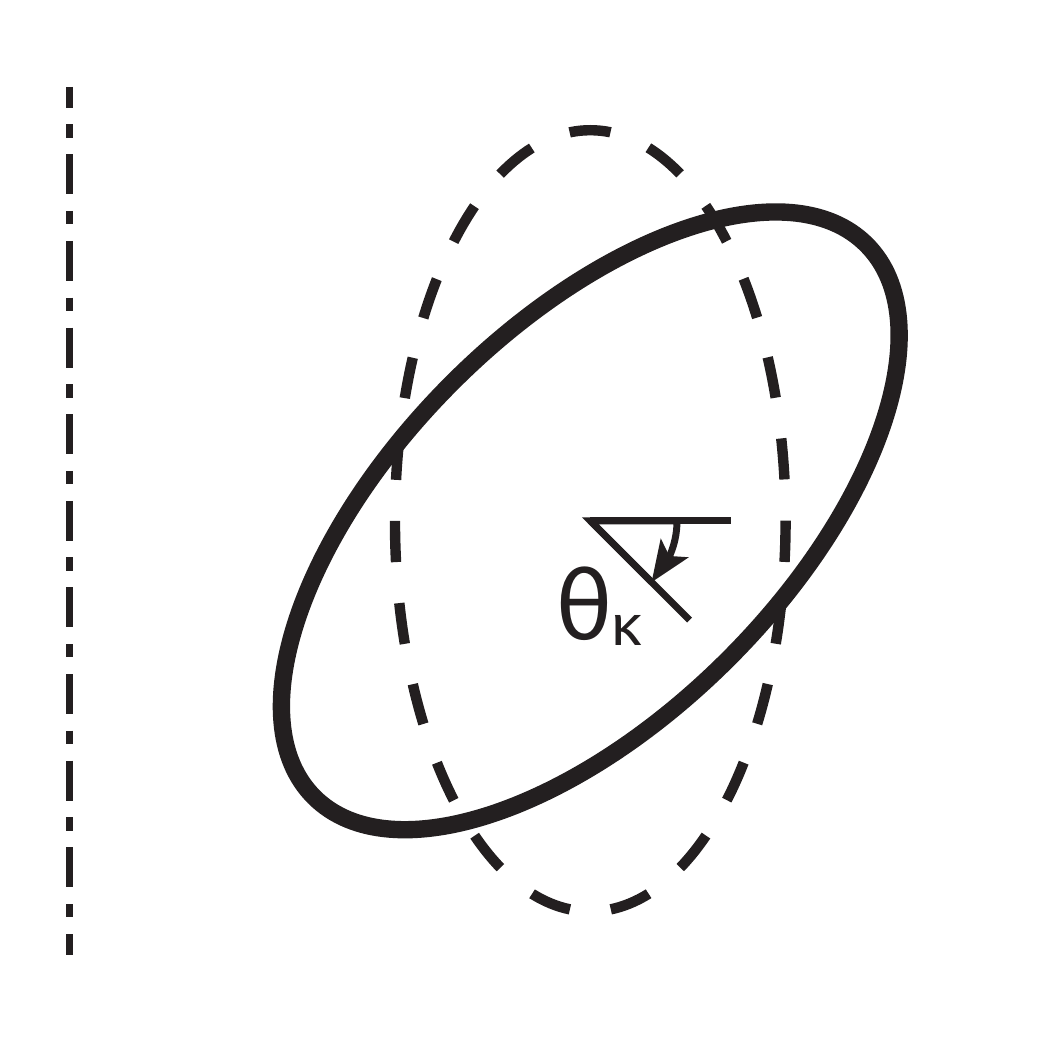}
 \hspace{0.1\textwidth}
 \includegraphics[width=0.4\textwidth]{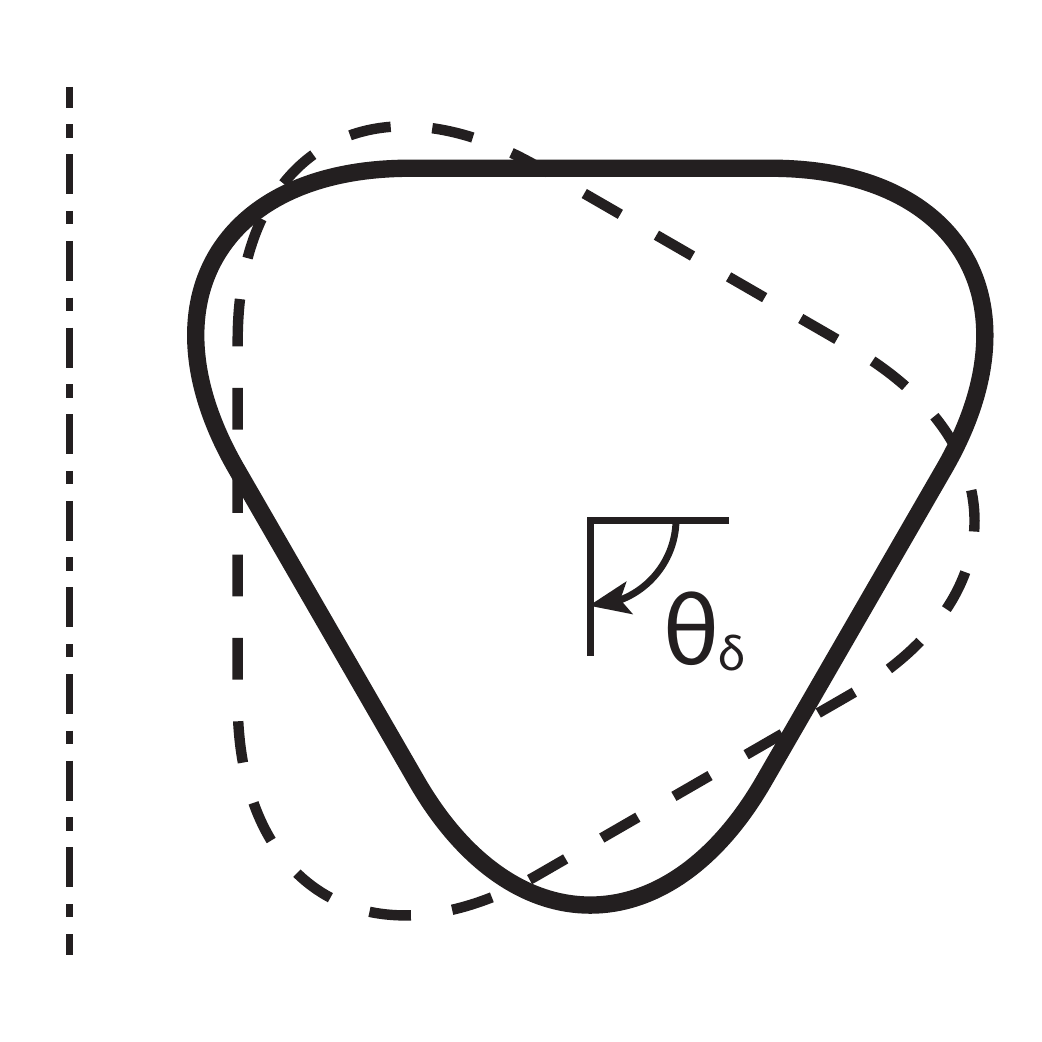}
 \caption{Definition of the (a) elongation tilt angle, $\theta_{\kappa}$, and (b) triangularity tilt angle, $\theta_{\delta}$, parameters.}
 \label{fig:tiltAngleSpecification}
\end{figure}

Originally, the GS2 Miller geometry \cite{MillerGeometry1998} input allowed for flux surface elongation and triangularity, but was not general enough to allow for tilted shapes. To support modeling up-down asymmetry four additional input parameters were added, given in table \ref{tab:newGS2InputParameters}. The elongation tilt angle and the triangularity tilt angle, shown in fig. \ref{fig:tiltAngleSpecification}, both have intuitively obvious definitions and can be varied independently. This allows significant additional flexibility in modeling unusual geometries, such as tilted comet-shaped flux surfaces \cite{KesnerCometTokamak1995}.

\begin{table}
  \centering
  \begin{tabular}{ r c l }
    Name & Definition & \texttt{GS2} Variable \\
    \hline
    Elongation tilt angle & $\theta_{\kappa}$ & \texttt{thetak} \\
    Elongation tilt angle derivative & $d \theta_{\kappa} / d r_{\psi N}$ & \texttt{thetakp} \\
    Triangularity tilt angle & $\theta_{\delta}$ & \texttt{thetad} \\
    Triangularity tilt angle derivative & $d \theta_{\delta} / d r_{\psi N}$ & \texttt{thetadp}
  \end{tabular}
  \caption{New GS2 input quantities and their corresponding variable names.}
  \label{tab:newGS2InputParameters}
\end{table}

Formerly, the Miller equilibrium flux surface shape was defined by
\begin{eqnarray}
   R_{N}^{old} \left( r_{\psi N}, \vartheta \right) &= R_{0 N} \left( r_{\psi N} \right) + r_{\psi N} \cos \left( \vartheta + \delta \left( r_{\psi N} \right) \sin \left( \vartheta \right) \right) \label{eq:oldGeoSpecR} \\
   Z_{N}^{old} \left( r_{\psi N}, \vartheta \right) &= r_{\psi N} \kappa \left( r_{\psi N} \right) \sin \left( \vartheta \right) , \label{eq:oldGeoSpecZ}
\end{eqnarray}
where $R_{N} \equiv R / l_{r}$, $Z_{N} \equiv Z / l_{r}$, and $l_{r}$ is an arbitrary normalization length. The angle $\vartheta$ is distinguished from the angle $\theta$, used in Section \ref{sec:MHD}, because it is not the usual cylindrical poloidal angle. From eqs. \refEq{eq:oldGeoSpecR} and \refEq{eq:oldGeoSpecZ}, the two neighboring flux surfaces were created using a Taylor expansion about the flux surface of interest $r_{\psi N} \equiv r_{\psi} / l_{r}$, where $r_{\psi} \equiv a \rho$ is a flux surface label. The definition of the neighboring flux surfaces is what necessitates providing input for the Shafranov shift, elongation derivative, and triangularity derivative.

\begin{figure}
 \centering
 \includegraphics[width=0.15\textwidth]{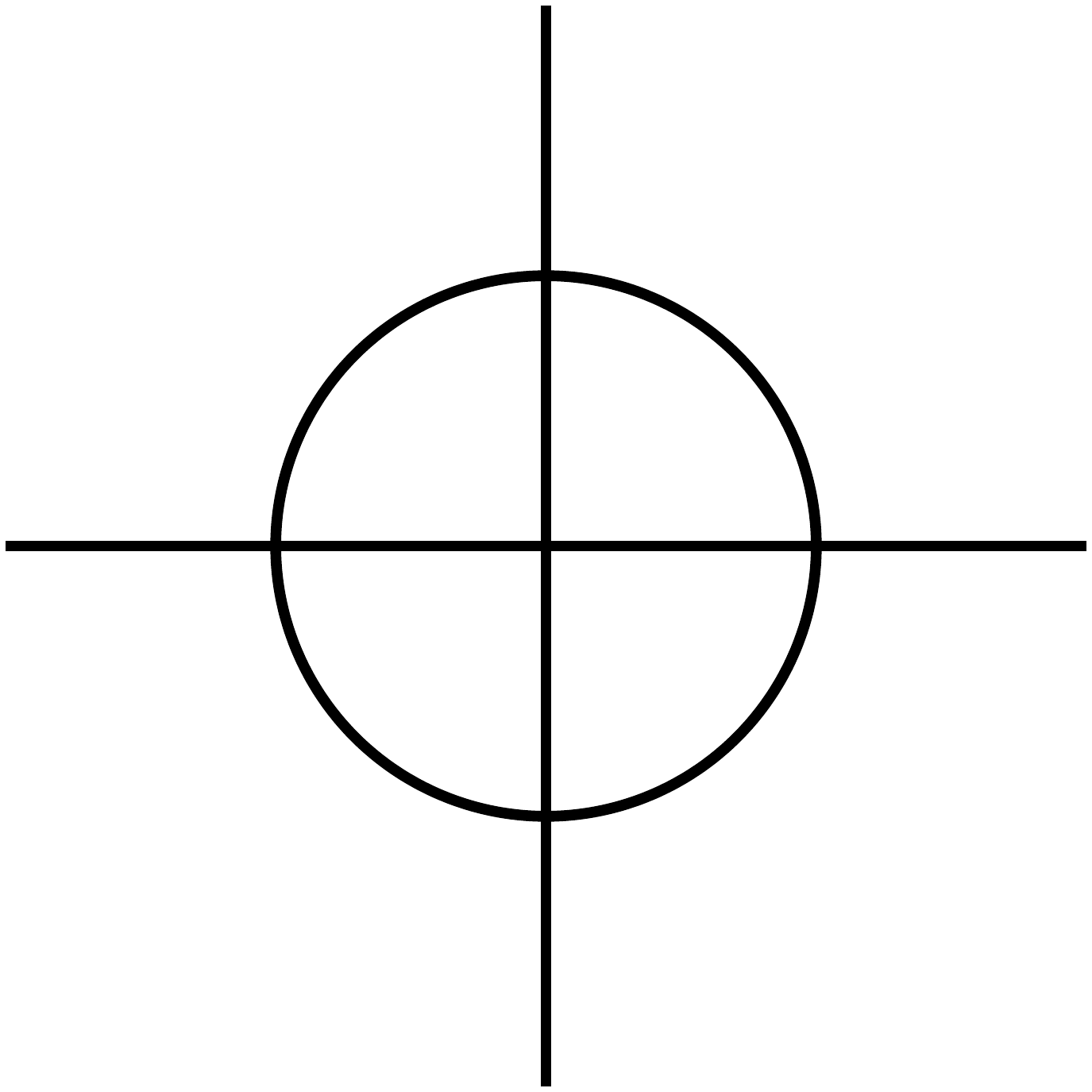}
 \includegraphics[width=0.15\textwidth]{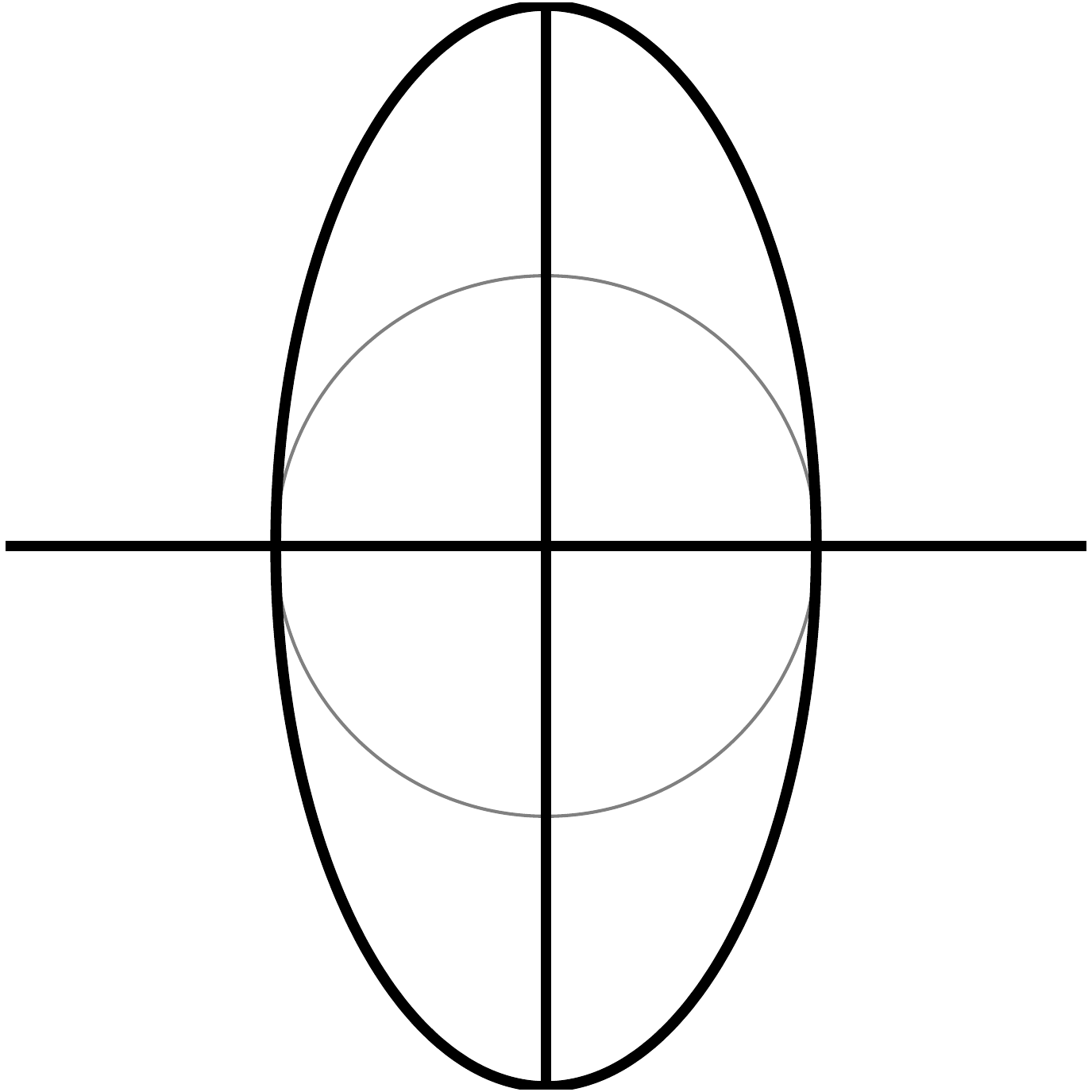}
 \includegraphics[width=0.15\textwidth]{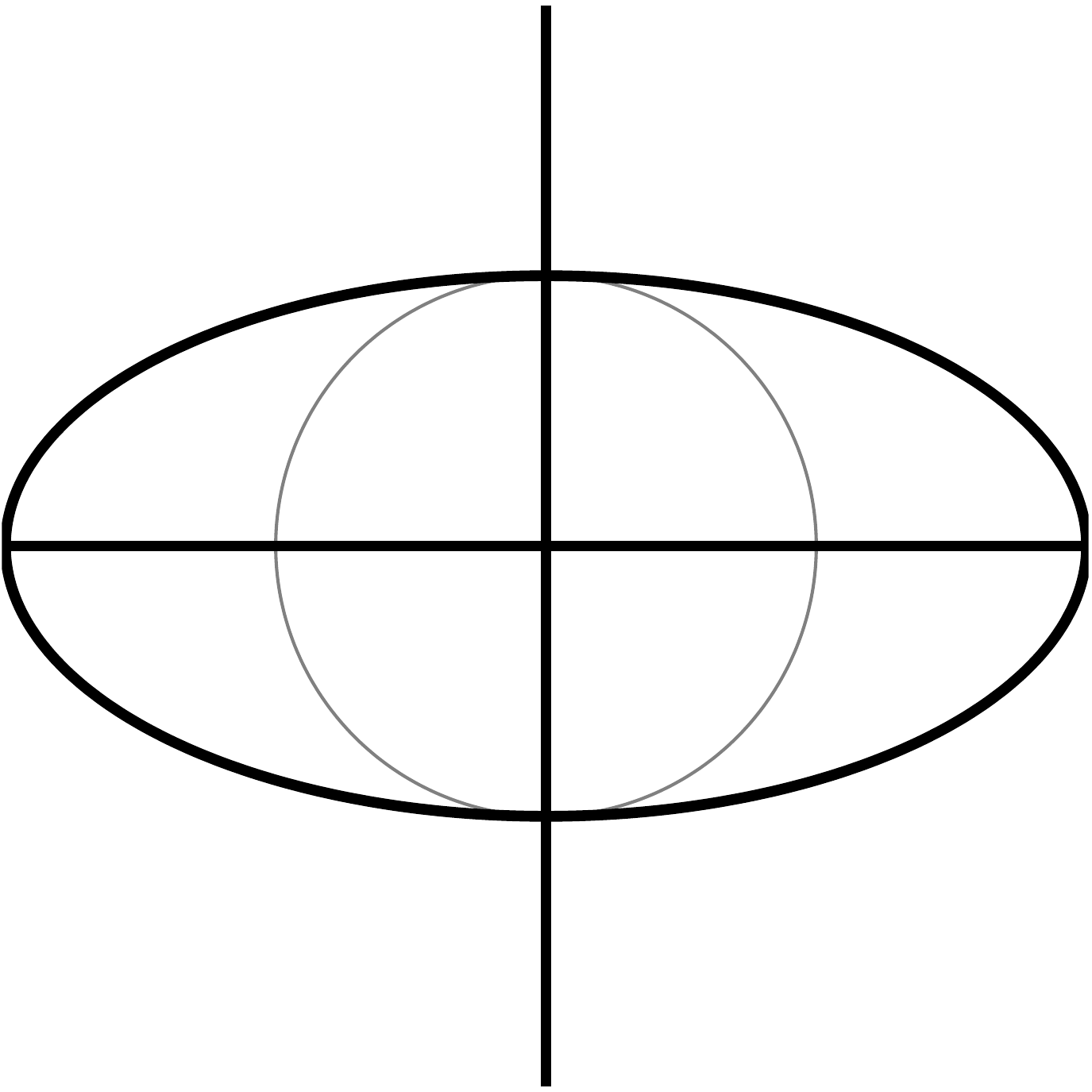}
 \includegraphics[width=0.15\textwidth]{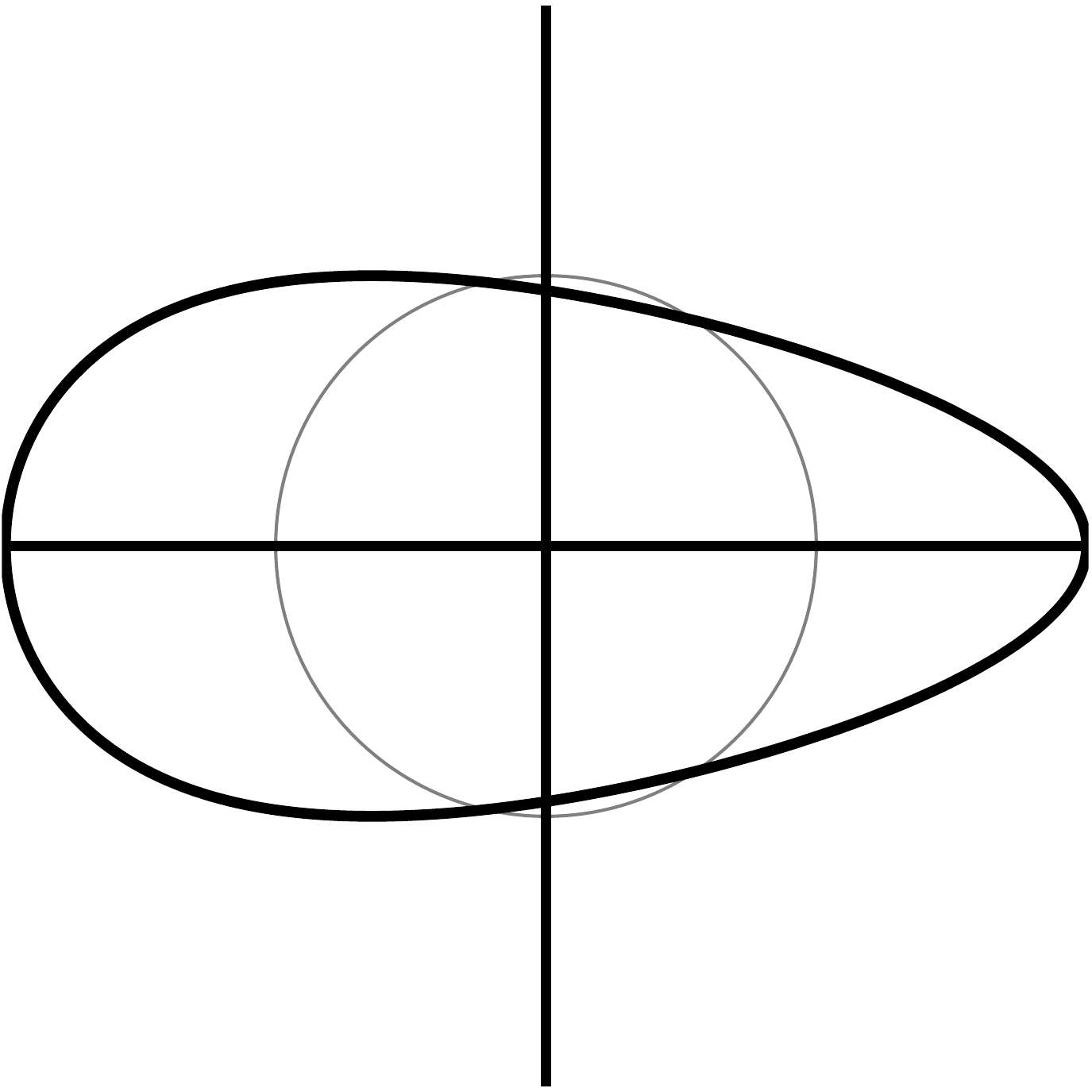}
 \includegraphics[width=0.15\textwidth]{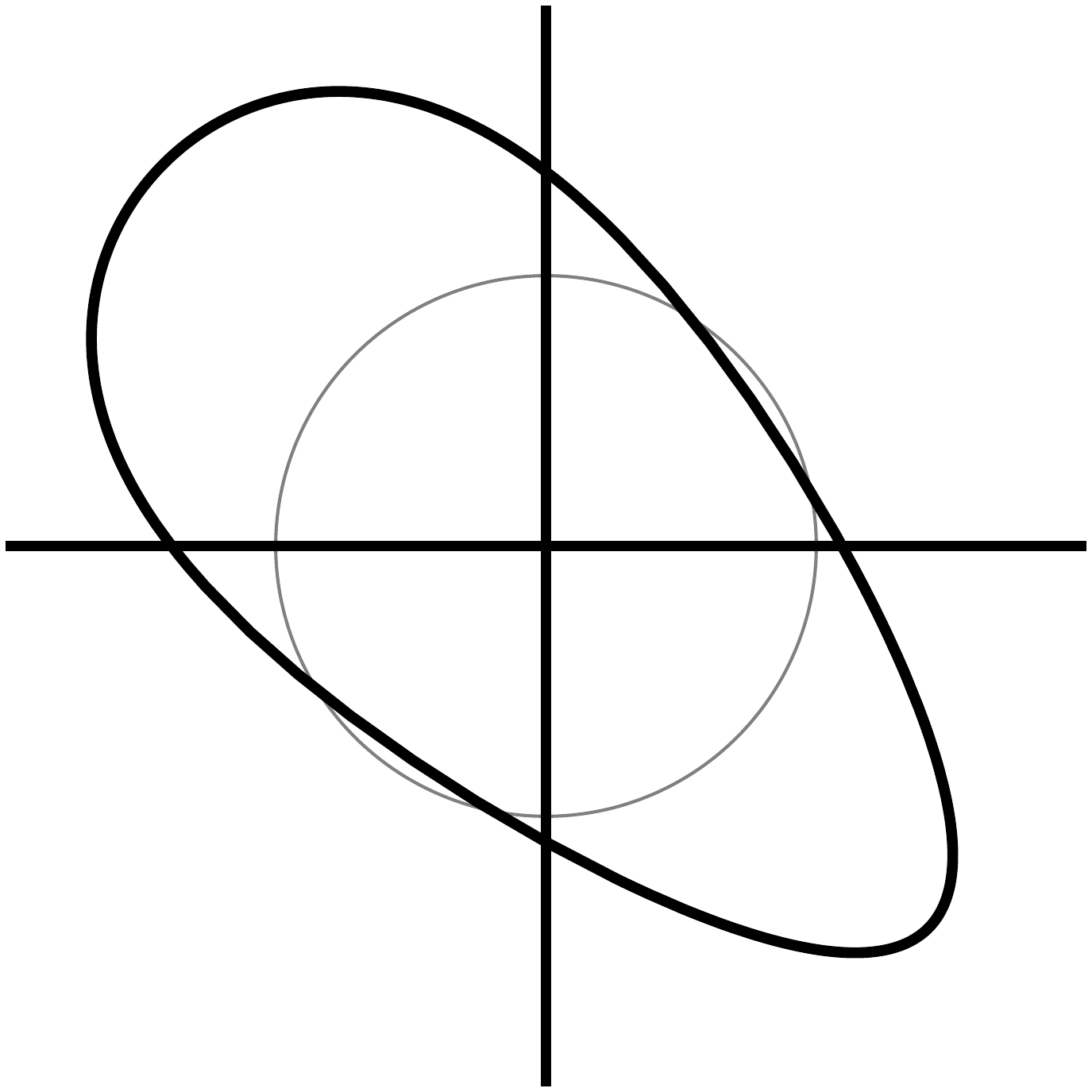}
 \includegraphics[width=0.185\textwidth]{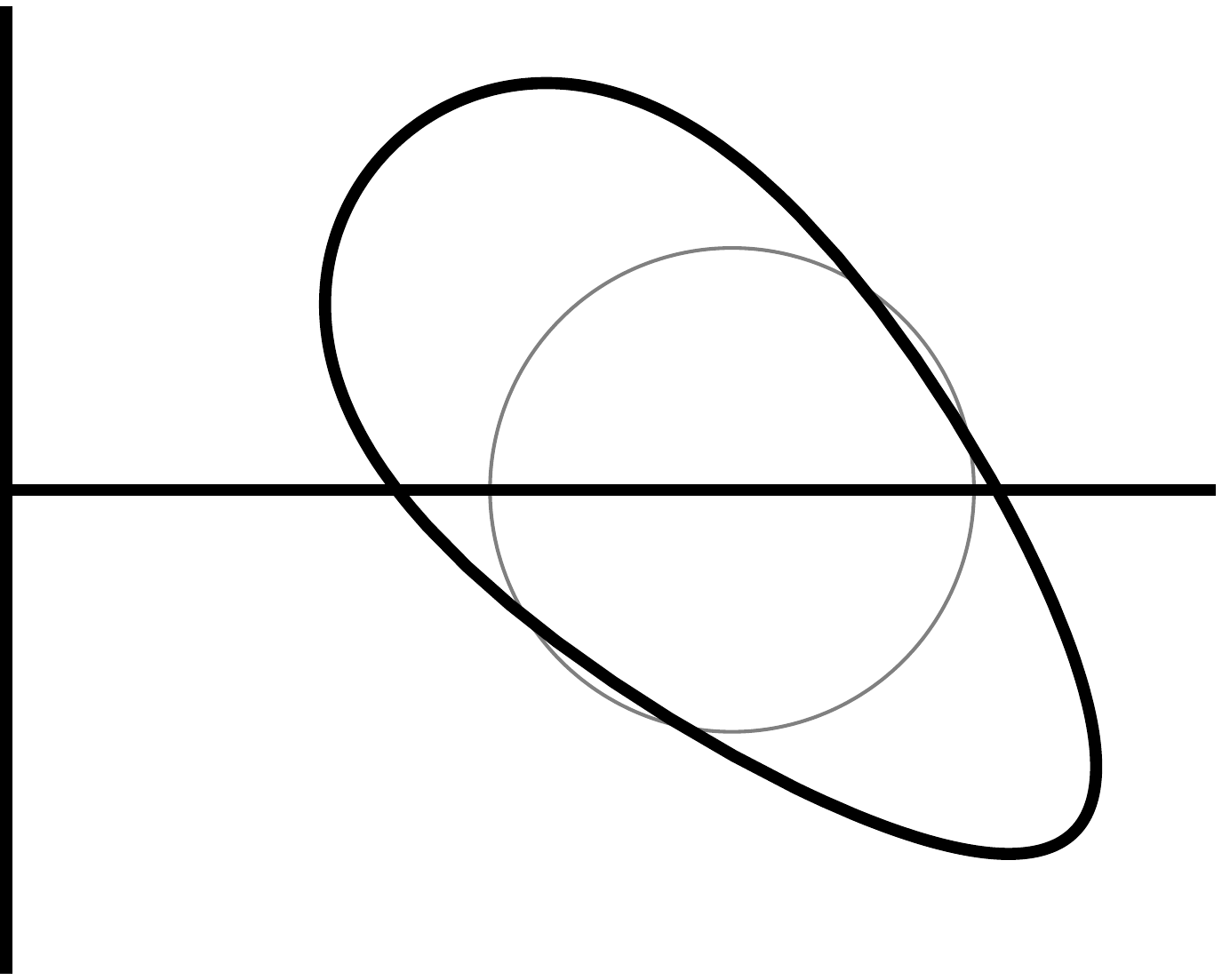}
 \caption{Demonstration of each stage of the iterative method (eqs. \refEq{eq:iterativeGeoSpecRcirc} through \refEq{eq:iterativeGeoSpecZnew}) to specify geometry with $R_{0N} = 3$, $r_{\psi N} = 1$, $\kappa = 2$, $\delta = 0.7$, $\theta_{\kappa} = 3 \pi / 4$, and $\theta_{\delta} = \pi / 4$.}
 \label{fig:geoTransformationExamples}
\end{figure}

The new, more general specification is done by adding each of the shaping effects in and tilting the appropriate angle. The new definition is
\begin{align}
   \vartheta' &\equiv \vartheta + \vartheta_{shift} , \\
   \nonumber \displaybreak[0] \\
   R_{c} \left( r_{\psi N}, \vartheta \right) &\equiv r_{\psi N} \cos \left( \vartheta' + \theta_{\kappa} \left( r_{\psi N} \right) - \theta_{\delta} \left( r_{\psi N} \right) \right) \label{eq:iterativeGeoSpecRcirc} , \\
   Z_{c} \left( r_{\psi N}, \vartheta \right) &\equiv r_{\psi N} \sin \left( \vartheta' + \theta_{\kappa} \left( r_{\psi N} \right) - \theta_{\delta} \left( r_{\psi N} \right) \right) , \\
   \nonumber \displaybreak[0] \\
   R_{\kappa} \left( r_{\psi N}, \vartheta \right) &\equiv R_{c}  \left( r_{\psi N}, \vartheta \right) , \\
   Z_{\kappa} \left( r_{\psi N}, \vartheta \right) &\equiv Z_{c}  \left( r_{\psi N}, \vartheta \right) + \left( \kappa \left( r_{\psi N} \right) - 1 \right) r_{\psi N} \sin \left( \vartheta' + \theta_{\kappa} \left( r_{\psi N} \right) - \theta_{\delta} \left( r_{\psi N} \right) \right) , \\
   \nonumber \displaybreak[0] \\
   R_{\kappa}^{tilt} \left( r_{\psi N}, \vartheta \right) &\equiv R_{\kappa}  \left( r_{\psi N}, \vartheta \right) \cos \left( \theta_{\kappa} \left( r_{\psi N} \right) - \theta_{\delta} \left( r_{\psi N} \right) \right) \nonumber \\
   &+ Z_{\kappa} \left( r_{\psi N}, \vartheta \right) \sin \left( \theta_{\kappa} \left( r_{\psi N} \right) - \theta_{\delta} \left( r_{\psi N} \right) \right) , \\
   Z_{\kappa}^{tilt} \left( r_{\psi N}, \vartheta \right) &\equiv Z_{\kappa} \left( r_{\psi N}, \vartheta \right) \cos \left( \theta_{\kappa} \left( r_{\psi N} \right) - \theta_{\delta} \left( r_{\psi N} \right) \right) \nonumber \\
   &- R_{\kappa} \left( r_{\psi N}, \vartheta \right) \sin \left( \theta_{\kappa} \left( r_{\psi N} \right) - \theta_{\delta} \left( r_{\psi N} \right) \right) , \\
   \nonumber \displaybreak[0] \\
   R_{\delta} \left( r_{\psi N}, \vartheta \right) &\equiv R_{\kappa}^{tilt} \left( r_{\psi N}, \vartheta \right) + r_{\psi N} \left[ \cos \left( \vartheta' + \delta \left( r_{\psi N} \right) \sin \left( \vartheta' \right) \right) - \cos \left( \vartheta' \right) \right] , \\
   Z_{\delta} \left( r_{\psi N}, \vartheta \right) &\equiv Z_{\kappa}^{tilt} \left( r_{\psi N}, \vartheta \right) , \\
   \nonumber \displaybreak[0] \\
   R_{\delta}^{tilt} \left( r_{\psi N}, \vartheta \right) &\equiv R_{\delta} \left( r_{\psi N}, \vartheta \right) \cos \left( \theta_{\delta} \left( r_{\psi N} \right) \right) + Z_{\delta} \left( r_{\psi N}, \vartheta \right) \sin \left( \theta_{\delta} \left( r_{\psi N} \right) \right) , \\
   Z_{\delta}^{tilt} \left( r_{\psi N}, \vartheta \right) &\equiv Z_{\delta} \left( r_{\psi N}, \vartheta \right) \cos \left( \theta_{\delta} \left( r_{\psi N} \right) \right) - R_{\delta} \left( r_{\psi N}, \vartheta \right) \sin \left( \theta_{\delta} \left( r_{\psi N} \right) \right) , \\
   \nonumber \displaybreak[0] \\
   R_{N}^{new} \left( r_{\psi N}, \vartheta \right) &= R_{0 N} \left( r_{\psi N} \right) + R_{\delta}^{tilt} \left( r_{\psi N}, \vartheta \right) , \label{eq:iterativeGeoSpecRnew} \\
   Z_{N}^{new} \left( r_{\psi N}, \vartheta \right) &= Z_{\delta}^{tilt} \left( r_{\psi N}, \vartheta \right) . \label{eq:iterativeGeoSpecZnew}
\end{align}
Fig. \ref{fig:geoTransformationExamples} shows each step of this geometry specification process. Note that $\delta \in \left( - \pi / 2, \pi / 2 \right)$, otherwise the flux surface cross-section can develop singular points. As before, calculating the poloidal magnetic field still requires the radial derivatives of the input parameters appearing in the flux surface specification. The translation of $\vartheta$ by $\theta_{\kappa} \left( r_{\psi N} \right) - \theta_{\delta} \left( r_{\psi N} \right)$ only serves to get the proper phase between the effects of elongation and triangularity. The $\vartheta_{shift}$ parameter ultimately determines the location of $\vartheta = 0$ and will be discussed in Section \ref{subsec:bouncePointTreatment}. In this work, all radial derivatives of quantities appearing in eqs. \refEq{eq:iterativeGeoSpecRcirc} through \refEq{eq:iterativeGeoSpecZnew} are set to zero.

\subsection{Numerical differentiation}
\label{subsec:numericalDifferentiation}

Within the Miller geometry module there are several numerical derivatives taken using the parameterized flux surfaces. Originally, these derivatives were taken over the $\vartheta \in \left[ 0,\pi \right]$ domain and later copied, with the proper symmetry, to the $\vartheta \in \left( - \pi, 0 \right]$ domain. For modeling up-down asymmetric geometries, these subroutines were converted to use the $\vartheta \in \left[ - \pi, \pi \right]$ domain throughout the entire algorithm.
 
\subsection{Treatment of bounce points}
\label{subsec:bouncePointTreatment}

A more subtle issue stems from GS2's treatment of trapped particles. The gridder is the portion of the code responsible for taking the input geometry and discretizing the spatial and velocity dimensions. The poloidal and velocity grids are structured so that particles have velocities such that they only ever bounce at grid points and not between grid points. To do this, the gridder assumes that the location of the maximum total magnetic field is at $\vartheta = \pm \pi$. However, this is not automatically the case for up-down asymmetric configurations.

Rather than modify the inner workings of the gridder, the definition of $\vartheta$ was translated by the quantity $\vartheta_{shift} \equiv \pm \pi - \vartheta_{B_{max}}$, where $\vartheta_{B_{max}}$ is the location of the maximum of $B$ in the flux surface. Therefore, the assumption is always satisfied. However, the location of the maximum magnetic field, $\vartheta_{B_{max}}$, in a general flux surface with separately tilted elongation and triangularity is not analytic. So, for the sake of convenience, this was implemented in GS2 as a two step process. First, the geometry is discretized and $B \left( \vartheta \right)$ is calculated with $\vartheta_{shift} = 0$, as was already the case. Then the code searches through all the values of $\vartheta$ and finds $\vartheta_{B_{max}}$. If $\vartheta_{B_{max}} = \pm \pi$, the code moves forwards. Otherwise, $\vartheta_{shift}$ is set accordingly and the initialization routine is started from the beginning a second time.

\subsection{Code benchmarking}

Several different tests were used to verify that the modifications to GS2 introduced no errors and that no further modifications were necessary to properly treat up-down asymmetry. First, a collisionless linear analytic solution to the gyrokinetic equation (with $k_{\alpha} = 0$) was found and compared to GS2 output. Also, the new input parameters added to the code allow a physical geometry to be specified in different ways. These different specifications were tested to ensure that they produced equivalent results. Lastly, all effects of the system geometry appear in the gyrokinetic equation as eight individual coefficients. All of these were calculated independently and compared against those calculated within GS2.

\subsubsection{Stationary mode test.}

The stationary mode test case is a comparison between an analytic calculation and GS2 results. The analytic calculation starts with the Fourier analyzed gyrokinetic equation (see eq. \refEq{eq:gyrokineticEq}). Now we choose to focus on modes with $k_{\alpha} = 0$ and ignore collisions. These two conditions can be enforced in GS2 by setting $\mathtt{aky} = 0$ and \texttt{collision\_model} = \texttt{`none'}. Next, we postulate that time-independent solutions for $h_{s}$ and $\phi$ exist and seek them by letting $\partial / \partial t = 0$. These simplifications, along with changing velocity space variables $\left( w_{||}, \mu, \varphi \right) \rightarrow \left( \mathscr{E}, \mu, \varphi \right)$, gives
\begin{eqnarray}
   w_{||} \left. \frac{\partial h_{s}}{\partial \vartheta} \right|_{\mathscr{E}, \mu} = i \frac{k_{\psi} I}{m_{s} \Omega_{s} B} \left( m_{s} w_{||}^{2} + \mu B \right) \frac{\partial B}{\partial \vartheta} h_{s} .
\end{eqnarray}

Solving for the nonadiabatic distribution function we find that
\begin{eqnarray}
   h_{s} \left( k_{\psi}, \vartheta, \mathscr{E}, \mu \right) = h_{s 0} \left( k_{\psi}, \mathscr{E}, \mu \right) \text{exp} \left( -i \frac{k_{\psi} w_{||}}{\Omega_{s}} I \right) , \label{eq:stationaryStateUnnormalizedDistFn}
\end{eqnarray}
where we choose the free function to be $h_{s 0} = Z_{N s} \left( \rho_{r} / l_{r} \right) F_{M s}$. The factor of $Z_{N s}$ is added for numerical reasons that will be discussed later and $\left( \rho_{r} / l_{r} \right) F_{M s}$ is chosen for proper GS2 normalization. Now we substitute this result into the quasineutrality equation, given by eq. \refEq{eq:gyroQuasineutrality}. Solving for the perturbed electric potential, using the identity
\begin{eqnarray}
   J_{0} \left( z \right) = \frac{1}{2 \pi} \oint_{0}^{2 \pi} d \varphi ~ \text{exp} \left( i z \sin \varphi \right)
\end{eqnarray}
and the change of integration variables $\left( w_{||}, \mu, \varphi \right) \rightarrow \left( w_{||}, w_{x}, w_{y} \right)$, we find
\begin{eqnarray}
   \phi = \left( \sum_{s} \frac{Z_{s}^{2} e n_{s}}{T_{s}} \right)^{-1} \frac{\rho_{r}}{l_{r}} \sum_{s} n_{s} Z_{s} Z_{N s} \text{exp} \left( - \frac{1}{2} \frac{k_{\psi}^{2}}{\Omega_{s}^{2}} R^{2} B^{2} \frac{T_{s}}{m_{s}} \right) . \label{eq:stationaryStatePhi}
\end{eqnarray}
Using eqs. \refEq{eq:kPsiDef} and \refEq{eq:kAlphaDef} with the definition of the complementary distribution function
\begin{eqnarray}
   g_{s} \equiv h_{s} - \frac{Z_{s} e}{T_{s}} J_{0} \left( \frac{k_{\perp} \sqrt{2 \mu B}}{\Omega_{s} \sqrt{m_{s}}} \right) \phi F_{M s} , \label{eq:compDistFnDef}
\end{eqnarray}
the distribution function that GS2 actually manipulates internally, we find
\begin{eqnarray}
   g_{N s} &= Z_{N s} \text{exp} \left( -i \frac{q}{r_{\psi N}} k_{x N} w_{|| N} R_{geo N} \frac{\sqrt{m_{N s} T_{N s}}}{Z_{N s} B_{N}} \right) \nonumber \\
   &- \frac{Z_{N s}}{T_{N s}} J_{0} \left( k_{\perp N} w_{\perp N} \frac{\sqrt{m_{N s} T_{N s}}}{Z_{N s} B_{N}} \right) \phi_{N} \label{eq:stationaryStateDistFnSol}
\end{eqnarray}
and
\begin{eqnarray}
   \phi_{N} = \left( \sum_{j} \frac{Z_{N j}^{2} n_{N j}}{T_{N j}} \right)^{-1} \sum_{k} n_{N k} Z_{N k}^{2} \text{exp} \left( - \frac{1}{4} \frac{q^{2}}{r_{\psi N}^{2}} k_{x N}^{2} R_{N}^{2} \frac{m_{N k} T_{N k}}{Z_{N s}^{2}} \right) . \label{eq:stationaryStatePotentialSol}
\end{eqnarray}

Therefore if we initialize the distribution function to eq. \refEq{eq:stationaryStateDistFnSol}, we expect the calculated potential at every grid point in $\vartheta$ to match eq. \refEq{eq:stationaryStatePotentialSol} and neither quantity to change in time. To quantify the time independence, at each poloidal grid point, we first calculate the fractional error between eq. \refEq{eq:stationaryStatePotentialSol} and the calculated potential distribution after 500 GS2 time steps of $0.1 l_{r} / v_{th r}$. The mean, $\mu_{err}$, is calculated from the fractional error at every $\vartheta$ grid point, producing a single number that indicates if a given GS2 run is treating geometrical effects correctly.

Initially, the factor of $Z_{s}$ was not included in the integration constant of eq. \refEq{eq:stationaryStateUnnormalizedDistFn}, causing the summations over species in eq. \refEq{eq:stationaryStatePotentialSol} to become a difference between the ion and electron terms. Depending on the argument of the exponent in eq. \refEq{eq:stationaryStatePhi}, this caused numerical errors to dominate and prevented all distribution functions from maintaining a stationary state. Introducing the factor of $Z_{s}$ into the integration constant of eq. \refEq{eq:stationaryStateUnnormalizedDistFn} keeps this cancellation from occurring and makes the problem better conditioned.

A total of 70 simulations were run for the test, consisting of five different geometries, each run at seven different radial wavenumbers, using both the original and updated versions of the code. All simulations were performed at very high spatial (with $\sim 128$ grid points in $\vartheta$) and velocity space ($\sim 32$ energy grid points and $\sim 20$ untrapped pitch angles moving in one direction along field line) resolution. Also, they were shaped variants of the Cyclone base case geometry given in table \ref{tab:simCaseParameters}. The Cyclone base case is a standard benchmark case used in tokamak simulations \cite{DimitsCycloneBaseCase2000} and is modeled after a particular DIII-D shot. Improperly treated up-down asymmetry was introduced into the original code as a control for the test.

\begin{figure}
 \centering
 (a) \hspace{0.8\textwidth}

 \includegraphics[width=0.55\textwidth]{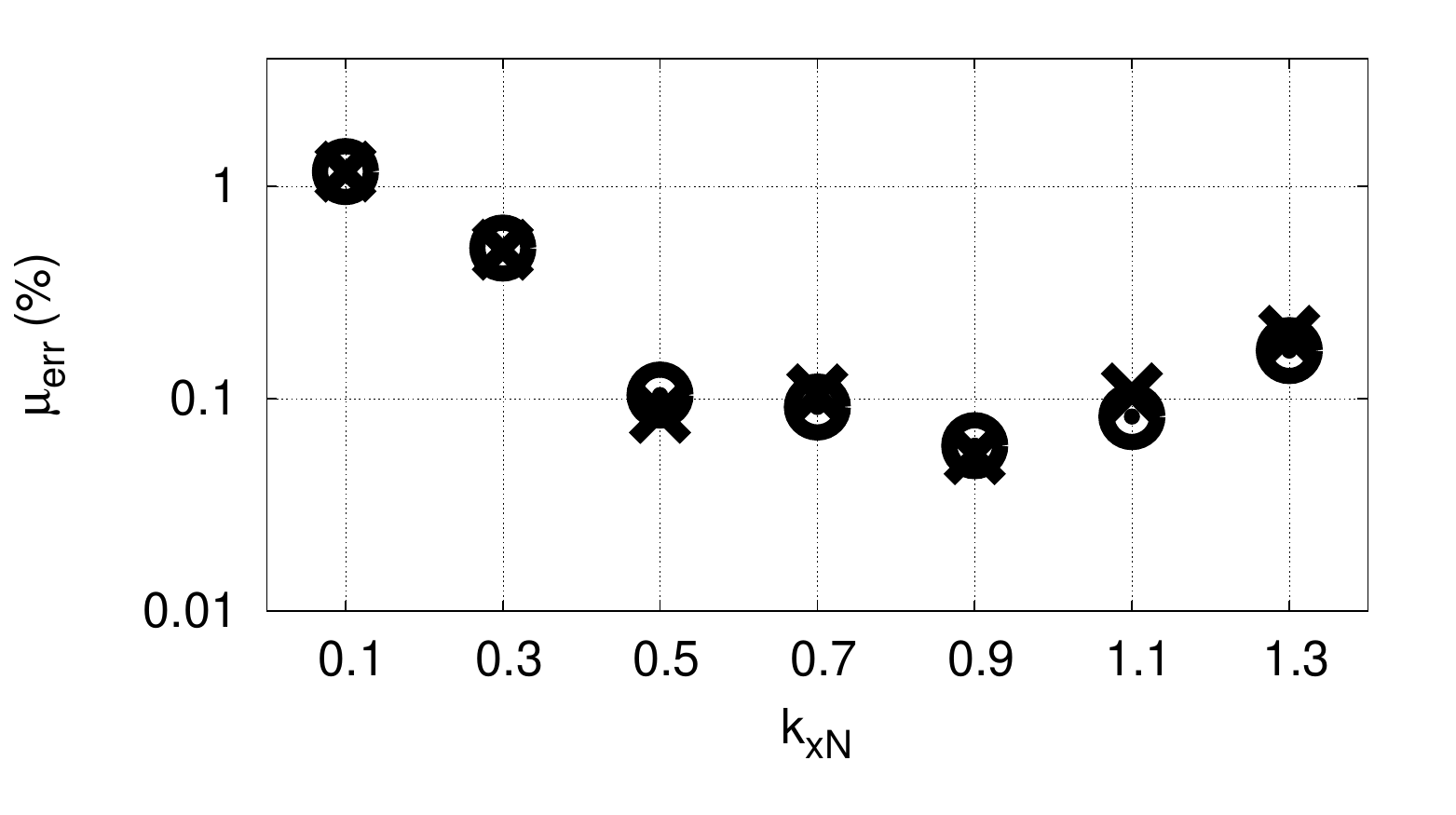}

 (b) \hspace{0.8\textwidth}

 \includegraphics[width=0.55\textwidth]{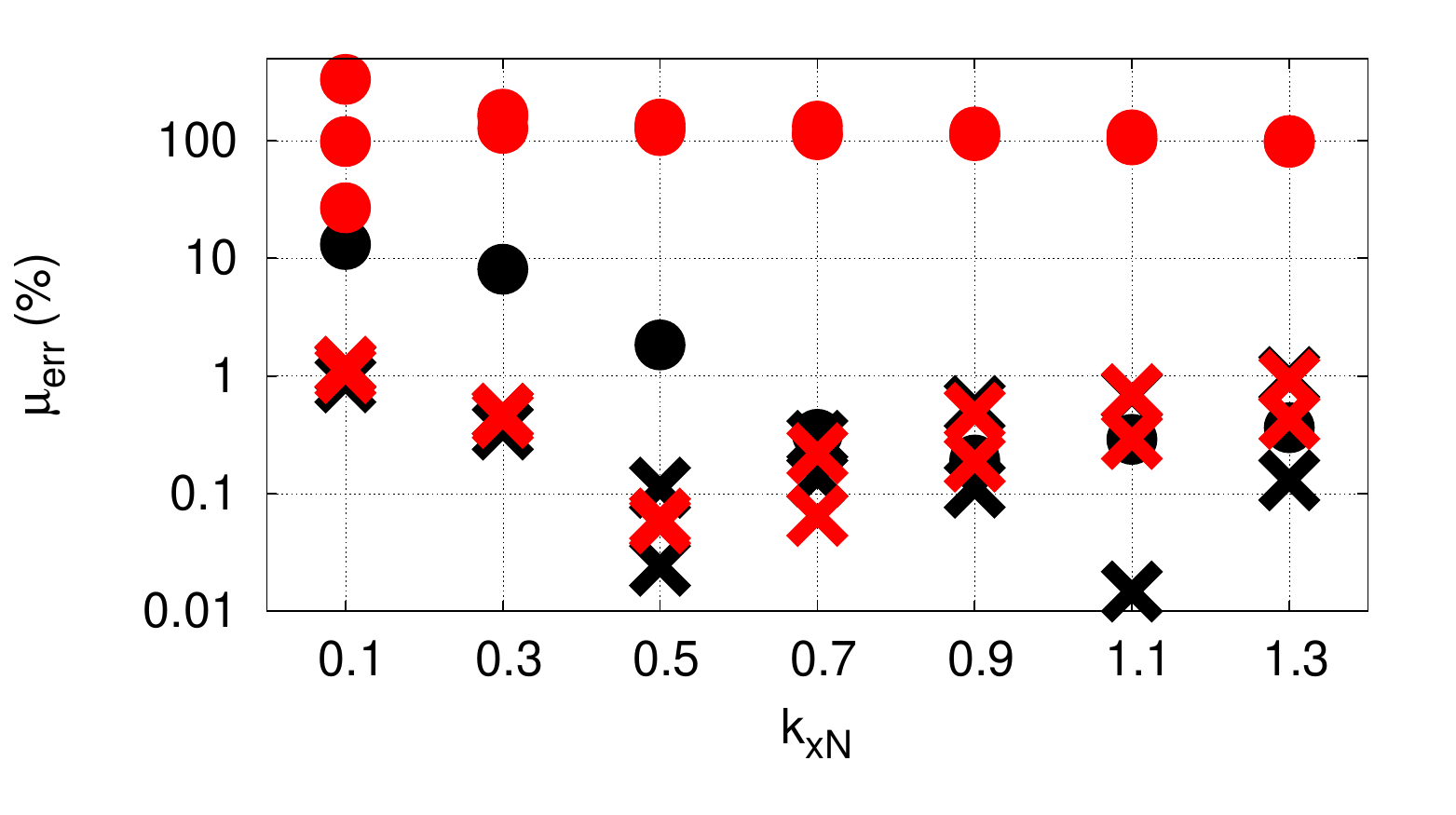}
 \caption{Stationary state test case error for both up-down symmetric (black) and up-down asymmetric (red) configurations performed using the original source code (circles) and the updated source code (crosses) for (a) circular flux surfaces or (b) shaped flux surfaces.}
 \label{fig:stationaryStateError}
\end{figure}

\begin{figure}
 \centering
 \includegraphics[width=0.55\textwidth]{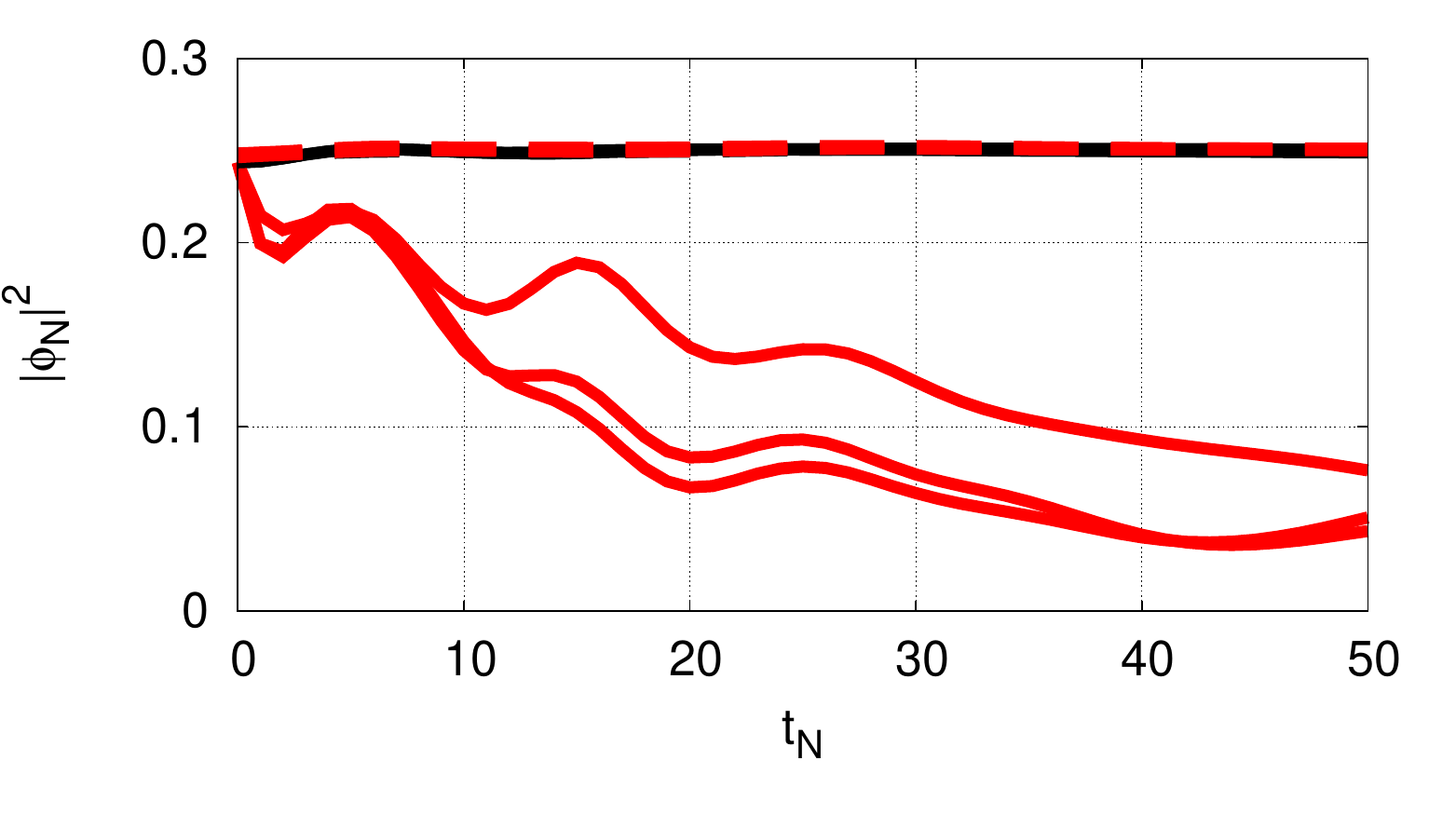}
 \caption{The potential amplitude with time for both up-down symmetric (black) and up-down asymmetric (red) configurations performed using the original source code (solid) and the updated source code (dotted). Only the test cases with $k_{x N} = 0.7$ are shown.}
 \label{fig:stationaryStatePhi2WithTime}
\end{figure}

The five simulation geometries modeled with the original source code consisted of a circular cross-section, an up-down symmetric triangular shape, and three up-down asymmetric shapes. We expect that these three up-down asymmetric geometries will fail to maintain the stationary state.

The five groups of simulations performed with the updated source code include a circular cross-section and four elongated shapes with $\kappa = 2$ and $\theta_{\kappa} \in \left\{ 0, \pi / 6, \pi / 3, \pi / 2 \right\}$. Therefore, two of these groups are up-down asymmetric, but, because of the updates, all should still maintain the stationary state.

The results, summarized in fig. \ref{fig:stationaryStateError}, were as expected. Fig. \ref{fig:stationaryStateError}a shows that the two codes produce nearly identical results for identical circular flux surfaces. Fig. \ref{fig:stationaryStateError}b shows a clear separation between the results of improperly treated up-down asymmetric runs using the original source code and all other runs. The up-down asymmetric runs using the updated version of the code have very similar error to the up-down symmetric runs. Furthermore, fig. \ref{fig:stationaryStatePhi2WithTime} shows that the improperly treated up-down asymmetric cases converge to different steady-state solutions than all of the other cases.


\subsubsection{Duplicate geometry test.}

\begin{figure}
 \centering
 \includegraphics[width=1.0\textwidth]{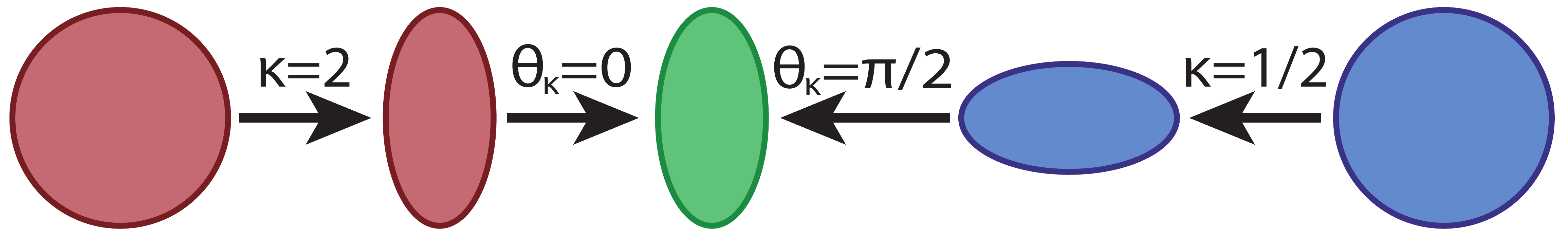}
 \caption{An example of two different GS2 specifications of the same physical geometry.}
 \label{fig:duplicateGeoSpecification}
\end{figure}

As illustrated in fig. \ref{fig:duplicateGeoSpecification}, the addition of the tilt parameters allows for multiple ways to specify the same physical geometry. Both these manners of specification should produce the same results. However, getting this test to work requires a comprehensive understanding of GS2's normalizations of input and output parameters. Given an arbitrary elongated configuration with no triangularity (indicated by a subscript 1), we can produce a physically identical configuration (indicated by a subscript 2) with a different GS2 specification using
\begin{eqnarray}
   \mathtt{akappa}_{2} &= \frac{1}{\mathtt{akappa}_{1}} , \\
   \mathtt{thetak}_{2} &= \mathtt{thetak}_{1} + \frac{\pi}{2} , \\
   \mathtt{rhoc}_{2} &= \mathtt{akappa}_{1} \mathtt{rhoc}_{1} , \\
   \mathtt{tprim}_{2} &= \frac{1}{\mathtt{akappa}_{1}} \mathtt{tprim}_{1} , \label{eq:tprimElongRotation} \\
   \mathtt{fprim}_{2} &= \frac{1}{\mathtt{akappa}_{1}} \mathtt{fprim}_{1} , \\
   \mathtt{y0}_{2} &= \frac{1}{\mathtt{akappa}_{1}} \mathtt{y0}_{1} , \\
   \mathtt{akx}_{2} &= \mathtt{akappa}_{1} \mathtt{akx}_{1} , \\
   \mathtt{aky}_{2} &= \mathtt{akappa}_{1} \mathtt{aky}_{1} , \\
   \mathtt{gds22}_{2} &= \left( \mathtt  {akappa}_{1} \right) ^{2} \mathtt{gds22}_{1} , \\
   \mathtt{gds21}_{2} &= \left( \mathtt{akappa}_{1} \right) ^{2} \mathtt{gds21}_{1} , \\
   \mathtt{gds2}_{2} &= \left( \mathtt{akappa}_{1} \right) ^{2} \mathtt{gds2}_{1} \label{eq:gds2ElongRotation}
\end{eqnarray}
where $\mathtt{tprim} \equiv 1 / L_{T N s} \equiv - \left( 1 / T_{s} \right) \partial T_{s} / \partial r_{\psi N}$ is the background temperature gradient, $\mathtt{fprim} \equiv 1 / L_{n N s} \equiv - \left( 1 / n_{s} \right) \partial n_{s} / \partial r_{\psi N}$ is the background density gradient, and $2 \pi \mathtt{y0}$ is the flux tube box size in the $\vec{\nabla} \alpha$ direction, while $\mathtt{gds22} \equiv \left( d q_{s} / d r_{\psi N} \right)^{2} \left| \vec{\nabla}_{N} \psi_{N} \right|^{2}$, $\mathtt{gds21} \equiv \left( d q_{s} / d r_{\psi N} \right) \left( d \psi_{N} / d r_{\psi N} \right)  \vec{\nabla}_{N} \psi_{N} \cdot \vec{\nabla}_{N} \alpha$, and $\mathtt{gds2} \equiv \left( d \psi_{N} / d r_{\psi N} \right)^{2} \left| \vec{\nabla}_{N} \alpha \right|^{2}$ are geometric coefficients that GS2 calculates internally. The factors of $\mathtt{akappa_{1}}$ arise in eqs. \refEq{eq:tprimElongRotation} through \refEq{eq:gds2ElongRotation} because GS2 chooses $\left| \vec{\nabla} \psi \right|$ at the midplane of the ellipse before tilting to normalize quantities such as $k_{\psi}$ and $k_{\alpha}$. As a result, the value of $\left| \vec{\nabla} \psi \right|$ used for normalizations is different for the two configurations. An analogous transformation exists for triangular flux surfaces with no elongation, given by
\begin{eqnarray}
   \mathtt{tri}_{2} &= -\mathtt{tri}_{1} , \\
   \mathtt{thetad}_{2} &= \mathtt{thetad}_{1} + \pi , \\
   \mathtt{tprim}_{2} &= \mathtt{tprim}_{1} , \\
   \mathtt{fprim}_{2} &= \mathtt{fprim}_{1} , \\
   \mathtt{y0}_{2} &= \mathtt{y0}_{1} , \\
   \mathtt{akx}_{2} &= \mathtt{akx}_{1} , \\
   \mathtt{aky}_{2} &= \mathtt{aky}_{1} , \\
   \mathtt{gds22}_{2} &= \mathtt{gds22}_{1} , \\
   \mathtt{gds21}_{2} &= \mathtt{gds21}_{1} , \\
   \mathtt{gds2}_{2} &= \mathtt{gds2}_{1} .
\end{eqnarray}

\begin{figure}
 \centering
 (a) \hspace{0.8\textwidth}

 \includegraphics[width=0.55\textwidth]{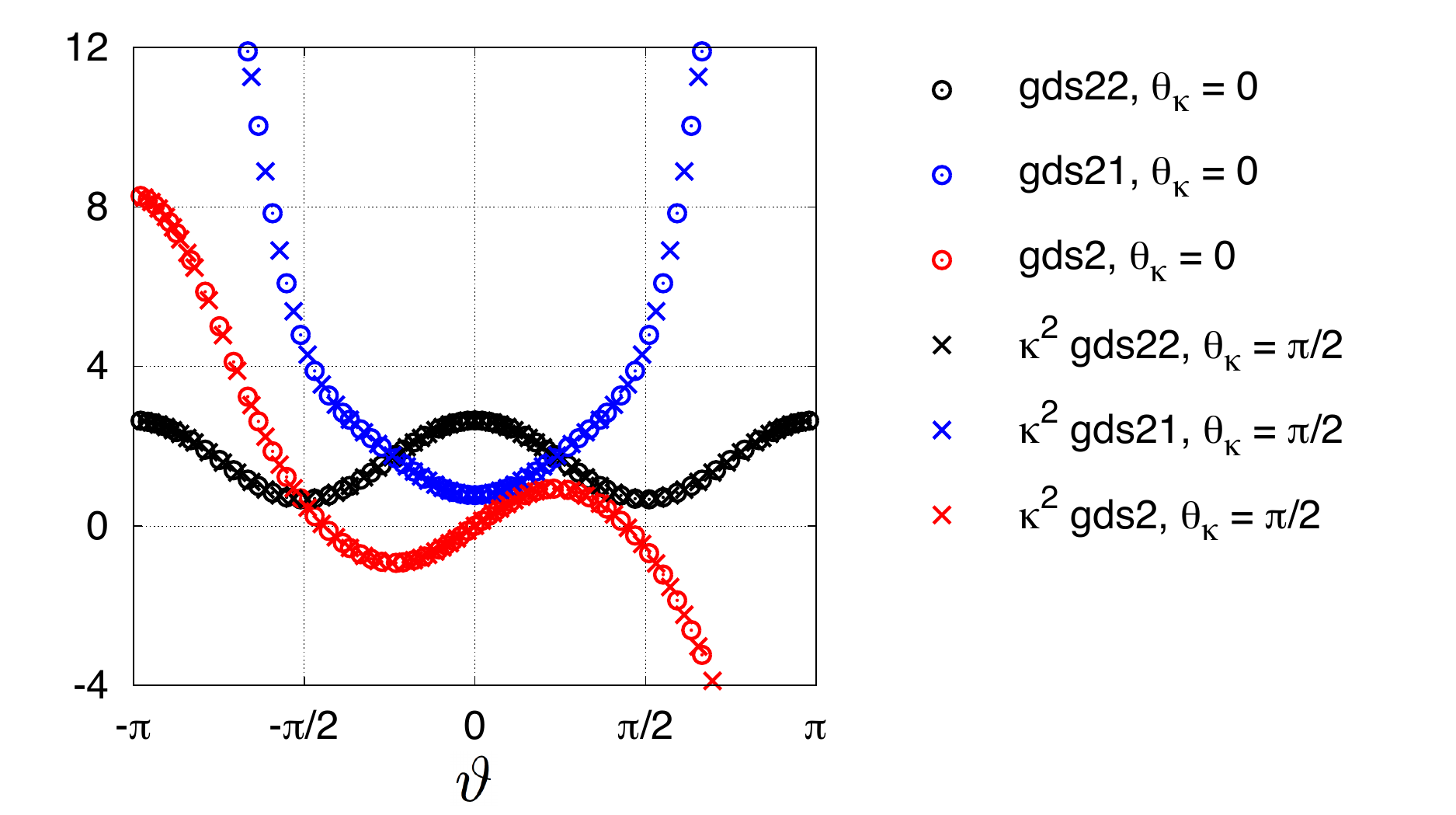}

 (b) \hspace{0.8\textwidth}

 \includegraphics[width=0.65\textwidth]{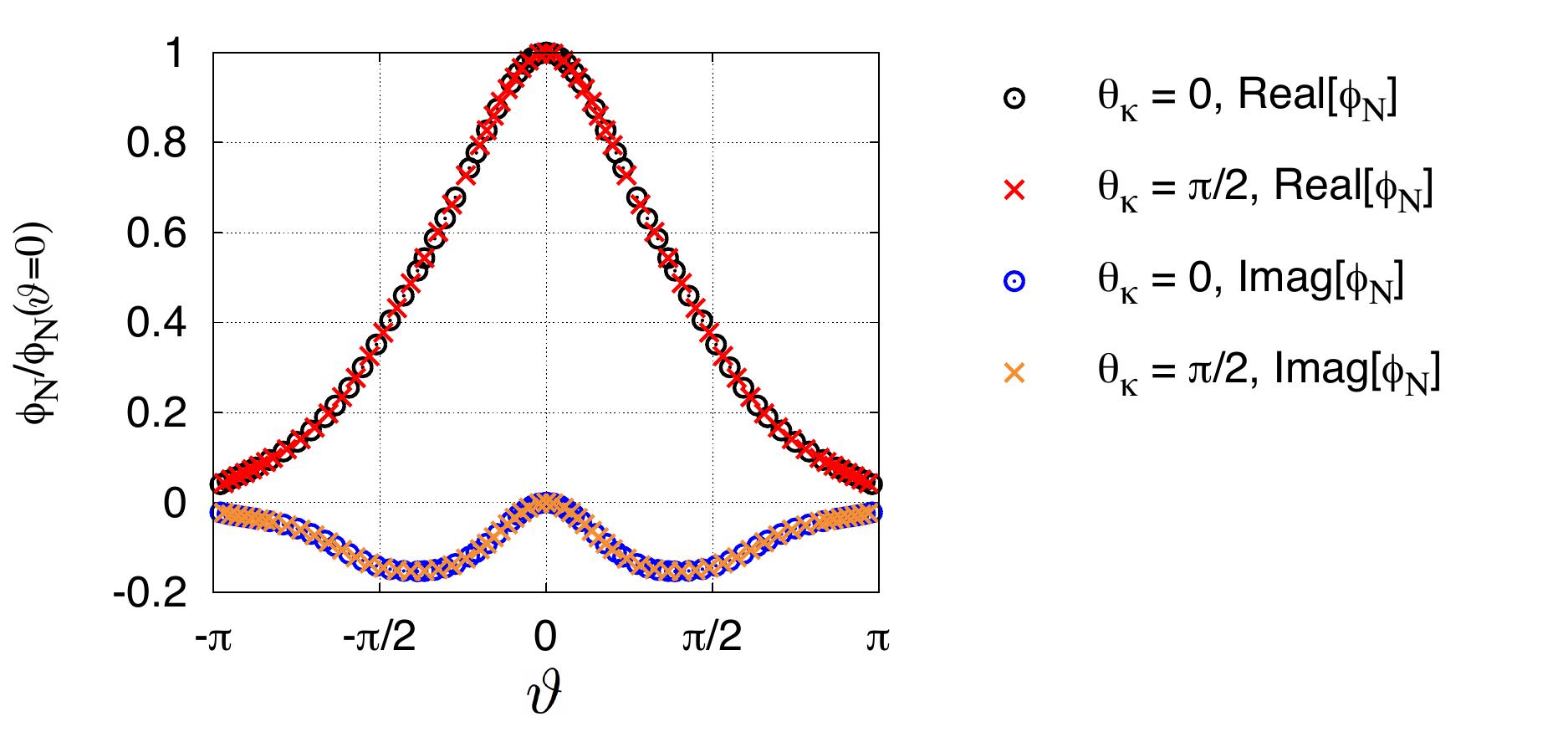}
 \caption{Comparison of (a) geometric coefficients and (b) potential for the two geometric specifications with $\omega_{N} = 0.2727 + 0.2907 i$ for $\theta_{\kappa} = 0$ and $\omega_{N} = 0.2727 + 0.2908 i$ for $\theta_{\kappa} = \pi/2$.}
 \label{fig:duplicateGeoBenchmark}
\end{figure}

Two elongated cases, one with no tilt and one with tilt, were run linearly for $k_{x} = 0$ and $k_{y} \neq 0$ in order to test a parameter space missed by the stationary state test (where $k_{y} = 0$). As expected, fig. \ref{fig:duplicateGeoBenchmark} shows that these two configurations both produce identical geometric coefficients as well as converge to the same linear growth rate and mode shape.

\begin{figure}
 \centering
 \includegraphics[width=0.6\textwidth]{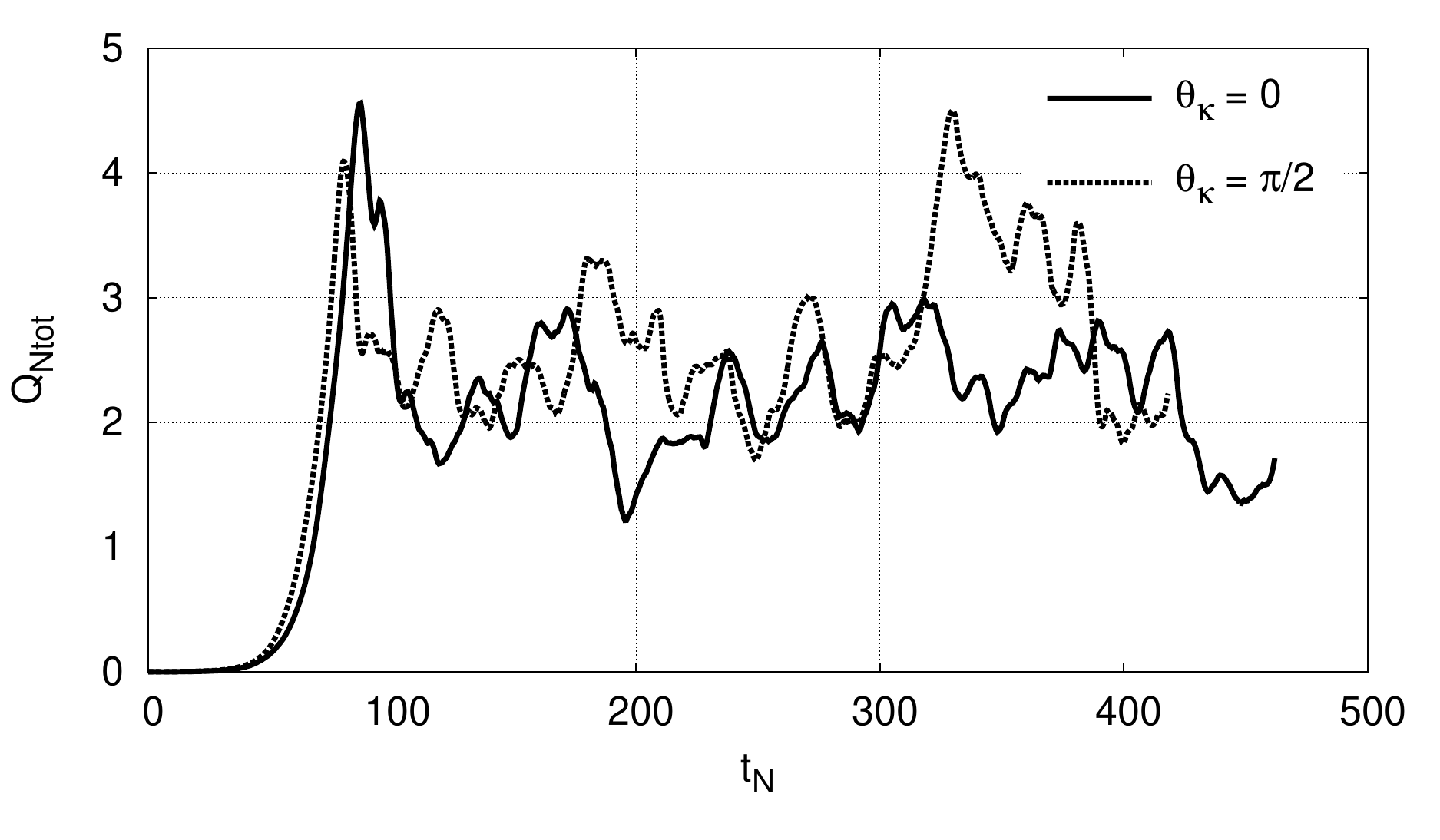}
 \caption{Comparison of total heat flux for the two geometric specifications.}
 \label{fig:duplicateGeoBenchmarkTotHeatFlux}
\end{figure}

Also, elongated test cases were run for a large number of nonlinearly interacting modes. Because of the fluctuating nature of nonlinear runs we only expect the two results to behave identically in the statistical sense. We can see in fig. \ref{fig:duplicateGeoBenchmarkTotHeatFlux} that the heat fluxes, normalized to their gyroBohm values, are identical through the linear growth phase (up to $t_{N} \sim 20$). Afterwards, during the nonlinear saturation, we see the results diverge, but still saturate at the same level, when averaged in time.

\subsubsection{Geometric coefficient test.}

\begin{figure}
 \centering
 (a) \hspace{0.8\textwidth}

 \includegraphics[width=0.8\textwidth]{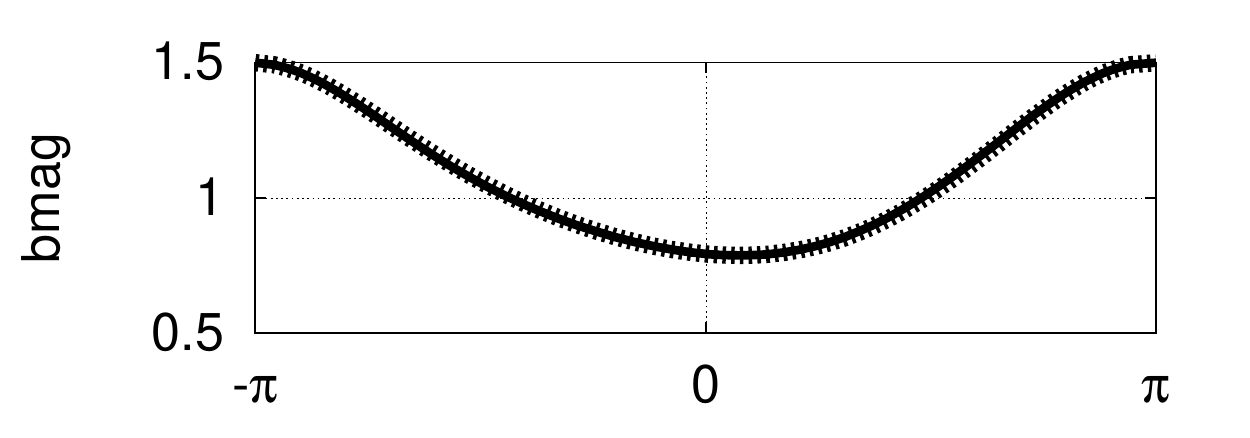}

 (b) \hspace{0.8\textwidth}

 \includegraphics[width=0.8\textwidth]{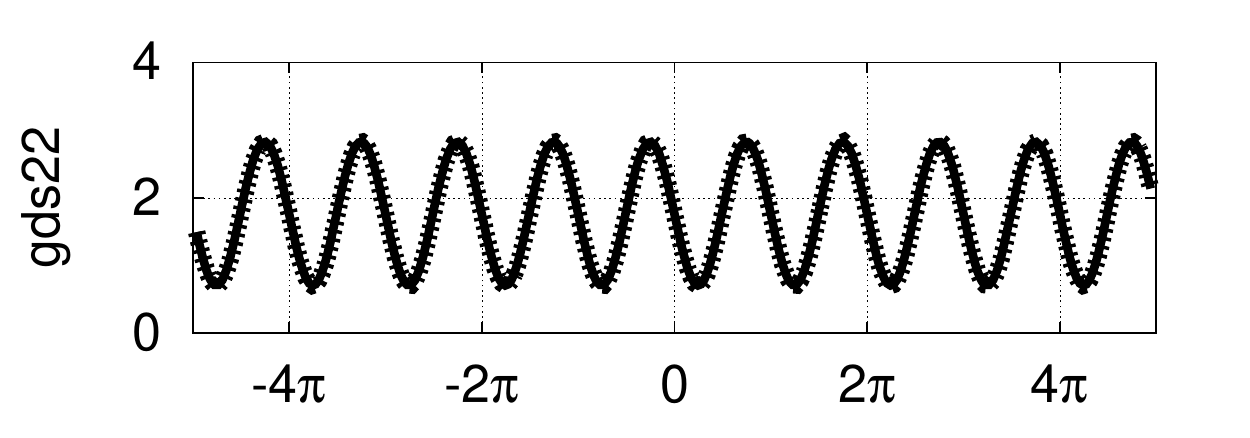}

 (c) \hspace{0.8\textwidth}

 \includegraphics[width=0.8\textwidth]{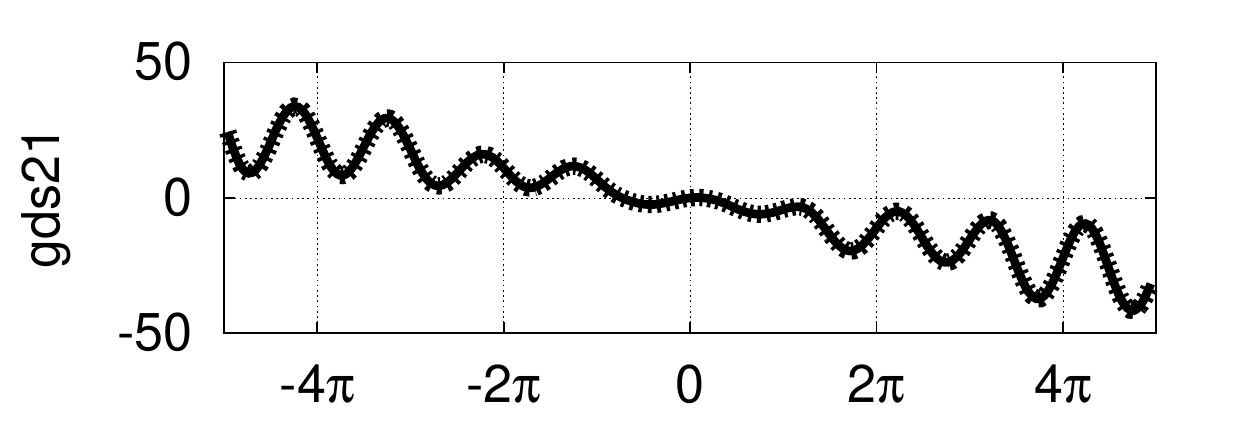}

 (d) \hspace{0.8\textwidth}

 \includegraphics[width=0.8\textwidth]{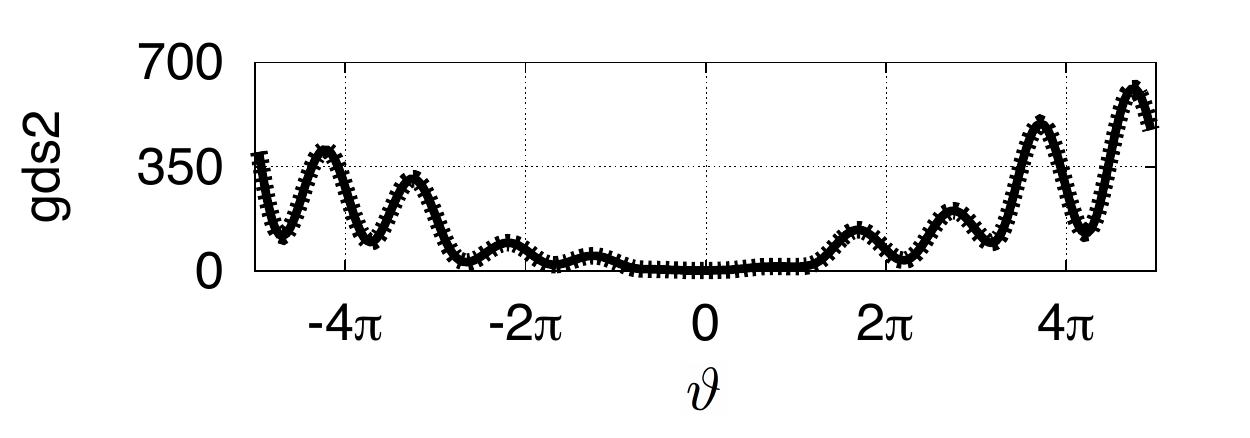}
 \caption{Geometrical coefficients output by GS2 (solid) and an independent numerical calculation (dotted) for elongated flux surfaces with $\theta_{\kappa} = \pi / 6$.}
 \label{fig:geoCoeffsBenchmark}
\end{figure}

Lastly, all effects of the system geometry appear in the equations governing gyrokinetics (eqs. \refEq{eq:gyrokineticEq} through \refEq{eq:radialEcrossBvel}) as eight coefficients that depend on $\vartheta$: $B$, $\partial B / \partial \vartheta$, $\partial B / \partial \psi$, $\hat{b} \cdot \vec{\nabla} \vartheta$, $\hat{b} \cdot \left( \vec{\nabla} \vartheta \times \vec{\nabla} \alpha \right)$, $\left| \vec{\nabla} \psi \right|^{2}$, $\vec{\nabla} \psi \cdot \vec{\nabla} \alpha$, and $\left| \vec{\nabla} \alpha \right|^{2}$. The final test performed was to verify that the geometric coefficients were correct for up-down asymmetric configurations. A numerical calculation, completely independent of GS2, was performed which found the coefficients using the Miller equilibrium model. Fig. \ref{fig:geoCoeffsBenchmark} shows four examples that reflect the excellent agreement of all the coefficients.

\section{Momentum transport}
\label{sec:momTransport}

In this section, we will investigate how turbulent momentum transport is affected by elongation tilt in several different geometries, which are all variants of the Cyclone base case. Furthermore, we will tilt the flux surfaces in several different manners, but see that it has little effect on certain features of momentum transport. The different elliptical shapes and plasma parameters (geometries) used are given in table \ref{tab:simCaseParameters} and the different ways to tilt these elliptical flux surfaces (transformations) used are given in table \ref{tab:simTransformations}.

Since GS2 only simulates a single flux surface and outputs turbulent fluxes, it is impossible to construct a rotation profile without a transport solver \cite{BarnesTrinityThesis2008,BarnesTrinity2010} and many GS2 runs. However, we will show that the turbulent fluxes from GS2 simulations (with $\Omega_{\zeta} = 0$ and $d \Omega_{\zeta} / d r_{\psi} = 0$) can be used to estimate the velocity gradients that would be possible in our geometries. Also, these velocity gradients, estimated from GS2 output, can be compared with an experimental study that looked at the connection between up-down asymmetry and intrinsic rotation.

\subsection{Elliptical geometries}

\begin{table}
  \centering
  \begin{tabular}{ r c c c c c c c c }
    \hline
    Name & $r_{\psi N}$ & $R_{0 N}$ & $q$ & $\hat{s}$ & $1/L_{T N s}$ & $1/L_{n N s}$ & $\kappa$ & $\delta$ \\
    \hline
    Base \cite{DimitsCycloneBaseCase2000} & $0.54$  & $3$ & $1.4$ & $0.8$ & $2.3$  & $0.733$ & $1$ & $0$   \\
    Elongated                 & $0.54$  & $3$ & $1.4$ & $0.8$ & $2.3$   & $0.733$ & $2$ & $0$   \\
    Elongated Extreme    & $0.54$ & $3$ & $1.4$ & $0.8$ & $3.45$ & $0.733$ & $2$ & $0$   \\
    Optimized Elongated & $1$      & $3$ & $1.4$ & $0.8$ & $2.5$   & $0.733$ & $2$ & $0$   \\
    Large Major Radius   & $0.54$ & $6$ & $1.4$ & $0.8$ & $2.3$   & $0.733$ & $2$ & $0$   \\
    Triangular Extreme    & $1$      & $3$ & $1.4$ & $0.8$ & $3.5$   & $1$        & $1$ & $0.7$
  \end{tabular}
  \caption{Normalized untilted input parameters for the geometry of each Cyclone base case variant, all with $m_{N i} = 1$, $m_{N e} = 2.7 \times 10^{-4}$, $T_{N s} = 1$ and $n_{N s} = 1$, where $s \in \left\{ i, e\right\}$.}
  \label{tab:simCaseParameters}
\end{table}

\begin{table}
  \centering
  \begin{tabular}{ r c c c c c }
    \hline
    Name & $R$ & $B$ & $| \vec{\nabla}_{N} \text{ln} ~ T_{N s} |$ & $| \vec{\nabla}_{N} \text{ln} ~ n_{N s} |$ & Size \\
    \hline
    Simplistic & $R_{0 N}$  & $B_{0 N}$ & $| \vec{\nabla}_{N} \text{ln} ~ T_{N s} | \left( r_{min} \right)$ & $| \vec{\nabla}_{N} \text{ln} ~ n_{N s} | \left( r_{min} \right)$ & $r_{\psi N}$ \\
    Sophisticated & $R_{min}^{LCFS}$  & $B_{\zeta N} \left( R_{min N} \right)$ & $| \vec{\nabla}_{N} \text{ln} ~ T_{N s} | \left( R_{max N} \right)$ & $| \vec{\nabla}_{N} \text{ln} ~ n_{N s} | \left( R_{max N} \right)$ & $r_{\psi N}$ \\
    Realistic & $R_{min}^{LCFS}$  & $B_{\zeta N} \left( R_{min N} \right)$ & $| \vec{\nabla}_{N} \text{ln} ~ T_{N s} | \left( r_{min} \right)$ & $| \vec{\nabla}_{N} \text{ln} ~ n_{N s} | \left( r_{min} \right)$ & $r_{\psi N}$ \\
    Constant-cost & $R_{min}^{LCFS}$  & $B_{\zeta N} \left( R_{min N} \right)$ & $| \vec{\nabla}_{N} \text{ln} ~ T_{N s} | \left( r_{min} \right)$ & $| \vec{\nabla}_{N} \text{ln} ~ n_{N s} | \left( r_{min} \right)$ & $V_{N}$
  \end{tabular}
  \caption{Summary of the parameters kept fixed during different transformations used to compare tilted elliptical configurations, where $s \in \left\{ i, e\right\}$ and $V_{N} = 2 \pi R_{0 N} \pi r_{\psi N}^{2} \kappa$.}
  \label{tab:simTransformations}
\end{table}

The results are composed of four sets of nonlinear simulations: the Elongated geometry with the Simplistic transformation (see fig. \ref{fig:geometryFixedR0cyclone}), the Elongated Extreme geometry with the Simplistic transformation (see fig. \ref{fig:geometryFixedR0}), the Optimized Elongated geometry with the Sophisticated transformation (see fig. \ref{fig:geometryFixedRmin}), and the Large Major Radius geometry with the Simplistic transformation (see fig. \ref{fig:geometryLAR}).

\subsubsection{Elongated geometry with the Simplistic transformation.}

\begin{figure}
 \centering
 \includegraphics[width=0.6\textwidth]{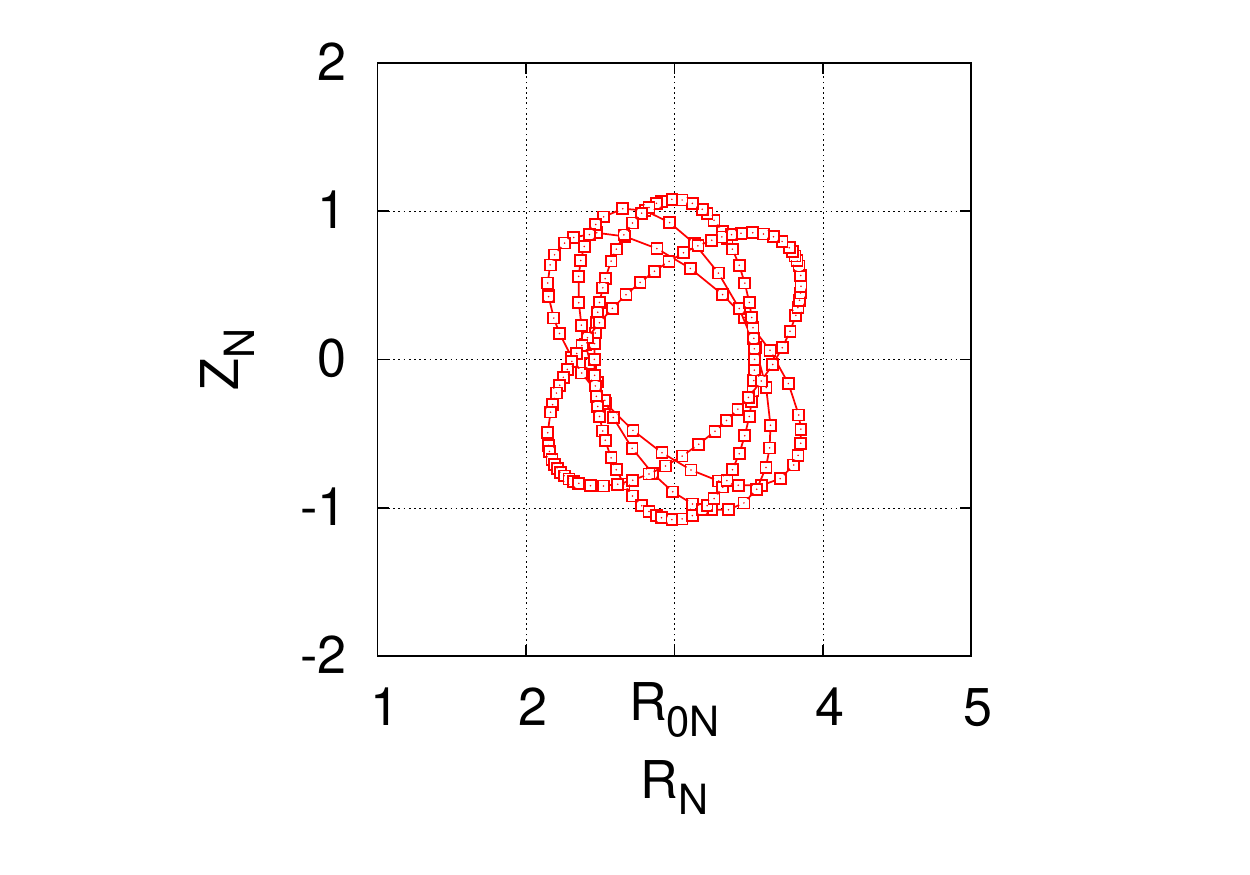}
 \caption{Elongated Cyclone base case at $\theta_{\kappa} = \left\{ - \pi / 4, - \pi / 8, 0, \pi / 4 \right\}$ with the Simplistic transformation.}
 \label{fig:geometryFixedR0cyclone}
\end{figure}

Fundamentally, when we compare the relative merits of different tokamak configurations, we want cost to be invariant. However, it is unclear how to translate cost into the global parameters appearing in tokamak design limits, let alone the local flux surface parameters that GS2 requires. If we take the GS2 input file for the Elongated cyclone base case geometry and change only the parameter \texttt{thetak} from $0$ to $\pi / 4$ we produce two of the configurations appearing in fig. \ref{fig:geometryFixedR0cyclone}. This transformation implicitly holds the major radius, the on-axis magnetic field, and the background gradients constant as the ellipse is tilted. In the GS2 input file, the background gradients are specified at the pre-tilt midplane (i.e. the location of the minimum minor radial position, $r_{min}$). This seems to be a fairly good method because it keeps the total plasma volume constant as well as the peak temperatures and densities (if we extrapolate the gradients to the magnetic axis). However it alters the amount of available space on the inboard side for structure and increases the required on-coil magnetic field. The maximum on-coil magnetic field increases because $B_{0}$ is fixed and the distance between $R_{0}$ and the inboard leg of the coil increases with elongation tilt.

\subsubsection{Elongated Extreme geometry with the Simplistic transformation.}

\begin{figure}
 \centering
 \includegraphics[width=0.6\textwidth]{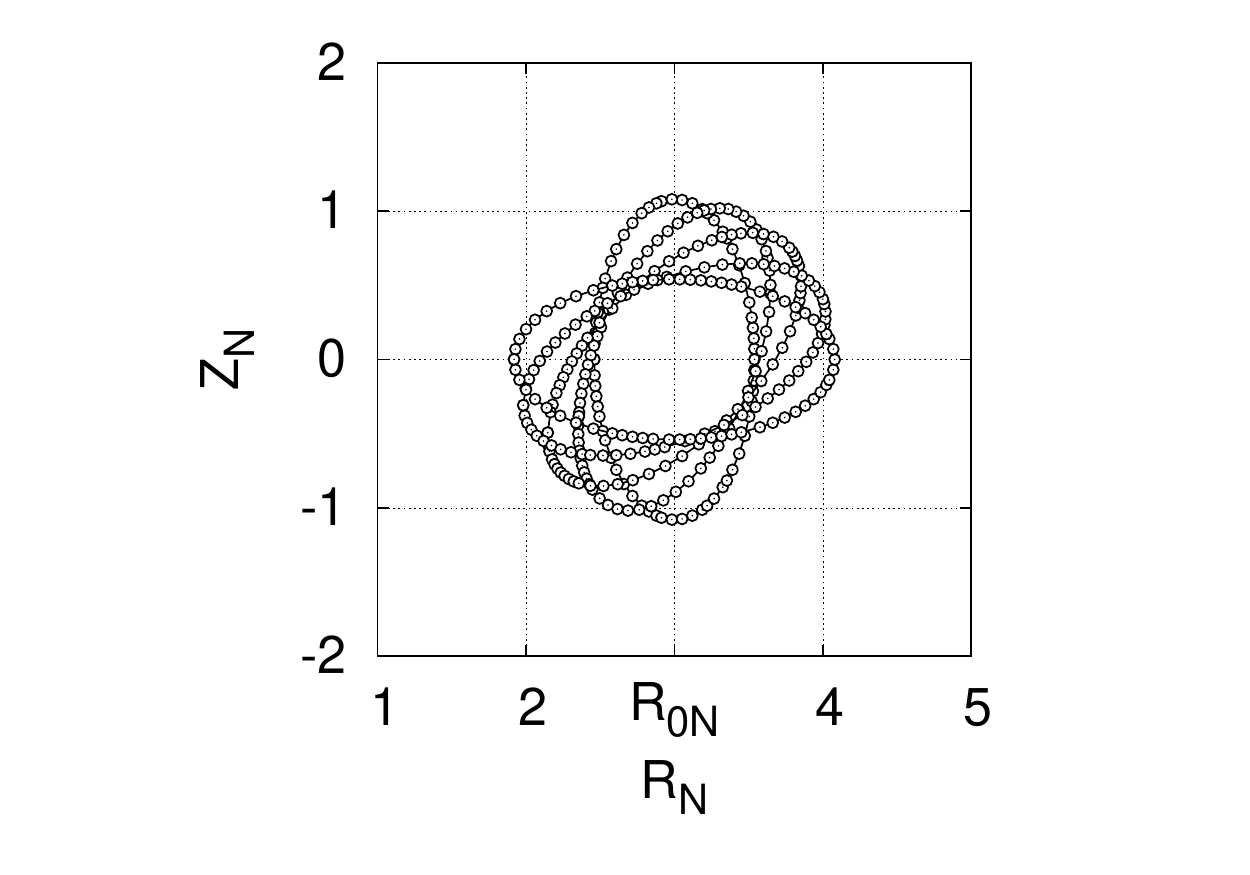}
 \caption{Elongated Extreme Cyclone base case at $\theta_{\kappa} = \left\{ 0, \pi / 8, \pi / 4, 3 \pi / 8, \pi / 2 \right\}$ with the Simplistic transformation.}
 \label{fig:geometryFixedR0}
\end{figure}

The addition of elongation to the Cyclone base was initially observed to significantly reduce turbulence, so the geometry was repeated with a $50\%$ increase in the background temperature gradient. This produced the Elongated Extreme geometry in fig. \ref{fig:geometryFixedR0}.

\subsubsection{Optimized Elongated geometry with the Sophisticated transformation.}

To reduce the reactor volume (which reduces cost) it is generally desirable to minimize $R_{0}$. However, the amount of necessary inboard space is dictated by technological limits, such as the required volumes of coil support structure, breeding blanket, and neutron shielding. Therefore $R_{min}^{LCFS}$, the minimum distance of the last closed flux surface from the axis of symmetry, should be considered the fixed parameter, not $R_{0}$. Also, the maximum allowable on-coil magnetic field is a material property of the magnet conductor and directly influences the magnet stresses. The choice of conductor material and amount of necessary magnet structure dramatically affects cost, which should stay fixed between designs. For this reason $B_{\zeta} \left( R_{min}^{LCFS} \right)$, the maximum on-coil magnetic field, should be fixed.

\begin{figure}
 \centering
 \includegraphics[width=0.3\textwidth]{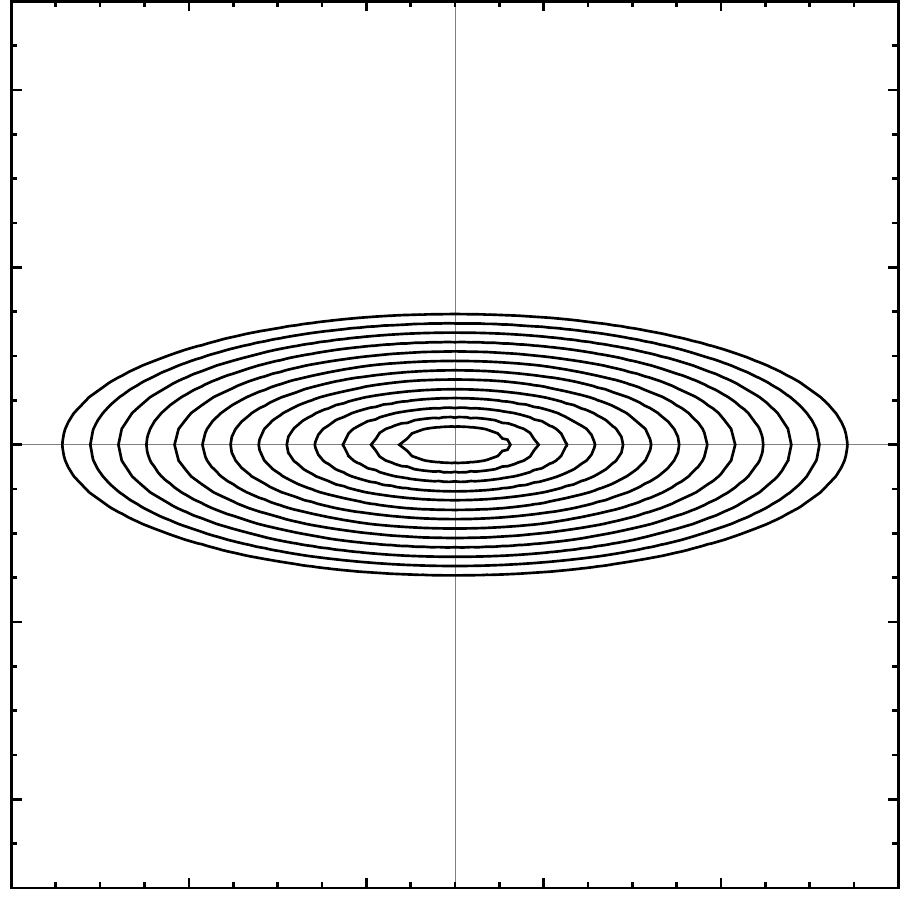}
 \caption{Elliptical flux surfaces with $\kappa = 3$ and $\theta_{\kappa} = \pi/2$, demonstrating that the background gradients along the vertical grid line are three times as steep as along the horizontal grid line.}
 \label{fig:shapingEffectOnGradients}
\end{figure}

\begin{figure}
 \centering
 \includegraphics[width=0.6\textwidth]{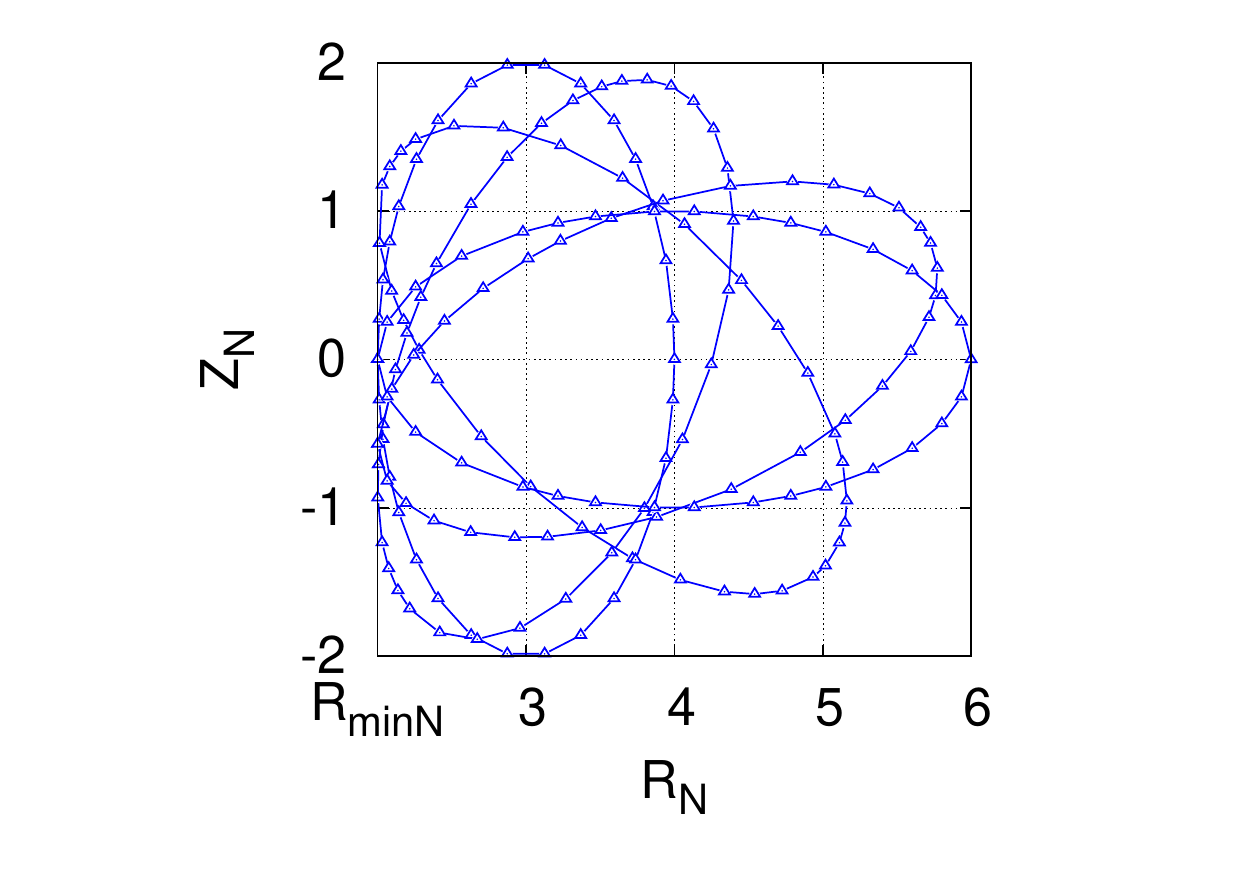}
 \caption{Optimized Elongated Cyclone base case at $\theta_{\kappa} = \left\{ - \pi / 4, 0, \pi / 8, 3 \pi / 8, \pi / 2 \right\}$ with the Sophisticated transformation.}
 \label{fig:geometryFixedRmin}
\end{figure}

In order to more closely approximate constant cost, a more realistic comparison fixes $R_{min}^{LCFS}$ and $B_{\zeta} \left( R_{min}^{LCFS} \right)$. In order to fix $B_{\zeta} \left( R_{min}^{LCFS} \right)$, $R_{geo N} = 3$ was kept constant during rotation. Previously, the geometry of fig. \ref{fig:geometryFixedR0cyclone} kept the global gradients constant, such that in a full reactor the peak temperature and density would be fixed. Instead we will attempt to keep the local turbulent drive of the mode constant. As a rough approximation, we would expect the mode to be centered around $R_{max}$ because it is the location of the strongest bad curvature and has the weakest toroidal magnetic field. The background gradients, which drive instability, are specified in GS2 input files at the pre-tilt midplane. Though the temperature and density are flux functions, the spacing between elliptical flux surfaces, parameterized by $| \vec{\nabla} r_{\psi} |$, changes with poloidal location (see fig. \ref{fig:shapingEffectOnGradients}). Therefore, to keep the local turbulent drive constant between different tilt angles, it is necessary to change the background gradients given to GS2. Fundamentally, the quantities $| \vec{\nabla}_{N} \text{ln} ~ T_{N s} | = \mathtt{tprim} ~ | \vec{\nabla}_{N} r_{\psi N} |$ and $| \vec{\nabla}_{N} \text{ln} ~ n_{N s} | = \mathtt{fprim} ~ | \vec{\nabla}_{N} r_{\psi N} |$ are held constant at $R_{max}$ to keep the local turbulent drive constant with tilt. Fixing $R_{min}^{LCFS}$, $B_{\zeta} \left( R_{min}^{LCFS} \right)$, and the background gradients at $R_{max}$ between different tilted geometries defines the Sophisticated transformation. It is important to realize that, if we extrapolate the background gradients to the magnetic axis, this transformation implies a change in the on-axis pressure. The $\theta_{\kappa} = \pi / 2$ case would have on-axis temperatures and densities that are a factor of $\kappa$ greater than in the $\theta_{\kappa} = 0$ case. Also, in these simulations (see fig. \ref{fig:geometryFixedRmin}), we chose the last closed flux surface in order to simulate a different aspect ratio from the Cyclone base case.

\subsubsection{Large Major Radius geometry with the Simplistic transformation.}

\begin{figure}
 \centering
 \includegraphics[width=0.6\textwidth]{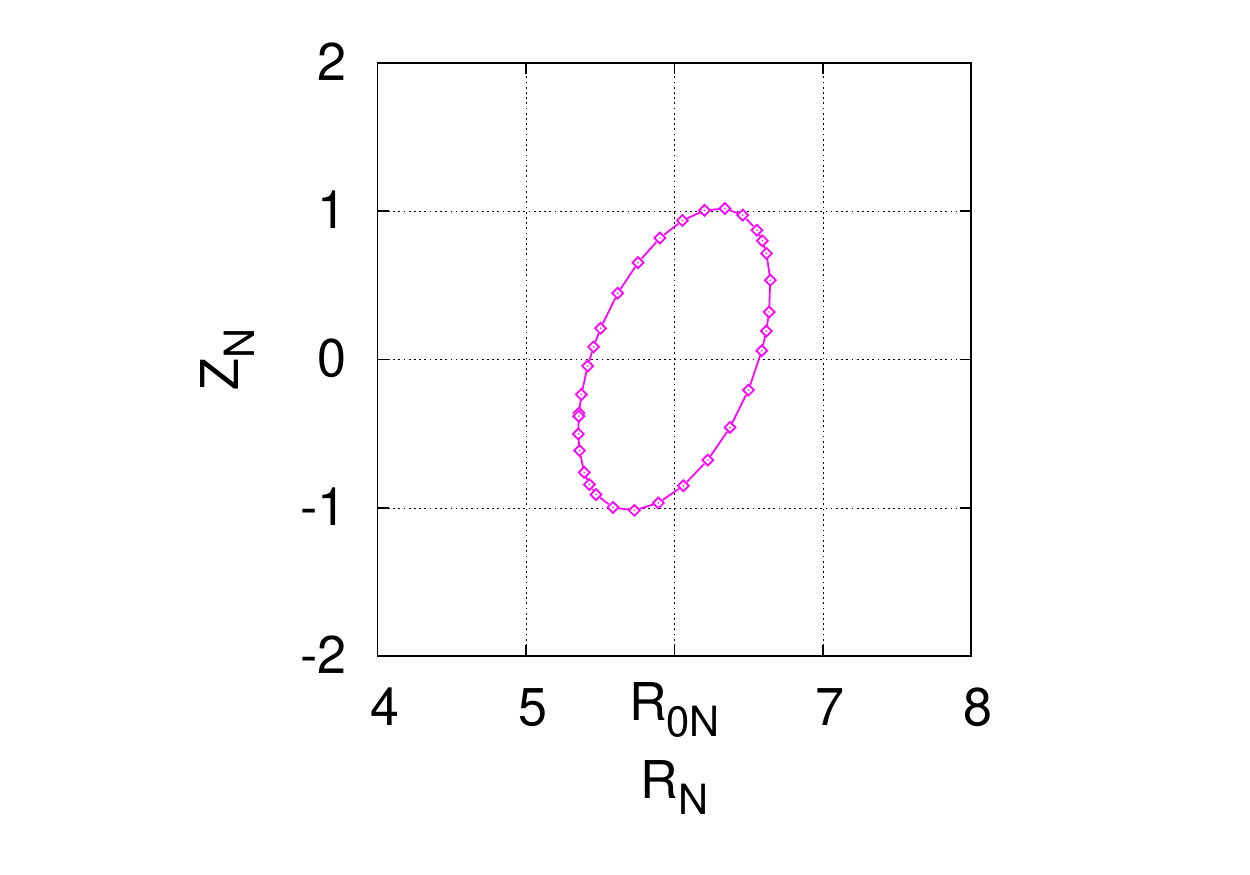}
 \caption{Large Aspect Ratio Cyclone base case at $\theta_{\kappa} = \left\{ \pi / 8 \right\}$ with the Simplistic transformation.}
 \label{fig:geometryLAR}
\end{figure}

The Large Major Radius geometry is identical to the Elongated geometry, except it has a major radius that is twice as large. A single simulation with a π/8 tilt was run in this geometry (see fig. \ref{fig:geometryLAR}) to demonstrate that the gyro-Bohm angular momentum flux (given by eq. \refEq{eq:gyroBohmMomFlux}) used for normalization in GS2 does not account for the natural scaling with major radius. In the analysis of this paper, we will see that $R_{0 N} \Pi_{gB r}$ is more fundamental normalization for the observed momentum transport. It adjusts for the fact that $l_{r}$ is typically interpreted as the minor radius, while angular momentum scales with the distance from the rotational axis, which is the major radius.

\subsubsection{Results.}
\label{subsubsec:Results}

\begin{figure}
 \centering
 \includegraphics[width=0.7\textwidth]{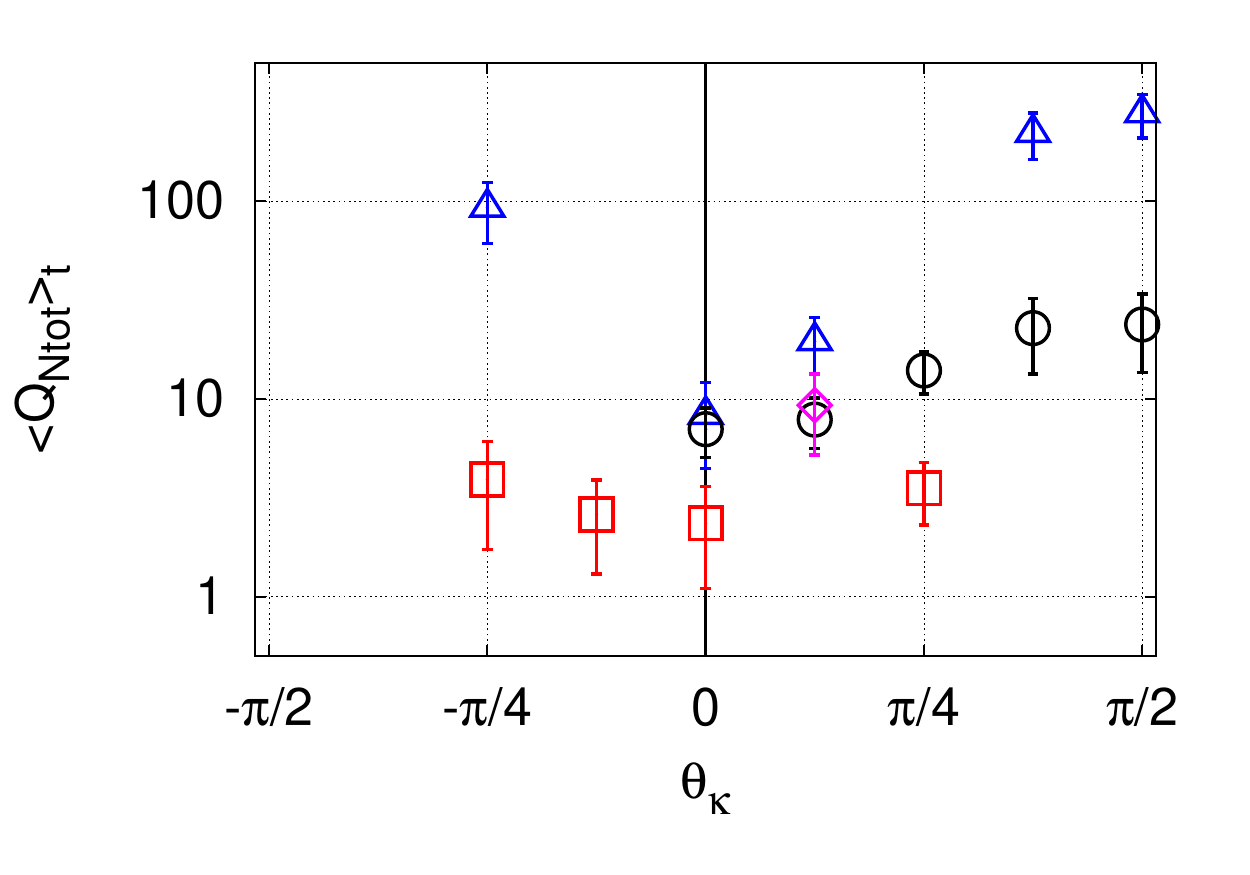}
 \caption{Time-averaged total nonlinear heat flux for the Elongated geometry with Simplistic transformation (red, squares), Elongated Extreme geometry with Simplistic transformation (black, circles), Optimized Elongated geometry with Sophisticated transformation (blue, triangles), and Large Aspect Ratio geometry with Simplistic transformation (magenta, diamonds).}
 \label{fig:nonlinearAvgTotHeatFlux}
\end{figure}

Fig. \ref{fig:nonlinearAvgTotHeatFlux} gives the time-averaged heat flux results for all of the above geometries. The time-average is given by $\langle \ldots \rangle_{t} \equiv \left( 1 / t_{corr} \right) \int_{t_{0}}^{t_{0}+t_{corr}} dt \left( \ldots \right)$, where $t_{corr}$ is much longer than the turbulent correlation time. We see from the blue markers that fixing the local gradients at $R_{max}$ to keep the turbulent drive constant was inappropriate because the heat flux now increases dramatically with tilt angle. From the results presented here it appears that tilting an elliptical flux surface increases turbulent energy transport, but further investigations have suggested that this is not always true. A thorough study of energy transport is the subject of future work.

\begin{figure}[ht]
 \centering
 \includegraphics[width=0.7\textwidth]{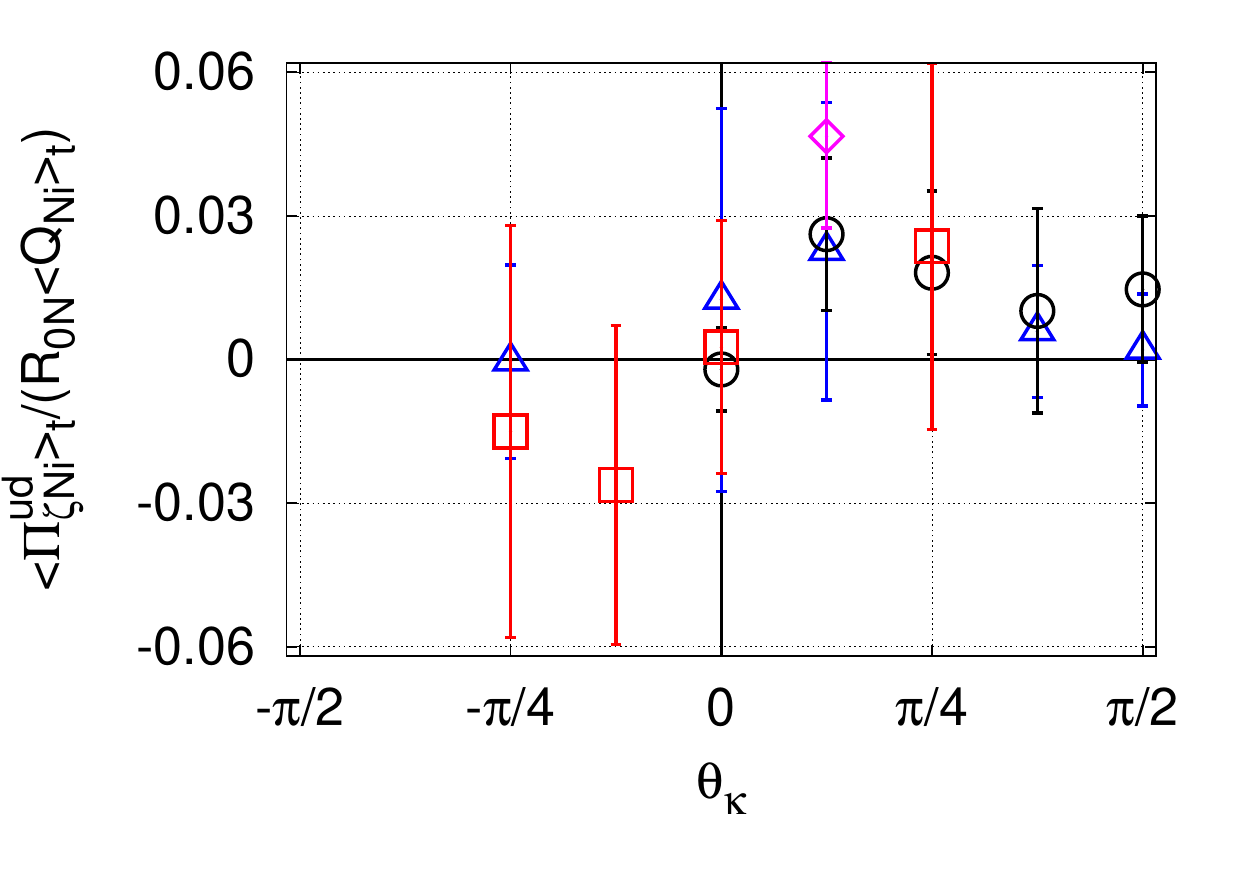}
 \caption{Time-averaged ratio of ion angular momentum and heat fluxes for the geometries and transformations of figs. \ref{fig:geometryFixedR0cyclone} (red, squares), \ref{fig:geometryFixedR0} (black, circles), \ref{fig:geometryFixedRmin} (blue, triangles), and \ref{fig:geometryLAR} (magenta, diamonds).}
 \label{fig:nonlinearAvgMomHeatRatio}
\end{figure}

Fig. \ref{fig:nonlinearAvgMomHeatRatio} shows the nonlinear momentum flux results for the three geometries and transformations. Since, to lowest order in $\sqrt{m_{e} / m_{i}}$, the electrons carry no momentum, we ignore their contribution to momentum transport. We plot the ratio between normalized momentum flux and normalized heat flux because this will turn out to be the relevant quantity for estimating the amount of rotation (see Section \ref{subsec:velocityGradEst}). Quantitatively, we see very similar behavior despite the differences in input parameters. This suggests that $\left\langle \Pi_{\zeta N i}^{ud} \right\rangle_{t} / \left( R_{0 N} \left\langle Q_{N i} \right\rangle_{t} \right)$ is relatively insensitive to the background gradients and aspect ratio.

As expected, the up-down symmetric cases at $\theta_{\kappa} = \left\{ 0, \pi / 2 \right\}$ have a vanishing momentum to heat flux ratio. The maximum of the curve is around $0.03$ and is located at $\theta_{\kappa} = \pi / 8$. Since vertical and horizontal elongation have very different effects on plasma turbulence, there is no reason to expect the curve to be symmetric about $\pi / 4$. The location of the peak is a positive result because it indicates that only a slight tilt of the standard vertically-elongated flux surfaces is required to induce significant rotation.

\subsection{Velocity gradient estimation from GS2 fluxes}
\label{subsec:velocityGradEst}

Using the local fluxes output by GS2, we can estimate the velocity gradient that is sustainable with a given temperature gradient. We start with the conservation equation for the flux surface averaged ion toroidal angular momentum density,
\begin{eqnarray}
   \frac{\partial}{\partial t} \left( R^{2} n_{i} m_{i} \Omega_{i} \right) = - \frac{1}{V'} \frac{\partial}{\partial \psi} \left( V' \Pi_{\zeta i} \right) + S_{\Pi i} . \label{eq:AngMomConservation}
\end{eqnarray}
Here $\Pi_{\zeta i}$ is the flux surface averaged flux of ion toroidal angular momentum density and $S_{\Pi i}$ is the flux surface averaged volumetric source of ion toroidal angular momentum density. Since we are interested in steady-state transport without external sources we arrive at
\begin{eqnarray}
    - \frac{1}{V'} \frac{d}{d \psi} \left( V' \left\langle \Pi_{\zeta i} \right\rangle_{t} \right) = 0 ,
\end{eqnarray}
where $V' \equiv dV /d \psi = \oint d \zeta d \theta \left( \vec{B} \cdot \vec{\nabla} \theta \right)^{-1}$. Forcing $\left\langle \Pi_{\zeta i} \right\rangle_{t}$ to be regular on axis gives
\begin{eqnarray}
    \left\langle \Pi_{\zeta i} \left( \Omega_{i}, \frac{d \Omega_{i}}{d r_{\psi}} \right) \right\rangle_{t} = 0 ,
\end{eqnarray}
which can be solved to find the radial rotation profile. Taking a Taylor expansion of this equation gives
\begin{eqnarray}
    \left\langle \Pi_{\zeta i} \right\rangle_{t} \approx \left\langle \Pi_{\zeta i} \left( 0,0 \right) \right\rangle_{t} + \frac{\partial \left\langle \Pi_{\zeta i} \right\rangle_{t}}{\partial \Omega_{i}} \Omega_{i} + \frac{\partial \left\langle \Pi_{\zeta i} \right\rangle_{t}}{\partial \left( d \Omega_{i} / d r_{\psi} \right)} \frac{d \Omega_{i}}{d r_{\psi}} \approx 0 ,
\end{eqnarray}
where $\Pi_{\zeta i}^{ud} \equiv \Pi_{\zeta i} \left( 0,0 \right)$ is the intrinsic momentum flux due to up-down asymmetry calculated with GS2 by setting $\Omega_{\zeta} = d \Omega_{\zeta} / d r_{\psi} = 0$, $P_{\Pi i} \equiv - \left( R_{0}^{2} n_{i} m_{i} \right)^{-1} \partial \left\langle \Pi_{\zeta i} \right\rangle_{t} / \partial \Omega_{i}$ is the angular momentum pinch coefficient, and $D_{\Pi i} \equiv - \left( R_{0}^{2} n_{i} m_{i} \right)^{-1} \partial \left\langle \Pi_{\zeta i} \right\rangle_{t} / \partial \left( d \Omega_{i} / d r_{\psi} \right)$ is the angular momentum diffusion coefficient. Making these substitutions, we find that
\begin{eqnarray}
    \left\langle \Pi_{\zeta i}^{ud} \right\rangle_{t} - P_{\Pi i} n_{i} m_{i} R_{0}^{2} \Omega_{i} - D_{\Pi i} n_{i} m_{i} R_{0}^{2} \frac{d \Omega_{i}}{d r_{\psi}} \approx 0 .
\end{eqnarray}
Using the method of integrating factors, the solution of this differential equation is found to be
\begin{eqnarray}
    \Omega_{i} \left( r_{\psi} \right) &= \int _{r_{\psi}}^{a} d r_{\psi}' \left( \frac{-1}{n_{i} \left( r_{\psi}' \right) m_{i} R_{0}^{2}} \frac{\left\langle \Pi_{\zeta i}^{ud} \left( r_{\psi}' \right) \right\rangle_{t}}{D_{\Pi i} \left( r_{\psi}' \right)} \text{exp} \left( \int_{r_{\psi}}^{r_{\psi}'} d r_{\psi}'' \frac{P_{\Pi i} \left( r_{\psi}'' \right)}{D_{\Pi i} \left( r_{\psi}'' \right)} \right) \right) \nonumber \\
   &+ \Omega_{i a} \text{exp} \left( \int_{r_{\psi}}^{a} d r_{\psi}' \frac{P_{\Pi i} \left( r_{\psi}' \right)}{D_{\Pi i} \left( r_{\psi}' \right)} \right) ,
\end{eqnarray}
where $\Omega_{i a} \equiv \Omega_{i} \left( a \right)$ is the edge boundary condition. Assuming the rotation is small near the wall, we can find the rotation gradient to be
\begin{eqnarray}
    \frac{d \Omega_{i}}{d r_{\psi}} &= \frac{1}{n_{i} \left( r_{\psi} \right) m_{i} R_{0}^{2}} \frac{\left\langle \Pi_{\zeta i}^{ud} \left( r_{\psi} \right) \right\rangle_{t}}{D_{\Pi i} \left( r_{\psi} \right)} \label{eq:rotationGrad} \\
   &+ \frac{P_{\Pi i}}{D_{\Pi i}} \int _{r_{\psi}}^{a} d r_{\psi}' \left( \frac{1}{n_{i} \left( r_{\psi}' \right) m_{i} R_{0}^{2}} \frac{\left\langle \Pi_{\zeta i}^{ud} \left( r_{\psi}' \right) \right\rangle_{t}}{D_{\Pi i} \left( r_{\psi}' \right)} \text{exp} \left( \int_{r_{\psi}}^{r_{\psi}'} d r_{\psi}'' \frac{P_{\Pi i} \left( r_{\psi}'' \right)}{D_{\Pi i} \left( r_{\psi}'' \right)} \right) \right) . \nonumber
\end{eqnarray}
Very roughly we expect $P_{\Pi i} \left( r_{\psi} \right) / D_{\Pi i} \left( r_{\psi} \right) \approx 3 / R_{0}$, meaning the exponential should be an $O \left( 1 \right)$ factor \cite{LeeDiamagneticMom2014, PeetersMomPinch2007}. More broadly, studying eq. \refEq{eq:rotationGrad}, we see the entire contribution of the pinch term can be considered as an $O \left( 1 \right)$ factor multiplying the first term. Therefore we make the estimation
\begin{eqnarray}
    \frac{d \Omega_{i}}{d r_{\psi}} \approx \frac{1}{n_{i} m_{i} R_{0}^{2}} \frac{\left\langle \Pi_{\zeta i}^{ud} \left( r_{\psi} \right) \right\rangle_{t}}{D_{\Pi i} \left( r_{\psi} \right)}
\end{eqnarray}
or equivalently
\begin{eqnarray}
    \left\langle \Pi_{\zeta i}^{ud} \right\rangle_{t} \approx D_{\Pi i} n_{i} m_{i} R_{0}^{2} \frac{d \Omega_{\zeta i}}{d r_{\psi}} . \label{eq:MomFluxEstimate}
\end{eqnarray}
We note that ignoring the pinch term is expected to lead to an underprediction of the rotation gradient, maybe by as much as a factor of 3.

The radial ion heat flux can be expressed as \cite{FreidbergFusionEnergy2007pg452}
\begin{eqnarray}
   \left\langle Q_{i} \right\rangle_{t} &\approx - D_{Q i} n_{i} \frac{\partial T_{i}}{\partial r_{\psi}} . \label{eq:HeatFluxEstimate}
\end{eqnarray}
Crucially, we note from ref. \cite{BarnesPrandtlNum2011} and fig. 8.2c in ref. \cite{HighcockManifold2012} that the turbulent ion Prandtl number $Pr_{i} \equiv D_{\Pi i} / D_{Q i}$ is approximately constant across tokamak parameters. This can be used to relate eqs. \refEq{eq:MomFluxEstimate} and \refEq{eq:HeatFluxEstimate}, giving the nondimensionalized form
\begin{eqnarray}
   \frac{\partial u_{\zeta N i} / \partial r_{\psi N}}{\partial T_{N i} / \partial r_{\psi N}} \approx \left( \frac{-1}{m_{N i} Pr_{i}} \right) \frac{\left\langle \Pi_{\zeta N i}^{ud} \right\rangle_{t}}{R_{0 N} \left\langle Q_{N i} \right\rangle_{t}} . \label{eq:gradEstimateFromFluxes}
\end{eqnarray}
An estimation of the ion Prandtl number was calculated using an untilted Elongated Cyclone base case simulation with $\mathtt{g\_exb} = 0.1$ to be
\begin{eqnarray}
   Pr_{i} \equiv \frac{D_{\Pi i}}{D_{Q i}} = \frac{- r_{\psi N} T_{N i}}{m_{N i} R_{0 N} q} \frac{\mathtt{tprim}}{\mathtt{g\_exb}} \frac{\left\langle \Pi_{\zeta N i} \right\rangle_{t}}{R_{0 N} \left\langle Q_{N i} \right\rangle_{t}} \approx 0.7 .
\end{eqnarray}
This means that for all simulations performed in this work (see table \ref{tab:simCaseParameters}), the estimated ratio of velocity and temperature gradients is given by
\begin{eqnarray}
    \frac{\partial u_{\zeta N i} / \partial r_{\psi N}}{\partial T_{N i} / \partial r_{\psi N}} \approx - 1.4 \frac{\left\langle \Pi_{\zeta N i}^{ud} \right\rangle_{t}}{R_{0 N} \left\langle Q_{N i} \right\rangle_{t}} . \label{eq:finalRotGradResult}
\end{eqnarray}

The fundamental conclusion is that the peak of $0.03$ in fig. \ref{fig:nonlinearAvgMomHeatRatio} corresponds to a velocity gradient, $\left( 1/v_{th i} \right) \partial u_{\zeta i} / \partial r_{\psi}$, that is roughly 5\% of the temperature gradient, $\left( 1/T_{i} \right) \partial T_{i} / \partial r_{\psi}$.

\subsection{Triangular geometry}

\begin{figure}[ht]
 \centering
 \includegraphics[width=0.6\textwidth]{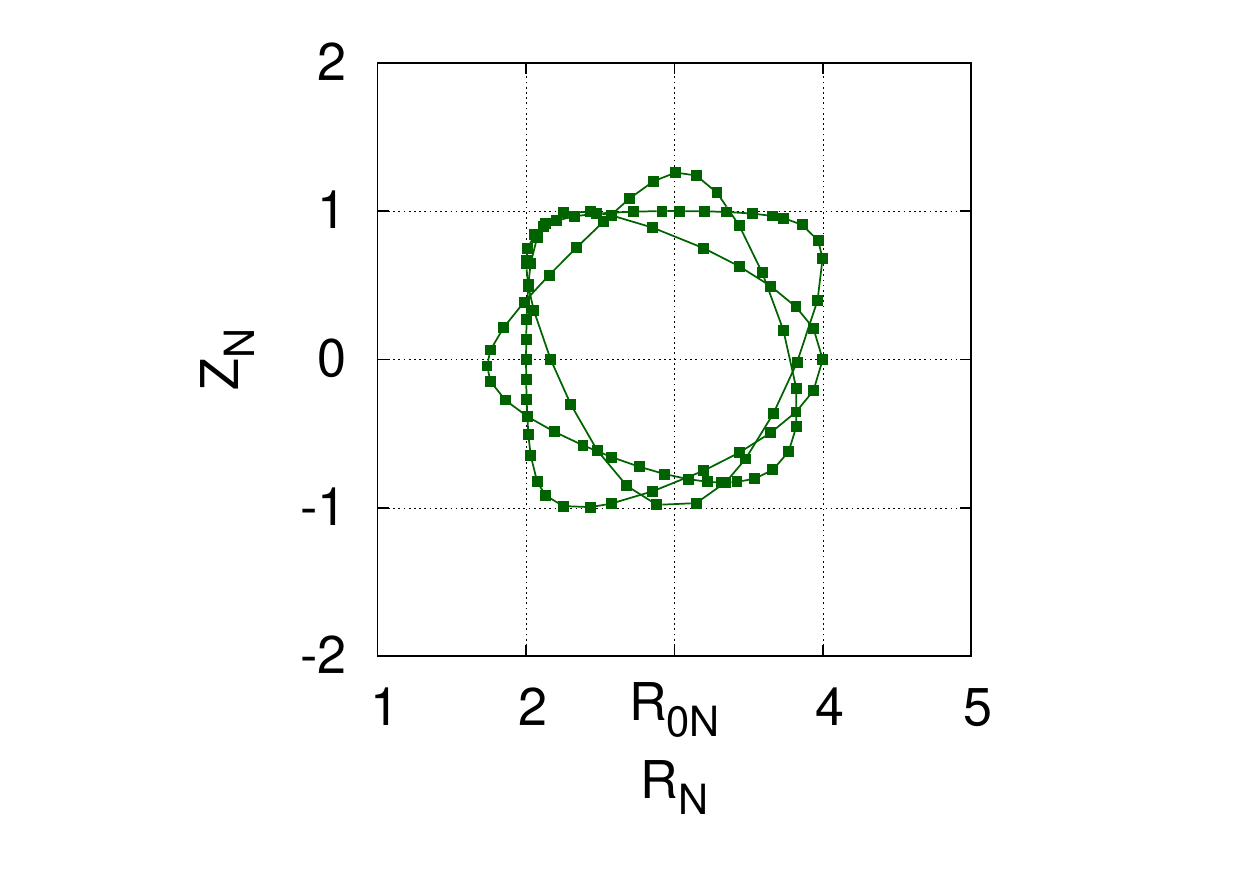}
 \caption{Triangular Extreme Cyclone base case (see table \ref{tab:simCaseParameters}) at $\theta_{\kappa} = \left\{ 0, \pi / 4, \pi / 2 \right\}$ with the Simplistic transformation.}
 \label{fig:geometryTri}
\end{figure}

Since the elliptical geometry showed such consistent momentum flux results across a range of input parameters, a triangular geometry was also simulated (see fig. \ref{fig:geometryTri}). As shown in Section \ref{sec:MHD}, triangularity has trouble penetrating to the magnetic axis in order to achieve up-down asymmetric flux surfaces throughout the plasma. As such, these simulations were not about advocating triangularity as a practical means to achieve high levels of intrinsic rotation. Rather, they were about showing that the magnitude momentum fluxes observed are characteristic of up-down asymmetry in general and are not a consequence of some peculiarity of elongated flux surfaces.

\begin{figure}[ht]
 \centering
 \includegraphics[width=0.6\textwidth]{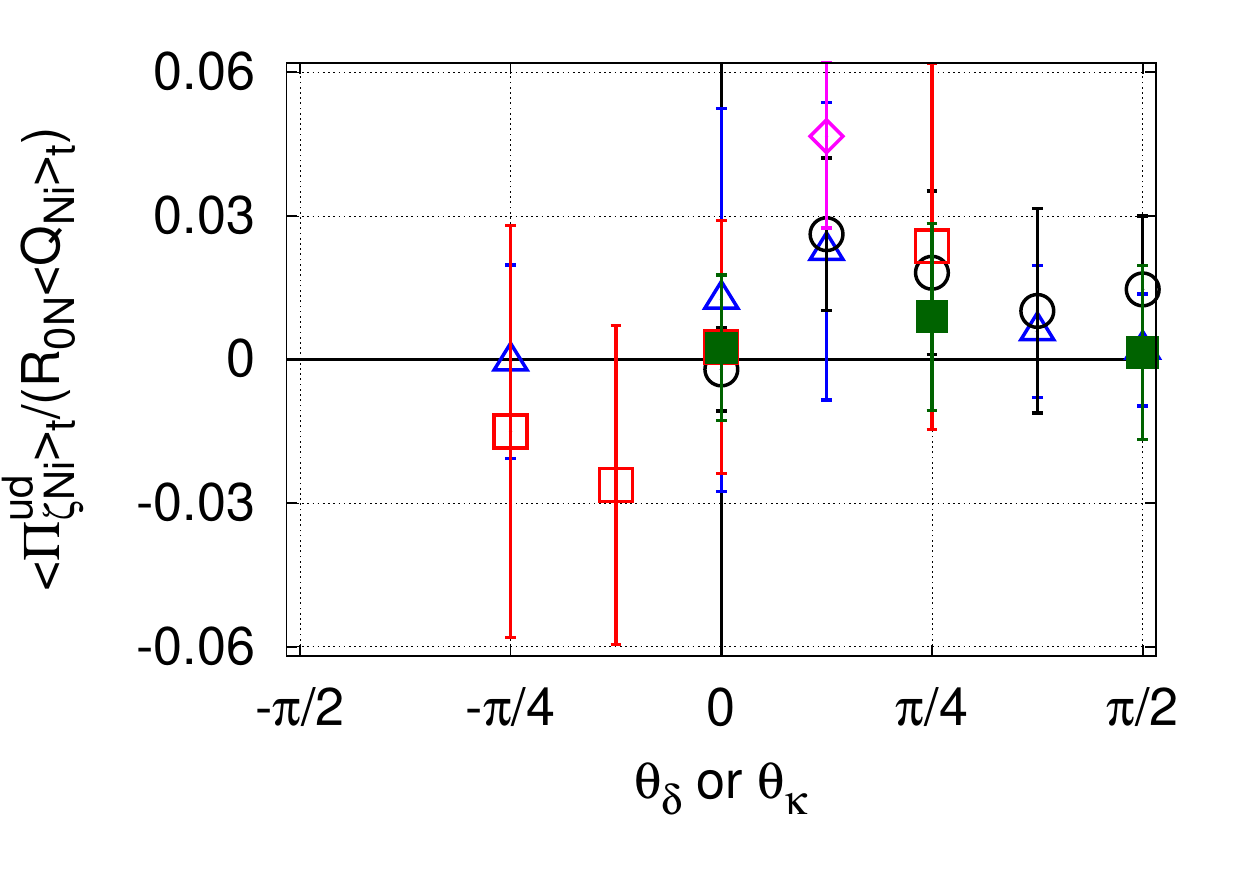}
 \caption{Time-averaged ratio of ion angular momentum and heat fluxes for the geometry of fig. \ref{fig:geometryTri} (green, filled squares) with the elongated results (empty shapes) shown for comparison.}
 \label{fig:nonlinearTriAvgIonMomHeatRatio}
\end{figure}

We see in fig. \ref{fig:nonlinearTriAvgIonMomHeatRatio} that the triangular flux surfaces caused much less momentum transport than the elongated surfaces. It is important to note that the $\theta_{\delta} = 0$ simulation is up-down symmetric, while the $\theta_{\delta} = \pi / 2$ simulation is asymmetric yet still has near zero momentum transport.

\subsection{TCV up-down asymmetry experiment \cite{CamenenTCVExp2010,CamenenPRLExp2010}}

In 2010, Camenen \emph{et al.} \cite{CamenenPRLExp2010} published the results of a TCV \cite{WessonTokamaks2004TCV,HofmannTCVOverview1994} experimental study of the effects of up-down asymmetry on intrinsic rotation. In order to isolate the effect of up-down asymmetry from $O \left( \rho_{\ast} \right)$ symmetry breaking mechanisms, the study used multiple shots to look for a differential effect on the rotation profile between two asymmetric magnetic configurations (see fig. 1 of ref. \cite{CamenenTCVExp2010}). The shots were made to be as identical as possible (see fig. 3 of ref. \cite{CamenenTCVExp2010}) except for key changes in the sign of three quantities: the equilibrium flux surface asymmetry, the toroidal magnetic field, and the plasma current. Switching the sign of any of these three quantities switches the sign of the intrinsic momentum flux, allowing experimenters to deduce the magnitude of the intrinsic momentum flux. In each shot, the toroidal rotation of the carbon impurity species was measured from Doppler shift of charge exchange radiation. The rotation of the main ion species, deuterium, is then calculated from the carbon rotation using a neoclassical physics code.

Ref. \cite{CamenenTCVExp2010} provides exactly enough information to allow comparison with our numerical results for $\left\langle \Pi_{\zeta N i}^{ud} \right\rangle_{t}/ \left( R_{0 N} \left\langle Q_{N i} \right\rangle_{t} \right)$. Here, in order to compare with experiment, we will interpret the reference macroscopic length, $l_{r}$, as the tokamak minor radius, $a$, implying that $r_{\psi N} = \rho \equiv r_{\psi} / a$. We can then invert eq. \refEq{eq:gradEstimateFromFluxes} to get
\begin{eqnarray}
   \frac{\left\langle \Pi_{\zeta N i}^{ud} \right\rangle_{t}}{R_{0 N} \left\langle Q_{N i} \right\rangle_{t}} \approx - m_{N i} Pr_{i} \frac{\partial u_{\zeta N i} / \partial \rho}{\partial T_{N i} / \partial \rho} .
\end{eqnarray}
Using GS2 normalizations we find that
\begin{eqnarray}
   \frac{\left\langle \Pi_{\zeta N i}^{ud} \right\rangle_{t}}{R_{0 N} \left\langle Q_{N i} \right\rangle_{t}} \approx - m_{i} Pr_{i} \frac{v_{th r}}{2} \frac{\partial u_{\zeta i} / \partial r_{\psi}}{\partial T_{i} / \partial r_{\psi}} . \label{eq:TCVCompHalfNormalized}
\end{eqnarray}

However, the TCV experiment measured a differential effect between two mirror opposite up-down asymmetric equilibrium. From inspection of fig. 1 of ref. \cite{CamenenTCVExp2010} we see that, primarily, the flux surfaces were elongated with $\theta_{\kappa} = \left\{ - \pi / 8, \pi / 8 \right\}$. That means we will recast eq. \refEq{eq:TCVCompHalfNormalized} as
\begin{eqnarray}
   \left. \frac{\left\langle \Pi_{\zeta N i}^{ud} \right\rangle_{t}}{R_{0 N} \left\langle Q_{N i} \right\rangle_{t}} \right|_{\theta_{\kappa} = \pi / 8} &- \left. \frac{\left\langle \Pi_{\zeta N i}^{ud} \right\rangle_{t}}{R_{0 N} \left\langle Q_{N i} \right\rangle_{t}} \right|_{\theta_{\kappa} = - \pi / 8} \approx \\
   &- m_{i} Pr_{i} \frac{v_{th r}}{2} \left( \left. \frac{\partial u_{\zeta i} / \partial r_{\psi}}{\partial T_{i} / \partial r_{\psi}} \right|_{\theta_{\kappa} = \pi / 8}  - \left. \frac{\partial u_{\zeta i} / \partial r_{\psi}}{\partial T_{i} / \partial r_{\psi}} \right|_{\theta_{\kappa} = - \pi / 8} \right) . \nonumber
\end{eqnarray}
Since the paper provides the difference in velocity, $\Delta u_{\zeta} \equiv \left. u_{\zeta} \right|_{\rho = 0.65} - \left. u_{\zeta} \right|_{\rho = 0.85}$, between two minor radial locations, we must discretize the derivatives about $r_{\psi}$ to get
\begin{eqnarray}
   \left. \frac{\left\langle \Pi_{\zeta N i}^{ud} \right\rangle_{t}}{R_{0 N} \left\langle Q_{N i} \right\rangle_{t}} \right|_{\theta_{\kappa} = \pi / 8} &- \left. \frac{\left\langle \Pi_{\zeta N i}^{ud} \right\rangle_{t}}{R_{0 N} \left\langle Q_{N i} \right\rangle_{t}} \right|_{\theta_{\kappa} = - \pi / 8} \approx \\
   &- m_{i} Pr_{i} \frac{v_{th r}}{2} \left( \frac{\left. \Delta u_{\zeta} \right|_{\theta_{\kappa} = \pi / 8} - \left. \Delta u_{\zeta} \right|_{\theta_{\kappa} = - \pi / 8}}{\Delta T_{i}} \right) , \nonumber
\end{eqnarray}
where $\Delta T_{i} \equiv \left. T_{i} \right|_{\rho = 0.65} - \left. T_{i} \right|_{\rho = 0.85}$ is defined analogously to $\Delta u_{\zeta}$. Using the upper left plot of fig. 3 of ref. \cite{CamenenTCVExp2010} we can estimate both $T_{i} = 400$ eV and $\Delta T_{i} = 400$ eV. Also, we approximate the difference in $\Delta u_{\zeta}$ between $\theta_{\kappa} = \pi / 8$ and $\theta_{\kappa} = - \pi / 8$ by averaging over the three sets of counter-current measurements listed in table 1 of ref. \cite{CamenenTCVExp2010} to get $\left. \Delta u_{\zeta} \right|_{\theta_{\kappa} = \pi / 8} - \left. \Delta u_{\zeta} \right|_{\theta_{\kappa} = - \pi / 8} \approx 8$ km/s. Now we use that $m_{i} = m_{D} = 2$ amu, $v_{th r} = \sqrt{2 T_{i} / m_{i}}$, and $Pr_{i} \approx 0.7$ to find the experimental value to be
\begin{eqnarray}
   \left( \left. \frac{\left\langle \Pi_{\zeta N i}^{ud} \right\rangle_{t}}{R_{0 N} \left\langle Q_{N i} \right\rangle_{t}} \right|_{\theta_{\kappa} = \pi / 8} - \left. \frac{\left\langle \Pi_{\zeta N i}^{ud} \right\rangle_{t}}{R_{0 N} \left\langle Q_{N i} \right\rangle_{t}} \right|_{\theta_{\kappa} = - \pi / 8} \right)_{exp} \approx 0.03 .
\end{eqnarray}

By using fig. \ref{fig:nonlinearAvgMomHeatRatio}, GS2 simulations give
\begin{eqnarray}
   \left( \left. \frac{\left\langle \Pi_{\zeta N i}^{ud} \right\rangle_{t}}{R_{0 N} \left\langle Q_{N i} \right\rangle_{t}} \right|_{\theta_{\kappa} = \pi / 8} - \left. \frac{\left\langle \Pi_{\zeta N i}^{ud} \right\rangle_{t}}{R_{0 N} \left\langle Q_{N i} \right\rangle_{t}} \right|_{\theta_{\kappa} = - \pi / 8} \right)_{sim} \approx 0.06 ,
\end{eqnarray}
which is consistant with the experimental value. In fact, it would be unreasonable to expect perfect agreement considering we ignored the pinch term, took the large aspect ratio limit, and averaged over the entire outer region of the plasma. Furthermore, the elongated Cyclone base case geometry used for the simulations is considerably different than the geometry of TCV. Still, this comparison shows that neither the simulations nor the experimental results appear unreasonable.

In TCV, the introduction of up-down asymmetry increased the up-down symmetric rotation profiles by roughly 50\%. The rotation present in the up-down symmetric case was due to effects that are formally small in $\rho_{\ast}$. However, in larger machines, $\rho_{\ast}$ is smaller meaning external momentum injection appears less feasible. This means that the effect of up-down asymmetry would likely be much more significant in these larger devices. It may not be possible to access higher levels of rotation using intrinsic rotation, but it does seem that the level of rotation seen in current machines can be generated in future, reactor-sized devices by using up-down asymmetry.

\subsection{Estimation of rotation in ITER}

The more ambitious ITER operational scenarios are expect to violate beta limits, leading to resistive wall modes that must be stabilized \cite{PolevoiITERscenario2002, ShimadaITER2007}. Toroidal rotation is able to stabilize these modes, but only when the Alfv\'{e}n Mach number is a few percent \cite{LiuITERrwmStabilization2004}. In TCV, the introduction of a $\pi/8$ tilt changed the core rotation by about 50\% \cite{CamenenTCVExp2010}. However, in TCV, $\rho_{\ast} \approx 1/50$, which allows formally small mechanisms to induce rotation that competes with the effects of up-down asymmetry. In ITER, $\rho_{\ast} \approx 1/400$, so we expect all sources of intrinsic rotation (except up-down asymmetry) to be significantly reduced. This means, if ITER could be given a $\pi/8$ tilt, we would expect the effects of up-down asymmetry to dominate the rotation profile.

Here we will apply the results of this paper to show that intrinsic rotation induced by up-down asymmetry may be enough to stabilize the resistive wall mode in ITER. Using eqs. \refEq{eq:MomFluxEstimate} and \refEq{eq:HeatFluxEstimate} and normalizing the fluxes, we find that
\begin{eqnarray}
   \frac{\partial \Omega_{\zeta i}}{\partial r_{\psi}} \approx \frac{-1}{m_{i} R_{0} Pr_{i}} \left( \sqrt{\frac{2 m_{i}}{T_{i}}} \frac{\left\langle \Pi_{\zeta N i}^{ud} \right\rangle_{t}}{R_{0 N} \left\langle Q_{N i} \right\rangle_{t}} \right) \frac{\partial T_{i}}{\partial r_{\psi}} .
\end{eqnarray}
We will assume that the edge temperature and rotation are zero and that the on-axis temperature is $T_{i 0} = 18$ keV \cite{AymarITERSummary2001}. Additionally, we will take $\left\langle \Pi_{\zeta N i}^{ud} \right\rangle_{t} / \left( R_{0 N} \left\langle Q_{N i} \right\rangle_{t} \right)$ to be constant in minor radius, which seems reasonable given the results presented in Section \ref{subsubsec:Results}. We can now integrate to find the on-axis rotation to be
\begin{eqnarray}
   \Omega_{\zeta i 0} \approx \frac{- 2}{R_{0} Pr_{i}} \sqrt{\frac{2 T_{i 0}}{m_{i}}} \frac{\left\langle \Pi_{\zeta N i}^{ud} \right\rangle_{t}}{R_{0 N} \left\langle Q_{N i} \right\rangle_{t}} .
\end{eqnarray}

We can calculate the on-axis Mach number to be
\begin{eqnarray}
   M_{S} \equiv \frac{\left| u_{\zeta i} \right|}{v_{th i}} \approx \frac{2}{Pr_{i}} \frac{\left\langle \Pi_{\zeta N i}^{ud} \right\rangle_{t}}{R_{0 N} \left\langle Q_{N i} \right\rangle_{t}} .
\end{eqnarray}
We see that the momentum and heat transport caused by turbulence fundamentally sets the Mach number, but we care about the Alfv\'{e}n Mach number for stabilization of the resistive wall mode. Using the definition of the Alfv\'{e}n velocity, $v_{A} \equiv B_{0} / \sqrt{\mu_{0} n_{i} m_{i}}$, we can calculate the on-axis Alfv\'{e}n Mach number and to be
\begin{eqnarray}
   M_{A} \equiv \frac{\left| u_{\zeta i} \right|}{v_{A}} \approx \frac{\sqrt{2}}{Pr_{i}} \sqrt{\frac{4 \mu_{0} n_{i 0} T_{i 0} }{B_{0}^{2}}} \frac{\left\langle \Pi_{\zeta N i}^{ud} \right\rangle_{t}}{R_{0 N} \left\langle Q_{N i} \right\rangle_{t}} \approx \frac{\sqrt{2 \beta_{T 0}}}{Pr_{i}} \frac{\left\langle \Pi_{\zeta N i}^{ud} \right\rangle_{t}}{R_{0 N} \left\langle Q_{N i} \right\rangle_{t}} ,
\end{eqnarray}
where $\beta_{T 0}$ is the on-axis toroidal plasma beta. For expected ITER parameters, we find that $M_{S} \approx 10\%$ and $M_{A} \approx 1\%$. Additionally, the effect of the pinch may be able to increase the magnitude of rotation by as much as a factor of 3.

\section{Conclusions}
\label{sec:conclusions}

This paper analyzed the equilibrium and momentum transport characteristics of tokamaks with up-down asymmetric poloidal cross-sections.

The results of MHD equilibrium analysis (see Section \ref{sec:MHD}) demonstrated that external PF coils only have direct control over the outermost flux surface. Inside the plasma the toroidal current distribution has a significant effect on modifying the flux surface shape. It was shown that hollow current profiles are optimal for supporting up-down asymmetry to the magnetic axis. Furthermore, ellipticity, the lowest harmonic shaping effect, penetrated to the magnetic axis most effectively.

Section \ref{sec:GS2} detailed the modification and testing of GS2 to support the modeling of up-down asymmetric tokamak configurations.

This newly modified code was applied to model the turbulent momentum transport in tilted elliptical tokamaks (see Section \ref{sec:momTransport}). The nonlinear momentum flux simulations, shown in fig. \ref{fig:nonlinearAvgMomHeatRatio}, give rough quantitative agreement with TCV experimental results. They both predict $\left( 1 / v_{th i} \right) \partial u_{\zeta i} / \partial \rho$ to be approximately 5\% of $\left( 1 / T_{i} \right) \partial T_{i} / \partial \rho$ for elliptical flux surfaces with a $\pi/8$ tilt. We have also shown that, given a $\pi/8$ tilt, up-down asymmetry may be enough to stabilize the resistive wall mode in ITER.

Turbulent energy transport in tilted elliptical tokamaks shows a complex dependence on the tilt angle that is currently under investigation.

\ack

J.R.B. and F.I.P. were partially supported by U.S. DoE Grant No. DE-SC008435, by the RCUK Energy Programme (grant number EP/I501045) and by the European Union’s Horizon 2020 research and innovation programme. P.R. and N.F.L. were supported by EURATOM, within the framework of the European Fusion Development Agreement. Funda\c{c}\~{a}o para a Ci\^{e}ncia e Tecnologia also supported IST activities through project Pest-OE/SADG/LA0010/2011, and N.F.L. through grants IF/00530/2013 and PTDC/FIS/118187/2010. The computing time was provided by the National Energy Scientific Computing Center, supported by the Office of Science of the U.S. Department of Energy under Contract No. DE-AC02-05CH11231, and by the Helios supercomputer at IFERC-CSC under project GKMSC. The views and opinions expressed herein do not necessarily reflect those of the European Commission.

\section*{References}
\bibliographystyle{unsrt}
\bibliography{references.bib}

\end{document}